%% file: article.tex
\documentclass[twocolumn,rmp,preprintnumbers,amsmath,amssymb,floatfix]{revtex4}

\usepackage{graphicx}
\usepackage{dcolumn}
\usepackage{bm}

\usepackage[utf8]{inputenc}
\usepackage{amsfonts}
\usepackage{siunitx}
\usepackage{mathrsfs}
\usepackage{units}
\usepackage[section]{placeins}

\def\clap#1{\hbox to 0pt{\hss#1\hss}}

\def\mathrlap{\mathpalette\mathrlapinternal}

\def\mathrlapinternal#1#2{%
	\rlap{$\mathsurround=0pt#1{#2}$}}

\begin{document}
\title{Semirelativity in Semiconductors: a Review}
\author{Wlodek Zawadzki}
  \email{zawad@ifpan.edu.pl}
\affiliation{Institute of Physics, Polish Academy of Sciences, Aleja Lotnikow 32/46, PL-02668 Warsaw, Poland}

\begin{abstract}
An analogy between behavior of electrons in narrow-gap semiconductors (NGS) and relativistic electrons in vacuum is reviewed. Energy band structures $\varepsilon(\mathbf{k})$ are considered for various NGS  materials and their correspondence to the energy-momentum relation in special relativity is emphasized. It is indicated that special relativity for vacuum is analogous to a two-band $\mathbf{k}\cdot\mathbf{p}$ description for NGS. The maximum  electron velocity in NGS is $u\simeq1\times 10^8\si{cm/s}$, which corresponds to the light velocity in vacuum. An effective mass of charge carriers in semiconductors is introduced, relating their velocity to quasimomentum and it is shown that this mass depends on electron energy (or velocity) in a way similar to the mass of free relativistic electrons. In $\mathrm{Hg}_{1-x}\mathrm{Cd\vphantom{g}}_{x}\mathrm{Te}$ alloys one can reach vanishing energy gap at which electrons and light holes become three-dimensional massless Dirac fermions.  A wavelength $\lambda_z$ is defined for NGS, in analogy to the Compton wavelength in relativistic quantum mechanics. It is estimated that $\lambda_z$ is on the order of tens of Angstroms in typical semiconducting materials which is experimentally confirmed in tunneling experiments on energy dispersion in the forbidden gap. Statistical properties of the electron gas in NGS are calculated and their similarity is demonstrated to those of the Juttner gas of relativistic particles. Interband electron tunneling in NGS is described and shown to be in close analogy to the  predicted but unobserved tunneling between  negative and positive energies resulting from the Dirac equation for free electrons.

It is demonstrated that the relativistic analogy holds for orbital and spin properties of electrons in the presence of an external magnetic field. In particular, it is shown that the spin magnetic moment of both  NGS electrons and  relativistic electrons approaches zero with increasing  energy. This conclusion is confirmed experimentally for NGS.  Electrons in crossed electric and magnetic fields are described theoretically and experimentally. It is only the two-band description for NGS, equivalent to the Dirac or Klein-Gordon equations for free particles, that gives  a correct account of experimental results in this situation. A transverse Doppler shift  in the cyclotron resonance observed in crossed fields in InSb indicates that there exists a time dilatation between an oscillating electron and an observer. The phenomenon of Zitterbewegung (ZB, trembling motion) for electrons in NGS is considered theoretically, following the original proposition of Schr\"{o}dinger for free relativistic electrons in vacuum.  The two descriptions are in close analogy, but the frequency of ZB for electrons in NGS is orders of magnitude lower and its amplitude orders of magnitude higher making  possible experimental observations in semiconductors considerably more favorable. Finally, graphene and carbon nanotubes, as well as topological insulators are considered in the framework of relativistic analogy. These systems, with their linear energy-quasimomentum  dispersions, illustrate the extreme semirelativistic regime. Experimental results for the energy dispersions and the Landau quantizations in the presence of a magnetic field are quoted and their analogy to behavior of free relativistic electrons is discussed. Approximations and restrictions of the relativistic analogy are emphasized. On the other hand, it is indicated that in various situations it is considerably easier to observe  semirelativistic effects in semiconductors than the relativistic effects in vacuum.
\end{abstract}

\maketitle

\input{introduction}
\input{band_structure}
\input{band_consequences}
\input{statistics_of_electron_gas}
\input{interband_tunelling}
\input{e_in_mag_field}
\input{e_in_x_fields}
\input{zwitterbewegung}
\input{graphene}
\input{nanotubes}
\input{insulators}
\input{discussion}

\begin{acknowledgments}
It is my pleasure to acknowledge elucidating  discussions with  Drs.~Krzysztof~Dybko, Pawel~Pfeffer and Tomasz~M.~Rusin. I am obliged to Dr~Olek~Michalski for help in the preparation of the manuscript.  
\end{acknowledgments}

\end{document}

%% file: introduction.tex
\section{Introduction}
Since the beginning of solid state physics,  theoretical description of movable electrons and holes used physical quantities borrowed from free particles. In practice it meant that the motion of electrons and holes was described in analogy to free charged particles, with the free mass replaced by an effective mass which reflected the influence of periodic lattice potential. A  theoretical basis for the effective-mass approach was provided by \textcite{Luttinger1955} followed by \textcite{Zak1966}. It turned out, somewhat counter-intuitively, that the effective masses $m^\ast$ of electrons in semiconductors are considerably smaller than the free electron mass. In the presence of external magnetic field, spin magnetic moments of charge carriers come into play,  characterized by the corresponding spin $g$-factors. Effective values of $g^\ast$ in semiconductors can be dramatically different from the free-electron value $g=+2$ due to the effect of spin-orbit interaction and small energy gap, as first demonstrated by \textcite{Roth1959}.

However, the effects of periodic potential are not limited to the effective values of $m^\ast$ and $g^\ast$, particularly in narrow gap semiconductors (NGS). \textcite{Kane1957} was the first to show that, in the energy bands of narrow-gap material InSb, one deals with a nonquadratic $\varepsilon(k)$ dependence, where $\varepsilon$ is the electron energy and $\hslash k$ is the absolute value of quasimomentum. This feature, called band’s nonparabolicity, also results from small value of the gap and can be quite well described by a so called two-band model of interacting conduction and light-hole bands. It was remarked by \textcite{Keldysh1964}, \textcite{Wolff1964}, \textcite{Zawadzki1966}, \textcite{Aronov1967} that the two-band (or two-level) model of the $\mathbf{k}\cdot\mathbf{p}$ theory is analogous to the Dirac equation for relativistic electrons in vacuum.  For over 50 years this analogy between the behavior of electrons in narrow gap semiconductors and that of relativistic electrons in vacuum has been traced for various physical situations, culminating in recent years with the discoveries of graphene and topological insulators. The present article reviews this effort, see also \textcite{Zawadzki1970, Zawadzki1997}.
 
We describe various “semirelativistic” effects for electrons in semiconductors and indicate corresponding features  for relativistic electrons with appropriate references. We mention cases in which, in contrast to observations in semiconductors,  relativistic effects in vacuum have not been observed.  Some of the basic derivations for semiconductors are given without going into details, others are simplified in order to make basic ideas more easily accessible. Limitations of the relativistic analogy are emphasized.  An effort has been made to quote the important semiconductor work on the subject.

The review is organized as follows. First, we introduce the relativistic analogy and follow its consequences for semiconductor electrons in absence of external fields. Next, semiconductor electrons are considered in the presence of a magnetic field and in crossed electric and magnetic fields. In the third part we describe separately graphene, carbon nanotubes, and topological insulators, as they are characterized by reduced dimensionalities and possess quite special properties. The review is concluded by a discussion.

%% file: band_structure.tex
\section{\label{sec:band_struct}Band Structures in Semiconductors. Relativistic Analogy}
We begin by considering charge carriers in semiconductors in the absence of external fields. In this case one can use for their description the $\mathbf{k}\cdot\mathbf{p}$ theory \cite{Luttinger1955}. We present it here in a simple form neglecting spin effects, in order to come across with the basic idea. The initial eigenvalue equation for an electron in a periodic potential reads
\begin{equation}
	\label{eq:init_eigenval}
	\left[\frac{\hat{p}^2}{2m_0} + V_0(\mathbf{r})\right]\Psi=\varepsilon\Psi
\end{equation}
where $\mathbf{\hat{p}}$ is the momentum operator, $m_0$ is the free electron mass, $V_0(\mathbf{r})$ is the periodic potential of a crystal lattice and $\varepsilon$ is the energy. It is well known that solutions of Eq.~(\ref{eq:init_eigenval}) are the Bloch functions
\begin{equation}
	\label{eq:Bloch_fun}
	\Psi_{n\mathbf{k}} = e^{i \mathbf{k}\cdot\mathbf{r}} u_{n\mathbf{k}}(\mathbf{r})
\end{equation}
in which $\hslash\mathbf{k}$ is the quasimomentum and $u_{n\mathbf{k}}(\mathbf{r})$ is the periodic Bloch amplitude for the n’th energy band. The Bloch states are expanded in terms of the Luttinger-Kohn (LK) functions
\begin{equation}
	\label{eq:LK_fun}
	{\mathpalette{\raisebox{\depth}{$\chi$}}\relax}_{l\mathbf{k}} = e^{i \mathbf{k}\cdot\mathbf{r}} u_{l0}(\mathbf{r})
\end{equation}
where $u_{l0}$ are the periodic LK amplitudes given by the Bloch amplitudes taken at the $\mathbf{k}$ value corresponding  band’s extremum  (we take $\mathbf{k}=0$ for simplicity). It then follows from Eq. (\ref{eq:init_eigenval}) that the LK amplitudes satisfy the eigenvalue equations
\begin{equation}
	\label{eq:eigenval_solution}
	\left[\frac{\hat{p}^2}{2m_0} + V_0(\mathbf{r})\right] u_{l0} = \varepsilon_{l0} u_{l0}
\end{equation}
where $\varepsilon_{l0}$ is the energy of the $l$’th band at $\mathbf{k}=0$. The amplitudes $u_{l0}$ are orthonormal
\begin{equation}
	\label{eq:amp_orthonorm}
	\frac{1}{\Omega}\langle u_{l\smash{'}0} \vert u_{l0}\rangle = \delta_{l'l}
\end{equation}
where $\Omega$ is the volume of the unit cell and the integration is over $\Omega$. One proves that the LK functions  represent a complete set, with $l$ running over all energy bands. This means that one can expand the Bloch state for the $n$th band in terms of the LK functions (index $n$ is omitted on both sides)
\begin{equation}
	\label{eq:Bloch_expansion}
	\Psi_{\mathbf{k}} = \sum\limits_{l} c_l(\mathbf{k}) u^{i \mathbf{k}\cdot\mathbf{r}} u_{l0}(\mathbf{r})
\end{equation}
where the sum runs over all the bands $l$. One performs the elementary operations following from Eqs. (\ref{eq:init_eigenval}) and (\ref{eq:Bloch_expansion}), and uses Eq. (\ref{eq:eigenval_solution}). Multiplying both sides by $u^\ast_{l'0}/\Omega$ for $l'=1,2,3,\ldots$ and integrating over $\Omega$, one obtains the $\mathbf{k}\cdot\mathbf{p}$ theory in the form
\begin{equation}
	\label{eq:kp_theory}
	\sum\limits_l \left[\left({\varepsilon_{l0}-\varepsilon+\frac{\hslash^2 k^2}{2m_0}}\right)\delta_{l\smash{'} l} + \frac{\hslash\mathbf{k}}{m_0}\mathbf{p}_{l\smash{'} l}\right]c_l=0
\end{equation}
where $l'=1,2,3,\ldots$ and the momentum matrix elements are
\begin{equation}
	\label{eq:momentum_matrix_elements}
	\mathbf{p}_{l\smash{'}l} = \frac{1}{\Omega} \langle u_{l\smash{'}0} \vert\mathbf{\hat{p}}\vert u_{l0}\rangle .
\end{equation}

Equation (\ref{eq:kp_theory}) represents the $\mathbf{k}\cdot\mathbf{p}$ theory for the energies $\varepsilon_n(k)$ and the expansion coefficients $c^{n}_{l}$ for the Bloch state $\Psi_{nk}$. The phenomenological parameters are $\varepsilon_{l0}$ and $\mathbf{p}_{l\smash{'} l}$. If all the bands are included, set (\ref{eq:kp_theory}) contains no approximations.

It can be seen that, apart from the free-electron term $\hslash^2k^2/2m_0$, the structure of Eq. (\ref{eq:kp_theory}) looks very much like the Dirac equation (DE): it has band-edge energies on the diagonal and quasiomomenta $\hslash\mathbf{k}$ off the diagonal (in DE one deals with the momenta). This similarity provides the \emph{basis for the relativistic analogy}.

In its full form (\ref{eq:kp_theory}), the $\mathbf{k}\cdot\mathbf{p}$ matrix has infinite dimensions, so finding its complete solutions is not a practical task. Following \textcite{Luttinger1955} one can separate the energy bands by using second-order perturbation theory and show that, in the parabolic $\varepsilon(\mathbf{k})$ approximation, each band is characterized by an effective mass. However, one can take a different route, first indicated by \textcite{Kane1957}. Namely, one can neglect distant bands and find approximate solutions for a finite number of bands separated by small energy gaps. This procedure corresponds to the perturbation theory for nearly degenerate energy levels. The simplest approximation without spin is the two-band model (or two-level model) for InSb-type materials in which one takes into account at $\mathbf{k}=0$ the conduction level of s-like symmetry and the valence triple-degenerate level of p-like symmetry, separated from each other by the energy gap $\varepsilon_g$. This approximation is justified if the gap $\varepsilon_g$ is distinctly smaller than energy separations from other bands. We choose the zero of energy in the middle of the gap. Thus, at $\mathbf{k}=0$, one has for the conduction and valence levels $\varepsilon_g/2$ and $-\varepsilon_g/2$, respectively, the LK amplitudes
\begin{gather}
	u_{10}=iS,\\[1.2ex]
	u_{20}=(X-iY)/\sqrt{2},\quad
	u_{30}=Z,\quad u_{40}=(X+iY)/\sqrt{2}.
	\label{eq:LK_amplitudes}
\end{gather}
Taking into account symmetry of the momentum matrix elements, one obtains the reduced $\mathbf{k}\cdot\mathbf{p}$ matrix in the form
\begin{widetext}
\begin{equation}
	\left[
		\begin{array}{cccc}
			-\varepsilon'+\varepsilon_g/2 & C\hslash k_{-} & C\hslash k_{z} & C\hslash k_{+} \\
			C\hslash k_{+} & -\varepsilon'-\varepsilon_g/2 & 0 & 0\\
			C\hslash k_{\mathrlap{z}\phantom{+}} & 0 & -\varepsilon'-\varepsilon_g/2 & 0\\
			C\hslash k_{-} & 0 & 0 & -\varepsilon'-\varepsilon_g/2
		\end{array}
	\right]\left[
		\begin{array}{c}
			c_1\\ c_2\\	c_3\\ c_4
		\end{array}
	\right] = 0\quad ,
\label{eq:reduced_kp_matrix}
\end{equation}
\end{widetext}
where $C=(-i/m_0)\langle S\vert \hat{p}_x\vert X\rangle$ is a real number, $\varepsilon'=\varepsilon-\hslash^2 k^2/2m_0$ and $k_{\pm}= (k_x\pm ik_y)/\sqrt{2}$. One solves set (\ref{eq:reduced_kp_matrix}) and obtains four energy roots
\begin{align}
	\label{eq:energy_roots1}
	\varepsilon'_{1,2} &= \pm \left[ {\left(\frac{\varepsilon_g}{2}\right)^2 + \varepsilon_g\frac{\hslash^2 k^2}{2m^\ast_0}} \right]^{\frac{1}{2}}\\
	\label{eq:energy_roots2}
	\varepsilon'_{3,4} &= -\varepsilon_g/2,\\[1.2ex]
	\quad\frac{1}{m^\ast_0} &= \frac{2 C^2}{\varepsilon_g}.
	\label{eq:energy_roots3}
\end{align}
The roots $\varepsilon'_{1,2}$ correspond to the conduction and light-hole bands, while $\varepsilon'_{3,4}$ correspond to two heavy-hole bands. It turns out that $1/m^\ast_0$ of Eq. (\ref{eq:energy_roots3}) is usually much larger than $1/m_0$, so that one can neglect the free-electron term in Eq. (\ref{eq:energy_roots1}) and put $\varepsilon'(k)\simeq\varepsilon(k)$ and $\varepsilon'_{3,4}\simeq-\varepsilon_g/2$. In this approximation we have finally for the conduction and light-hole bands
\begin{equation}
	\varepsilon_{1,2} = \pm \left[ {\left(\frac{\varepsilon_g}{2}\right)^2 + \varepsilon_g	\frac{\hslash^2 k^2}{2m^\ast_0}} \right]^{\frac{1}{2}}
	\label{eq:lh_bands_conduction_approx}
\end{equation}
where $m^\ast_0$, given by Eq. (\ref{eq:energy_roots3}), is the effective mass of electrons and light holes. In the above model the resulting bands are spherical, i.e. the energy depends on the absolute value of $k$, not on its direction.

\begin{figure}[!t]
	\includegraphics[width=0.5\textwidth]{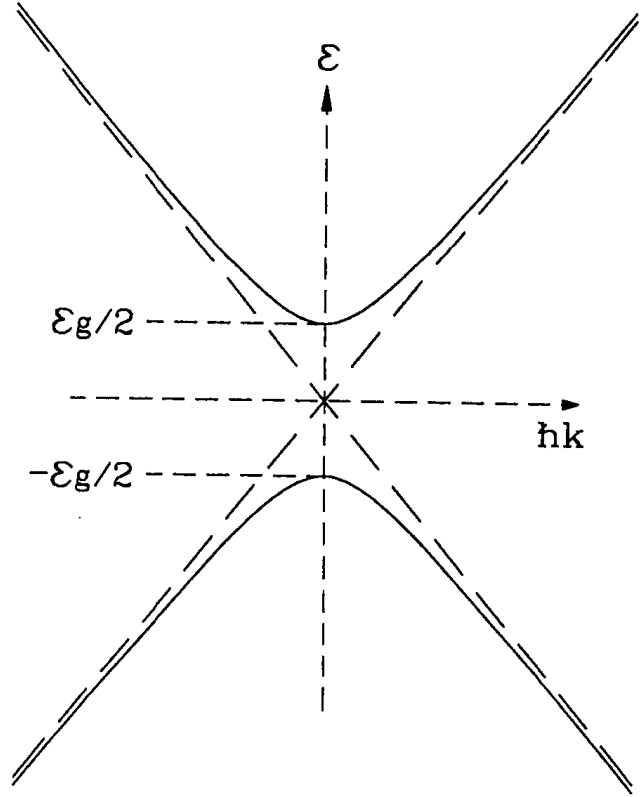}
	\caption{\label{fig:Dir-NGS}Energy versus quasimomentum for the two-band model of a semiconductor
	(schematically). For the Dirac equation the gap is $2m_0c^2$ and the “bottom of the band” mass is $m_0$.}
\end{figure}

Equation (\ref{eq:lh_bands_conduction_approx}) presents an important result. First, for not too large $k$ values one can expand the square root and take the first two terms. This gives $\varepsilon_{1,2}=\pm(\varepsilon_g/2+\hslash^2 k^2/2m^\ast_0)$, which is the standard parabolic approximation to the conduction and light-hole bands. Second, for sufficiently large $k$ values one can neglect the first term in the square root and obtain $\varepsilon_{1,2}=\pm(\varepsilon_g/2m^\ast_0)^{\nicefrac{1}{2}}\hslash k$. This linear $k$ approximation is valid for not too large $k$ values since it is known that, near the Brillouin zone boundaries, the $\varepsilon(k)$ dependence must be horizontal. Third, and this is our essential point, \emph{formula (\ref{eq:lh_bands_conduction_approx})  has the “relativistic” character}. The energy-momentum relation resulting from the Dirac equation for relativistic electrons in vacuum is \cite{Dirac1958}
\begin{equation}
	\varepsilon = \pm \left[{\left(\frac{2m_0c^2}{2}\right)^2+2m_0c^2\frac{p^2}{2m_0}}\right]^\frac{1}{2}.
	\label{eq:Dirac_rel_e}
\end{equation}

It is seen that the dispersion relations (\ref{eq:lh_bands_conduction_approx}) and (\ref{eq:Dirac_rel_e}) have identical forms with the following correspondence
\begin{align}
	\label{eq:disp_rel1}
	2m_0c^2 &\rightarrow \varepsilon_g\\
	\label{eq:disp_rel2}
	m_0 &\rightarrow m^\ast_0\\
	\label{eq:disp_rel3}
	\mathbf{p} &\rightarrow \hslash\mathbf{k}
\end{align}

The correspondence  expressed in Eqs.~(\ref{eq:lh_bands_conduction_approx}) and (\ref{eq:Dirac_rel_e}) is illustrated in Fig.~\ref{fig:Dir-NGS}. Energy gaps in  semiconductors are many orders of magnitudes smaller than $2m_0c^2 \simeq \SI{1}{\mega\electronvolt}$ and the effective masses $m^\ast_0$ are considerably smaller than $m_0$. Expression (\ref{eq:disp_rel3}) suggests that the quasimomentum of semiconductor electrons corresponds to the momentum of electrons in vacuum. We emphasize that, while the quasimomentum has indeed many properties of the momentum, this is not always the case, as we will show below in the description of Zitterbewegung. The quasimomentum appears for electrons in solids instead of momentum because in the presence of a periodic potential the Bloch states are characterized by this quantity, see Eq.~(\ref{eq:LK_fun}) and \textcite{Zawadzki2013}.

One can use the same procedure for HgTe-type of materials: $\alpha$-Sn, HgSe, HgS, in which the p-like triple-degenerate level is above the s-like level, see \textcite{Groves1963}. One obtains solutions of this problem  by simply replacing $\varepsilon_g$ by $-\varepsilon_0$, where $\varepsilon_0$ is positive. In this case the relativistic analogy still holds for the conduction and light-hole bands. It is seen that, as far as the nonparabolicity of the bands is concerned, the role of $\varepsilon_g$ is now played by $\varepsilon_0$, often called “the interaction gap”.  The main difference from the previous case is that now at $\mathbf{k}=0$ the conduction band is degenerate with two heavy-hole bands. For this reason the HgTe-type materials are often called  zero-gap semiconductors.

We are now in a position to enumerate approximations and restrictions necessary to use the semirelativistic analogy of Eqs.~(\ref{eq:disp_rel1})~--~(\ref{eq:disp_rel3}). As to approximations, we have neglected distant  bands keeping only one s-like and one  p-like level. This results in the two-band model, the heavy holes also appearing but not being a part of the picture. We have also neglected the free electron term in Eq.~(\ref{eq:lh_bands_conduction_approx}). As to restrictions, the two-band model is valid for not too large $k$ values, approximately up the inflection points on the complete $\varepsilon(k)$ curve. Second, the sphericity of the bands described in Eq.~(\ref{eq:lh_bands_conduction_approx}) is related to the fact that in III-V InSb-like and II-VI HgTe-like materials the band extrema of our interest occur in the center of the Brillouin zone (the $\Gamma$ point). In other materials, for example in IV-VI lead chalcogenides, this may not be the case. Finally, as we show below,  in semiconductors one does not deal in reality with relativistic electron velocities, so that the relativistic analogy has a rather formal character. Still, it leads to many physical similarities.

\begin{figure}[h]
	\includegraphics[width=0.5\textwidth]{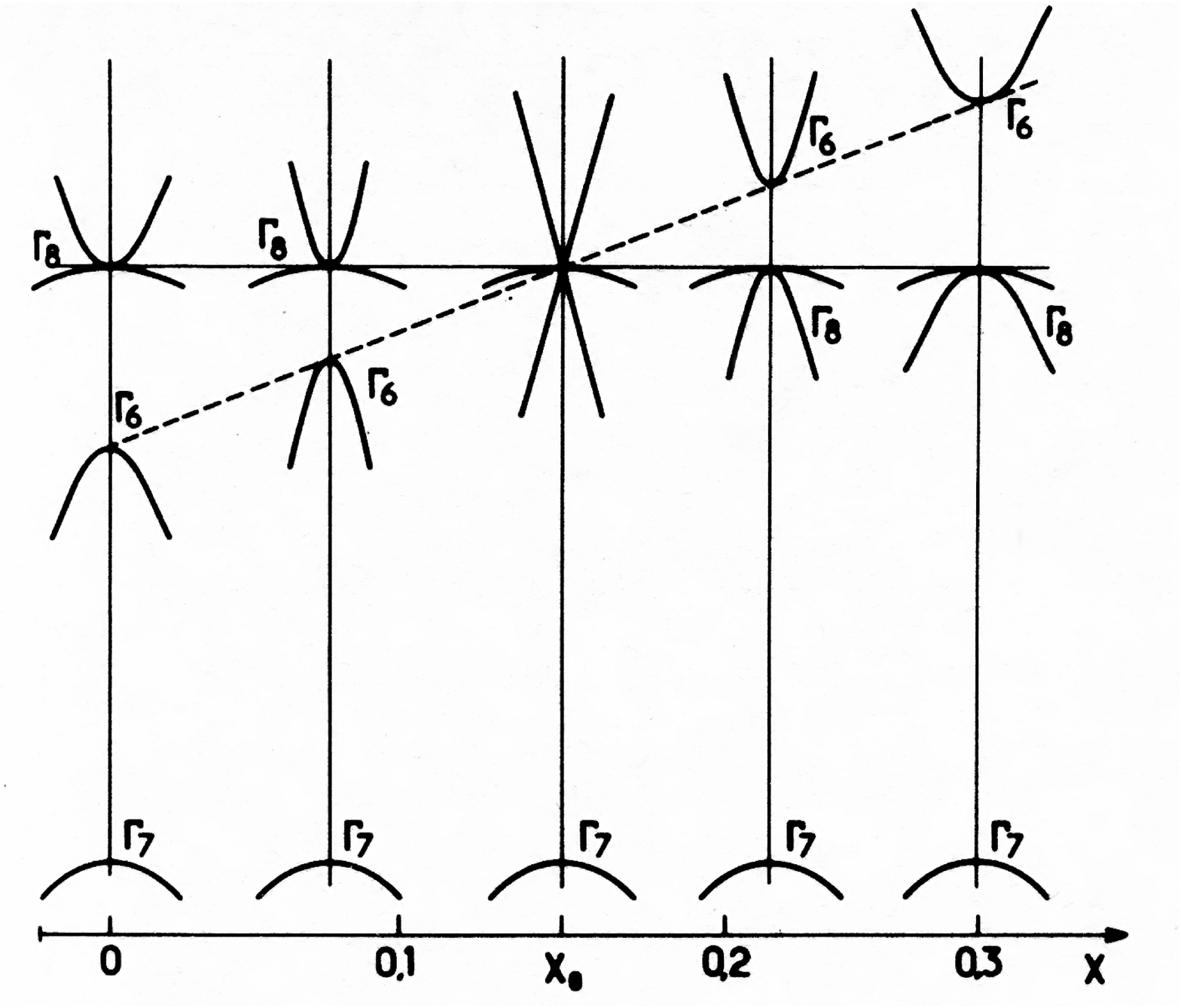}
	\caption{\label{fig:bnstrHCT}Band structures of mixed crystals $\mathrm{H\smash{g}}_{1-x}\mathrm{Cd}_{x}\mathrm{Te}$ for increasing composition $x$ which change from the “reversed” band ordering in HgTe to the “straight” ordering in CdTe (schematically). For zero interaction gap between $\Gamma_8$ and $\Gamma_6$ bands one deals with linear $\varepsilon(k)$ dispersions and three-dimensional “massless” electrons and light holes.}
\end{figure}

Ternary alloys $\mathrm{Hg}_{1-x}\mathrm{Cd\vphantom{g}}_{x}\mathrm{Te}$ offer an interesting possibility which also can be interpreted in terms of the relativistic analogy. CdTe has the “straight” band ordering of InSb-type while HgTe has the “inverted” band ordering. By producing ternary alloys of these two materials in different chemical proportions one can go continuously  from one band structure to the other. In particular, one can reach the situation in which the energy separation between s-like and p-like levels, i.e. the interaction gap, is exactly zero, see e.g. \textcite{Dornhaus1976}. This occurs for $x=0.165$ at \SI{300}{\kelvin}. The two-band model describes this case very well. For $\varepsilon_g=0$ the first term under the square root in Eq.~(\ref{eq:lh_bands_conduction_approx}) vanishes. In the second term the gap cancels out, see Eq.~(\ref{eq:energy_roots3}), and the dispersion becomes
\begin{equation}
	\label{eq:band_dispersion}
	\varepsilon(k) = \pm C \hslash k
\end{equation}
i.e. it is linear in $k$. In terms of the relativistic analogy this means that one deals with the “extreme relativistic" situation. The transition from straight band ordering to the inverted one in the HgCdTe system is illustrated in Fig.~\ref{fig:bnstrHCT}.

In narrow gap lead chalcogenides PbTe, PbSe, PbS and their ternary alloys the conduction band minima and valence band maxima are at the L points of the Brillouin zone. As a consequence, the $\varepsilon(\mathbf{k})$ dispersion relations are not spherical but spheroidal, the complete bands having cubic symmetry. However, the two-band $\mathbf{k}\cdot\mathbf{p}$ description is still valid for not too high electron energies. Choosing the zero of energy in the middle of the gap one has, see  \textcite{Dimmock1964}
\begin{equation}
	\label{eq:2band_kp}
	\varepsilon=\pm\left[{\left(\frac{\varepsilon_g}{2}\right)^2
	+\varepsilon_g\left({\frac{\hslash^2k_\perp^2}{2m^\ast_{\perp 0}}
	+\frac{\hslash^2k_\parallel^2}{2m^\ast_{\parallel 0}}}\right)}\right]^{\frac{1}{2}}\quad ,
\end{equation}
where $m^\ast_{\perp0}$ and $m^\ast_{\parallel0}$ effective masses describe the band ellipsoids. One can obtain the spherical description in $k$ by performing a transformation of $k_\perp$ and $k_\parallel$ components, see \textcite{Aronov1967}.

Ternary alloys of PbSnSe and PbSnTe also offer a possibility of going through zero energy gap, see \textcite{Strauss1967}, \textcite{Nimtz}. As follows from Eq.~(\ref{eq:2band_kp}), in this situation one also deals with the linear dispersion $\varepsilon(k)$. Since in lead salts the band structure in each material depends strongly on the temperature, one can reach vanishing gaps as functions of $T$.

We conclude this section by remarking that the band descriptions of Eqs.~(\ref{eq:lh_bands_conduction_approx}) and (\ref{eq:band_dispersion}) offer better approximations to real $\varepsilon(k)$ dispersions than may appear at first glance, see \textcite{Zawadzki1964}. Suppose one deals with a spherical energy band of an almost arbitrary form $\varepsilon(k)$ having a minimum at the $k=0$ point. This occurs, for example, when one takes into account more bands in the $\mathbf{k}\cdot\mathbf{p}$ theory. In this situation one can expand quite generally the $k^2(\varepsilon)$ dependence in a power series of energy
\begin{equation}
	\label{eq:power_series_energy}
	k^2(\varepsilon)= \lambda_0+\lambda_1\varepsilon
	+\lambda_2\varepsilon^2+\lambda_3\varepsilon^3+\ldots\;.
\end{equation}
It is seen that, taking the first two terms of this expansion, one obtains the standard parabolic description of an energy band. However, taking the first three terms and solving the resulting quadratic equation for the energy one obtains an $\varepsilon(k)$ relation of the type given by Eq.~(\ref{eq:lh_bands_conduction_approx}). The conclusion of this reasoning is that the first nonparabolic approximation to any $\varepsilon(k)$ dependence is always of the relativistic type.  Also the linear $\varepsilon(k)$ dispersions are accounted for by taking $\lambda_0=\lambda_1=0$ and $\lambda_2\neq0$. Thus, the above two-band formulas are just special cases of the general rule. It is clear that the above phenomenological expansion is valid also for two-dimensional and one-dimensional systems. 

%% file: band_consequences.tex
\section{Consequences of Band Structure}
We now follow the first physical consequences of the semirelativistic $\varepsilon(k)$ relation given by Eq.~(\ref{eq:lh_bands_conduction_approx}).
\subsection{Effective Mass}
We begin with the important concept of the effective mass of charge carriers. In standard textbooks on semiconductors, the effective mass of electrons or holes is defined by a relation between force and acceleration. This leads to the inverse mass given by the second derivative of the energy with respect to quasimomentum, see below. However, we show that a considerably more useful definition of the  mass $m^\ast$ comes from a relation between the electron velocity $\mathbf{v}$ and its quasimomentum $\hslash\mathbf{k}$. Thus we define the “velocity” effective mass
\begin{equation}
	\label{eq:vel_eff_mass}
	\hat{m}^\ast\mathbf{v}=\hslash\mathbf{k},
\end{equation}
where $\hat{m}^\ast$ is generally a tensor. There is
\begin{equation}
	\label{eq:vel_rel_1}
	v_i = \frac{\partial\varepsilon}{\partial\hslash k_i}
	= \frac{1}{\hslash}\frac{d\varepsilon}{dk}\frac{k_i}{k}
	= \frac{1}{\hslash k}\frac{d\varepsilon}{dk}\sum\limits_jk_j\delta_{ij},
\end{equation}
where the summation is over the coordinates:  $j = 1, 2, 3$. On the other hand, there is
\begin{equation}
	\label{eq:vel_rel_2}
	v_i = \hslash\sum\limits_jk_j\left(m^\ast\right)^{-1}_{ij}
\end{equation}
in which $\left(m^\ast\right)^{-1}_{ij}$ is the inverse mass tensor. Comparing  Eqs.~(\ref{eq:vel_rel_1}) and (\ref{eq:vel_rel_2}) one obtains
\begin{equation}
	\label{eq:scalar_mass_exp}
	\left(m^\ast\right)^{-1}_{ij} = \frac{1}{\hslash^2k}\frac{d\varepsilon}{dk}\delta_{ij},
\end{equation}
where $\delta_{ij}$ is the Kronecker delta. This means that the inverse tensor of the effective mass for a spherical energy band is a scalar. As a consequence, Eq.~(\ref{eq:scalar_mass_exp}) becomes
\begin{equation}
	\label{eq:scalar_mass}
	\frac{1}{m^\ast} = \frac{1}{\hslash^2k}\frac{d\varepsilon}{dk}.
\end{equation}

The constitutive equality (\ref{eq:vel_eff_mass}) with the scalar effective mass $m^\ast$ takes the form
\begin{equation}
	\label{eq:scalar_vel_eff_mass}
	{m}^\ast\mathbf{v}=\hslash\mathbf{k}.
\end{equation}
By writing down the three components of the above equation, squaring them , adding and taking the square root of both sides, one shows that the relation (\ref{eq:scalar_vel_eff_mass}) holds also for the absolute values of velocity  and quasimomentum, i.e.
\begin{equation}
	\label{eq:absolute_vel_eff_mass}
	m^\ast v = \hslash k
\end{equation}
The above velocity mass appears in the cyclotron resonance and, most important, in the transport theory. On the other hand, the mass relating force to acceleration: $\mathbf{F}=\hat{M}^\ast\mathbf{a}$, is given by
\begin{equation}
	\left({M^\ast}\right)^{-1}_{ij}
	= \frac{1}{\hslash^2}\frac{\partial^2\varepsilon}{\partial k_i\partial k_j}.
\end{equation}
The mass $M^\ast$ is less useful than the mass $m^\ast$ defined above. In particular, it is not a scalar for a spherical energy  band.

We can now determine the maximum electron velocity in energy bands described by the two-level model. One can use Eq.~(\ref{eq:lh_bands_conduction_approx}), calculate the velocity from Eq.~(\ref{eq:vel_rel_1}) and take its limit for large $k$ values. But a simpler way is to use directly the correspondence of Eqs.~(\ref{eq:disp_rel1}) and (\ref{eq:disp_rel2}) and obtain the same result. According to relativity, the maximum electron velocity is $c$. We can write
\begin{equation}
	\label{eq:max_electron_vel}
	c = \left({2m_0c^2/2m_0}\right)^\frac{1}{2}
	\rightarrow \left({\varepsilon_g/2m^\ast_0}\right)^\frac{1}{2} = u.
\end{equation}
Clearly, one obtains the same for the light holes. The maximum velocity $u$ in a material can be estimated  by using experimentally determined values of $\varepsilon_g$ and $m^\ast_0$. It turns out that the resulting value is $u\simeq\SI{e8}{\centi\meter/\second}$. Further, this value  is very similar for various semiconducting materials. Thus $u$ is about 300 times smaller than the light velocity $c$ indicating that we do not deal in semiconductors with truly relativistic velocities. The value of $u$ is reasonable since typical electron velocities in semiconductors are of the order of \SI{e7}{\centi\meter/\second}. By using $u$ one can write the dispersion relation (\ref{eq:lh_bands_conduction_approx}) in the form
\begin{equation}
	\varepsilon = \pm \left[{(m^\ast_0u^2)^2+u^2(\hslash k)^2}\right]^\frac{1}{2},
\end{equation}
which is even more reminiscent of the relativistic formula (\ref{eq:Dirac_rel_e}), with $u$ replacing $c$ and $\hslash k$ replacing $p$.
It is known that in special relativity one also defines the mass by the relation between velocity and momentum: $\mathbf{p} = m\mathbf{v}$. This velocity mass is a scalar and it depends on the velocity (or energy). The relativistic increase of the electron mass with growing velocity has been observed in many experiments, see e.g. \textcite{Guye1915}, \textcite{Rogers1940}, \textcite{Kaplan1955}. Objections concerning the use of velocity-dependent mass in the theory of relativity have been expressed in the literature, as such a mass does not have proper transformation properties, see e.g. \textcite{Ginzburg1979}. However, in semiconductors the energy-dependent effective mass is an important physical quantity.

One can determine an explicit form of the velocity mass within the two-band model. Calculating $d\varepsilon/dk=\varepsilon_g\hslash^2k/(\varepsilon2m^\ast_0)$ and using Eq.~(\ref{eq:max_electron_vel}) for $u$, we get
\begin{equation}
	\label{eq:explicit_vel_mass}
	m^\ast(\varepsilon) = m^\ast_0(2\varepsilon/\varepsilon_g)=\varepsilon/u^2.
\end{equation}
Thus the mass is proportional to the energy counted from the middle of the gap. It then
follows
\begin{equation}
	\label{eq:mass_energy_proportion}
	\varepsilon=m^\ast u^2.
\end{equation}
This formula is in a \emph{direct analogy with the famous relation of special relativity:} $\mathscr{E}=mc^2$. In both expressions the energy is given by the velocity mass multiplied by the square of maximum velocity.

In the theory of relativity one often quotes the dependence of the mass  on velocity: $m(v)=m_0(1-v^2/c^2)^{-\nicefrac{1}{2}}$. We can derive an analogous formula for the effective mass of carriers in a spherical energy band within the two-band model. Equation~(\ref{eq:lh_bands_conduction_approx}) gives: $v_i = (u^2/\varepsilon)\hslash k_i$, from  which one easily gets
\begin{equation}
	v^2\varepsilon^2/u^4 = (\hslash k)^2.
\end{equation}
On the other hand
\begin{equation}
	\frac{v^2\varepsilon^2}{u^2}= {m^\ast_0}^2v^2+\frac{(\hslash k)^2v^2}{u^2}.
\end{equation}
Combing this with Eq.~(\ref{eq:mass_energy_proportion}) and using the definition (\ref{eq:absolute_vel_eff_mass}) for the effective mass $m^\ast$, we obtain
\begin{equation}
	m^\ast_0v(1-v^2/u^2)=\hslash k,
\end{equation}
from which it follows
\begin{equation}
	m^\ast = m^\ast_0(1-v^2/u^2)^{-\frac{1}{2}},
\end{equation}
which is in direct analogy with the relativistic formula quoted above.

\begin{figure}
	\includegraphics[width=0.5\textwidth]{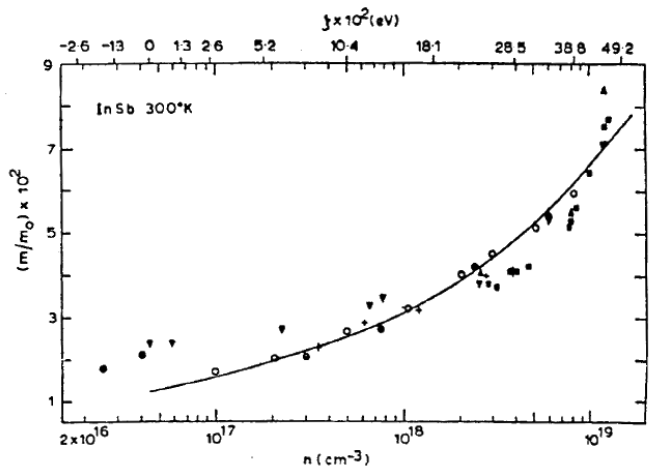}
	\caption{\label{fig:mInSb}Electron effective mass in InSb versus free electron density, as measured at room temperature by various authors. Solid line, calculated for the two-band model, represents  mass values at the Fermi energy. The latter, counted from the band edge, is indicated on the upper abscissa. After \textcite{Zawadzki1974}.}
\end{figure}

Now we give a few experimental examples for the energy-dependent effective masses of electrons in semiconductors which illustrate their semirelativistic behavior. At low temperatures the mass is usually measured at the Fermi energy within the band. As a consequence, one often plots the mass as a function of the electron density which determines the Fermi energy.  Figure \ref{fig:mInSb} shows the effective masses of electrons in InSb, as measured by various authors,  plotted versus the electron density $n$. The corresponding values of the Fermi energy, as counted from the conduction band edge, are indicated  on the upper  abscissa. It is seen that the experimental values are in good agreement with the predictions of the two-band model. Somewhat too high experimental values at low densities result from the fact that, at the room temperature, the electron gas is not degenerate, so the average measured masses are higher than those at the Fermi energy. At the highest densities $n = \SI{e19}{cm^{-3}}$ the Fermi energy is about \SI{0.5}{eV} above the band edge while the gap value at \SI{300}{K} is around \SI{0.18}{eV}, so that one is well within the  nonparabolic (or semirelativistic) regime.

The situation is qualitatively similar for electrons in GaAs. This material is a medium-gap semiconductor, having $\varepsilon_g\simeq \SI{1.5}{eV}$. As the electron density $n$ increases from \num{e16} to \SI{e20}{cm^{-3}}, the electron effective mass $m^\ast$ at the Fermi energy increases from $0.065\,m_0$ to $0.13\,m_0$, see \textcite{Pfeffer1990}. This example illustrates the fact that, what matters for the semirelativity is not so much an absolute value of the gap, but rather the ratio of electron energy above the band edge to the gap value. Electrons in any material will exhibit semirelativistic properties if excited sufficiently high into the band.

\begin{figure}
	\includegraphics[width=0.5\textwidth]{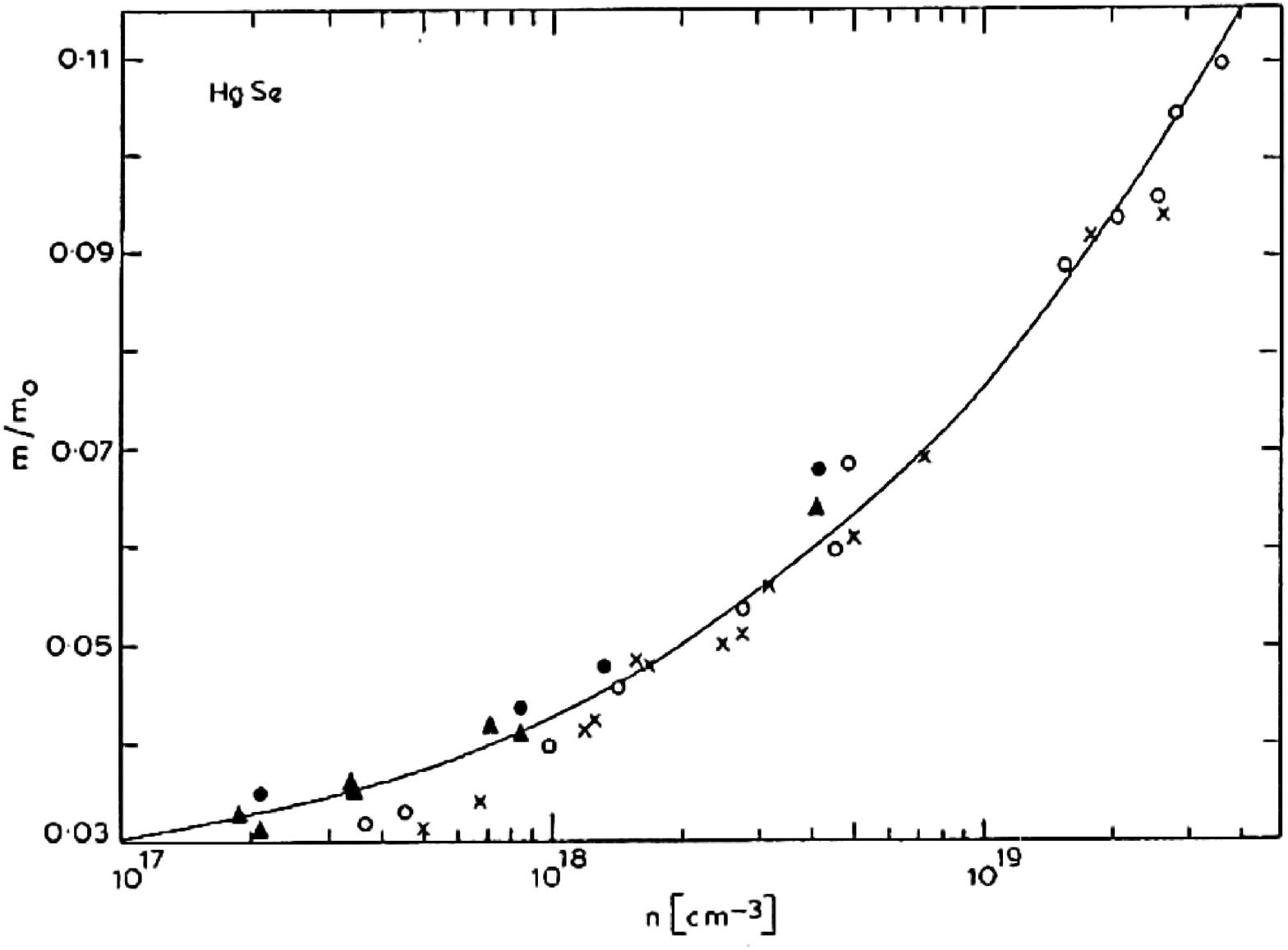}
	\caption{\label{fig:mHgSe}Electron effective mass in HgSe versus free electron density. The solid line is calculated for the “inverted” band model. After \textcite{Konczykowski}, see also \textcite{Zawadzki1974}. }
\end{figure}

\begin{figure}
	\includegraphics[width=0.5\textwidth]{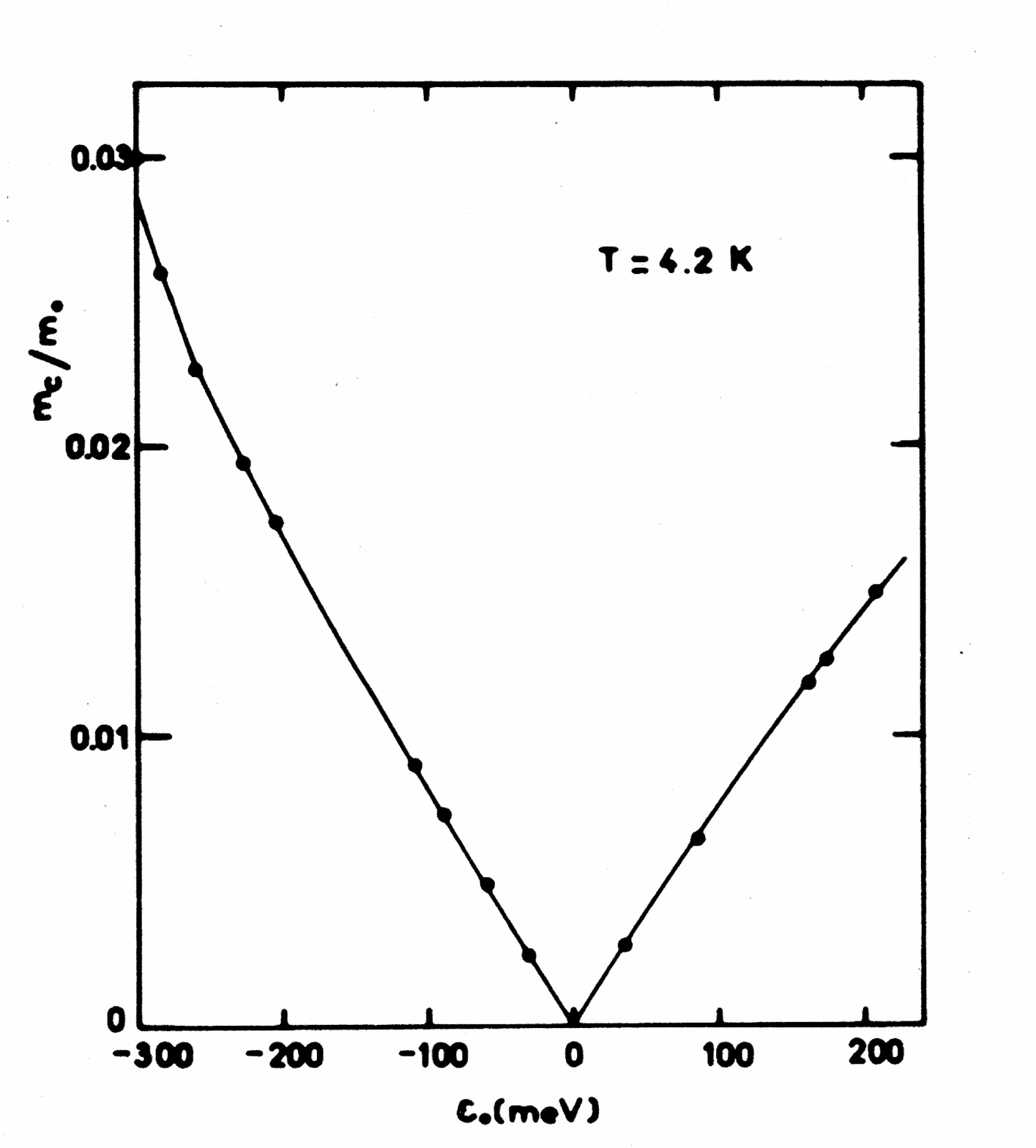}
	\caption{\label{fig:mHCTrig}Band-edge electron effective mass at $T = \SI{4.2}{K}$ versus the interaction gap in $\mathrm{H\smash{g}}_{1-x}\mathrm{Cd}_{x}\mathrm{Te}$ alloy system. At $x =0.165$ the mass goes to zero. The value of $\varepsilon_0=\SI{-300}{meV}$ corresponds to HgTe. After \textcite{Guldner1977}, see also \textcite{Rigaux}.}
\end{figure}

\begin{figure}
	\includegraphics[width=0.5\textwidth]{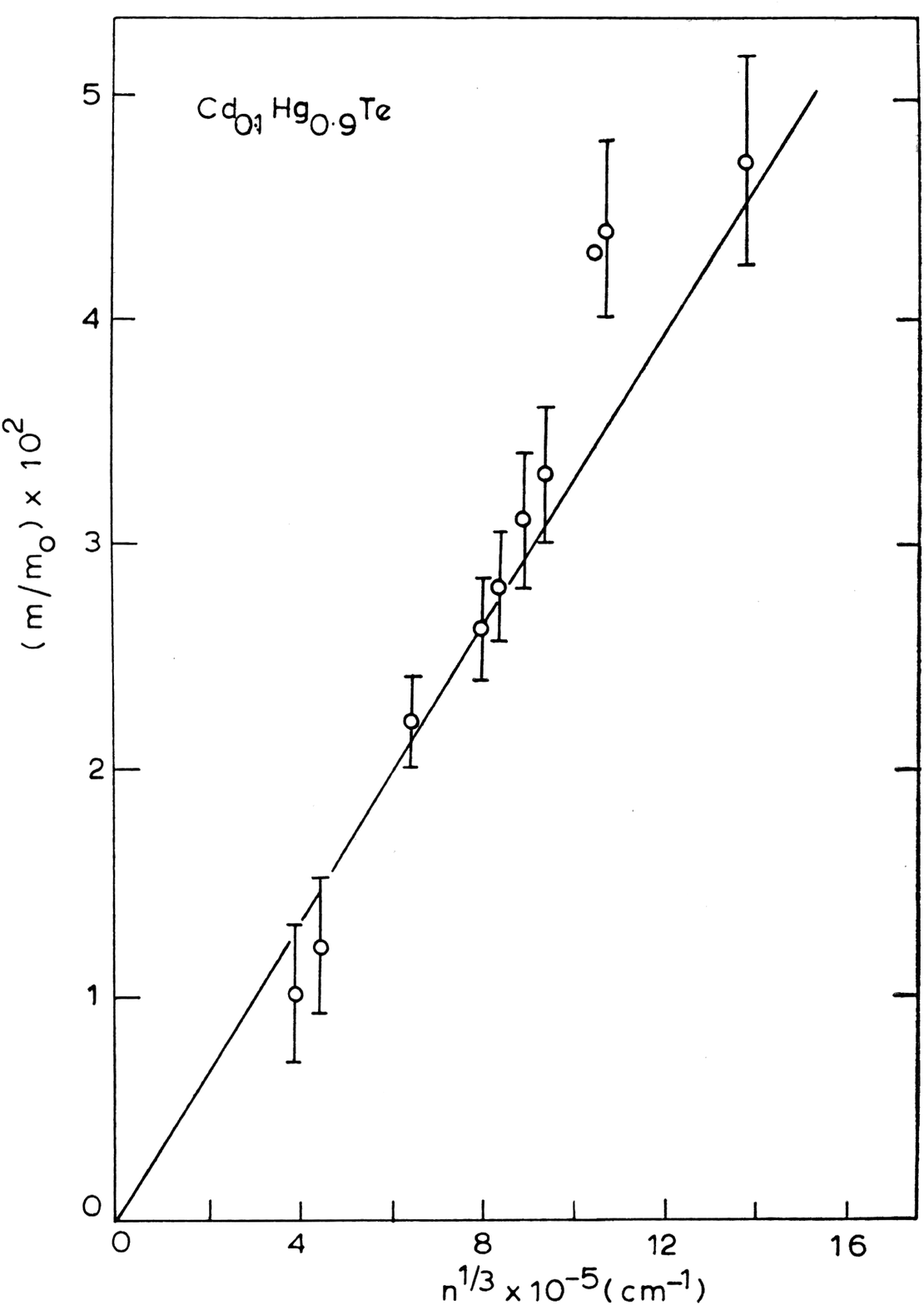}
	\caption{\label{fig:GalSos}Electron effective mass in $\mathrm{H\smash{g}}_{0.9}\mathrm{Cd}_{0.1}\mathrm{Te}$ at $T = \SI{300}{K}$ versus electron density $n^{\nicefrac{1}{3}}$ demonstrating linear variation of the mass with quasimomentum, see Eq.~(\ref{eq:eff_mass}). This corresponds to the “extreme relativistic” regime. After \textcite{Galazka1967}, see also \textcite{Zawadzki1974}.}
\end{figure}

The situation is quite similar for the “inverted” band structure of HgTe-type semiconductors, as discussed above. Figure \ref{fig:mHgSe} illustrates the increase of electron mass in HgSe. In this material one cannot reach electron densities below \SI{2e17}{cm^{-3}}.

As we mentioned above, in HgCdTe one can have a vanishing interaction gap, see Fig.~\ref{fig:Dir-NGS}. For the resulting “ultrarelativistic” linear $\varepsilon=u\hslash k$ dependence, see Eq.~(\ref{eq:band_dispersion}), one obtains from Eq.~(\ref{eq:absolute_vel_eff_mass}) the effective mass
\begin{equation}
	\label{eq:eff_mass}
	m^\ast=\varepsilon/u^2=\hslash k/u.
\end{equation}
It follows that for $\varepsilon=0$ the mass vanishes. Thus, at the band edge one deals with three-dimensional “massless fermions”. However, as the energy increases the mass is nonzero. From Eq. (\ref{eq:eff_mass}) we  obtain again the relation $\varepsilon=m^\ast u^2$. For the vanishing gap the mass is a linear function of quasimomentum $\hslash k$. This relation can be directly verified.

Equation (\ref{eq:eff_mass}) indicates that the effective mass at the band edge is proportional to the interaction gap $\varepsilon_g$ (or $\varepsilon_0$). Figure~\ref{fig:mHCTrig} shows the electron mass $m^\ast_0$ in $\mathrm{Hg}_{1-x}\mathrm{Cd\vphantom{g}}_{x}\mathrm{Te}$ at $T=\SI{4.2}{K}$ versus the interaction gap  going through zero for chemical composition $x \simeq 0.165$. On the other hand, Fig.~\ref{fig:GalSos} illustrates  dependence of the mass on the wave vector $k$ for HgCdTe in the zero-gap situation, as given by Eq.~(\ref{eq:eff_mass}). For the degenerate electron gas the electron density is $n=(1/3\pi^2)k^3$, so that $k=(3\pi^2n)^{\nicefrac{1}{3}}$. It is seen that, indeed, in agreement with Eq.~(\ref{eq:eff_mass}), the mass is linear in $n^{\nicefrac{1}{3}}$, i.e. it is proportional to $k$.

\begin{figure}
	\includegraphics[width=0.5\textwidth]{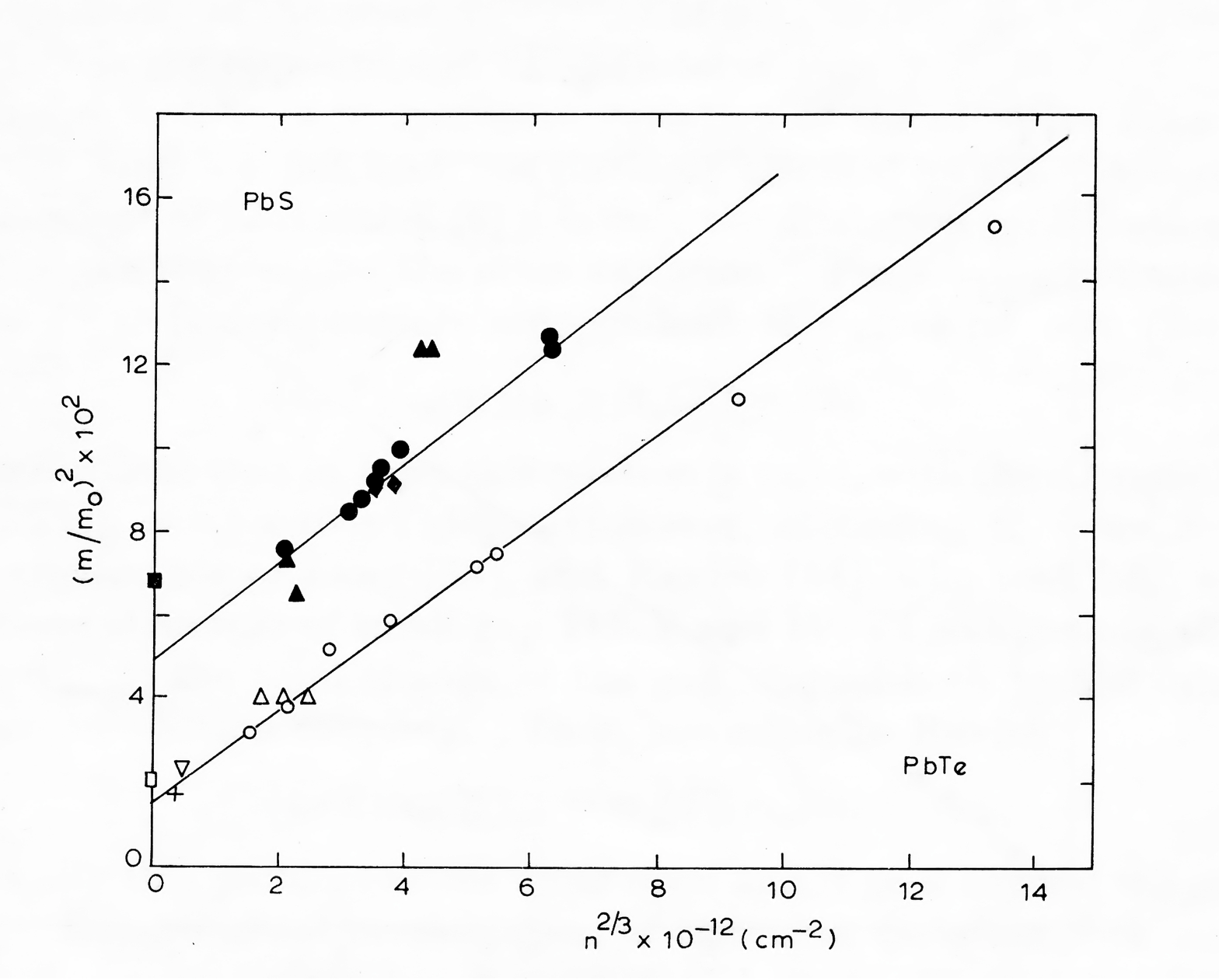}
	\caption{\label{fig:mPbSe}Density-of-states mass ${m^\ast_d}^2$ in PbS and PbTe at $T=\SI{77}{K}$ versus  free electron density $n^{\nicefrac{2}{3}}$.  Experimental data taken by various authors, straight lines calculated for the generalized two-band model. After \textcite{Zawadzki1974}.}
\end{figure}

As we mentioned before, in lead chalcogenides and their mixed crystals the bands are ellipsoidal and one deals with transverse and parallel effective masses, see Eq.~(\ref{eq:2band_kp}). It follows from formula (\ref{eq:2band_kp}) that both  masses depend on the energy the same way. It is important that also in this case one can define  the effective masses relating electron velocity to its quasimomentum \cite{Zukotynski1963, Zawadzki1974}. In order to follow the energy dependence of the masses one defines the density-of-states effective mass $m^\ast_d=4^{\nicefrac{2}{3}}(m^2_\perp m_\parallel)^{\nicefrac{1}{3}}$, see \textcite{Zawadzki1974}. Figure \ref{fig:mPbSe} shows the concentration dependence of $m^\ast_d$ masses in PbS and PbTe, as measured by various authors. The straight lines are calculated for the spheroidal two-band model. It is seen that the latter describes the data very well.  Similar agreement is obtained for PbSe \cite{Zawadzki1974}. This demonstrates that the semirelativity generalized to the spheroidal bands works for the narrow-gap lead
chalcogenides.
\subsection{Dispersion in the Gap}
One of important physical quantities in the relativistic quantum mechanics for electrons is the Compton wavelength $\lambda_c=\hslash/m_0c=\SI{3.8e-3}{\angstrom}$, see e.g. \textcite{Bjorken1964}. There exists a corresponding quantity $\lambda_z$ in the two-band $\mathbf{k}\cdot\mathbf{p}$ description of electrons in NGS which is defined with the use of relativistic analogy \cite{Zawadzki1997, Zawadzki2005}:
\begin{equation}
	\label{eq:NGS_electrons_rel_analogy}
	\lambda_z=\frac{\hslash}{m^\ast_0u}
	=\hslash\left(\frac{2}{m^\ast_0\varepsilon_g}\right)^\frac{1}{2}.
\end{equation}
In NGS one can have $m^\ast_0 \simeq \num{5e-2} m_0$ and since $u\simeq c/300$, one obtains $\lambda_z\simeq2\times10^4\lambda_c\simeq \SI{50}{\angstrom}$. This is a sizable length for semiconductor nanostructures.

\begin{figure}
	\includegraphics[width=0.5\textwidth]{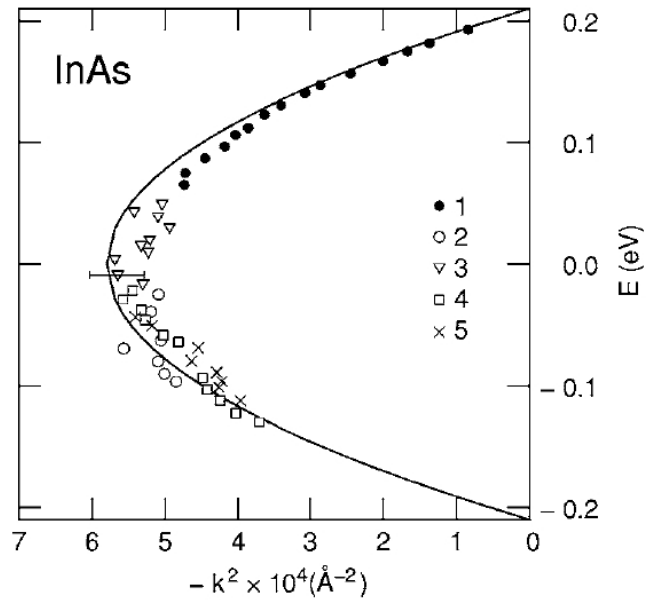}
	\caption{\label{fig:dispgap}Energy versus imaginary wave vector squared in the forbidden gap of InAs. Experimental data are after \textcite{Parker1968}, solid line is a theoretical fit after Eq.~(\ref{eq:disp_rel_ul}).  The adjusted parameters are $\lambda_z = \SI{41.5}{angstrom}$ and $u =\SI{1.33e8}{cm/s}$. After \textcite{Zawadzki2005}.}
\end{figure}

The dispersion relation (\ref{eq:lh_bands_conduction_approx}) can be rewritten in terms of $u$ and $\lambda_z$ in the form, see \textcite{Zawadzki2005}
\begin{equation}
	\label{eq:disp_rel_ul}
	\varepsilon(k)=\pm\hslash u\left({\lambda^{-2}_z+k^2}\right)^\frac{1}{2}.
\end{equation}
Let us consider the case $k_x=k_y=0$. For $k_z^2 \geq 0$, Eq.~(\ref{eq:disp_rel_ul}) describes the conduction and light hole bands. For $k_z^2 < 0$, that is for imaginary values of $k_z=i\kappa_z$, this equation describes the \emph{dispersion in the forbidden energy gap}. The length $\lambda_z$ can be determined experimentally. The gap region is classically forbidden, but  it can become accessible through one-dimensional quantum tunneling. Figure \ref{fig:dispgap} shows the dispersion in the gap of InAs obtained from tunneling experiments with double Schottky barriers. The solid line in Fig.~\ref{fig:dispgap} indicates the fit obtained using Eq.~(\ref{eq:disp_rel_ul}). The value of $\lambda_z$ is determined directly by $k_{z0}$ for which the energy vanishes: $\lambda_z^{-2}=k_{z0}^2$. This gives $\lambda_z=\SI{41.5}{\angstrom}$  and $u=\SI{1.33e8}{cm/s}$, in good agreement with the above estimations. Similar experiments were carried in GaAs and general picture of the dispersion in the gap is similar to that shown in Fig.~\ref{fig:dispgap}, see \textcite{Padovani1966}, \textcite{Conley1967}, \textcite{Pfeffer1990}.

Coming back to the relativistic electrons in vacuum, one obtains from Eq.~(\ref{eq:Dirac_rel_e}) the dispersion analogous to Eq.~(\ref{eq:disp_rel_ul})
\begin{equation}
	\mathscr{E}=\pm c\left(\hslash^2\lambda_c^{-2}+p^2\right)^\frac{1}{2}.
\end{equation}
For energies in the gap, i.e. for $p_z^2 < 0$, it takes the one-dimensional form similar to that shown in Fig.~\ref{fig:dispgap}, properly scaled. To our knowledge, there have been no attempts to investigate this relation for relativistic electrons in vacuum.

%% file: statistics_of_electron_gas.tex
\section{Statistical Properties of Electron Gas}
Description of statistical and thermodynamic properties of carriers in the relativistic-like energy bands of narrow-gap and middle-gap semiconductors requires a specialized mathematical formalism, as reviewed by \textcite{Zawadzki1974}. Here we present limited results which can be directly compared to those for the relativistic Juttner gas, see \textcite{Juttner1911}, \textcite{Synge1957}, \textcite{Arzelies1968}.  We begin with the simplest calculation of a three dimensional electron concentration $n$ in an arbitrary spherical energy band $\varepsilon(k)$. It is assumed that the electrons are governed by the Fermi-Dirac statistics $f_0=[\exp(z-\eta)+1]^{-1}$, where $z=\varepsilon/k_0T$ and $\eta=\zeta/k_0T$ are the reduced energy and the reduced Fermi energy, respectively. We have
\begin{equation}
	\label{eq:statistical_gas_prop1}
	n=(1/3\pi^2)\langle 1 \rangle,
\end{equation}
where in general
\begin{equation}
	\label{eq:statistical_gas_prop}
	\langle A\rangle=
	\int\limits_0^{\infty}\left(-\dfrac{\partial f_0}{\partial\varepsilon}\right)
	Ak^3(\varepsilon)\,d\varepsilon.
\end{equation}
In order to obtain Eq.~(\ref{eq:statistical_gas_prop1}) one writes $d^3k=k^2\,dk\sin\theta\,d\theta\,d\varphi$, integrates over the angles $\theta$ and $\varphi$, and integrates over $dk$ by parts noticing that at $k=0$ there is $\varepsilon=0$ and at $k=\infty$ there is $f_0=0$. It turns out that all statistical properties of the electron gas can be expressed by the integrals (\ref{eq:statistical_gas_prop}) with reasonably behaving $A(\varepsilon)$ functions. Since $-\partial f_0/\partial\varepsilon$ is nonzero only in a limited range of energies around the Fermi level, integrals (\ref{eq:statistical_gas_prop}) can be computed  numerically without much difficulty. In particular, at $T=0$ one deals with the complete degeneracy of the electron gas, so that $-\partial f_0/\partial\varepsilon= \delta(\varepsilon-\zeta)$, where $\delta$ is  the Dirac delta  function. The integral (\ref{eq:statistical_gas_prop}) is then reduced to one point at $\varepsilon=\zeta$. This gives $k_\zeta=(3\pi^2)^{\nicefrac{1}{3}}n^{\nicefrac{1}{3}}$. The above relation was used in Fig.~\ref{fig:GalSos} to show that the effective mass in HgCdTe for vanishing energy gap is proportional to $k$.

\begin{figure}
	\includegraphics[width=0.5\textwidth]{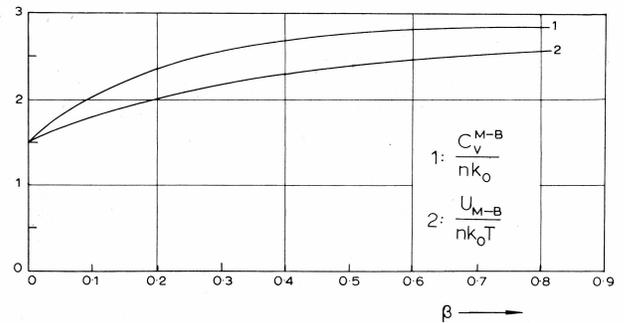}
	\caption{\label{fig:spec-heat}Mean internal energy and mean specific heat of the nondegenerate electron gas calculated  for a nonparabolic energy band versus $\beta=k_0T/\varepsilon_g$. For $\beta=0$ the “classical” values of 3/2 are obtained. After \textcite{Zawadzki1964}, see also \textcite{Zawadzki1974}.}
\end{figure}

However, we shall be interested here in the nondegenerate electron gas obeying the Maxwell- Boltzmann (MB) statistics. In this limit there is $f_0=-\partial f_0/\partial z$. Let us calculate internal energy of the electron gas $U_{MB}$ for this situation \cite{Zawadzki1964, Zawadzki1974}. Carrying integration by parts one obtains
\begin{equation}
	U_{MB}=\frac{2}{(2\pi)^3}\int f_0\varepsilon\,d^3k
	=nk_0T\left(\,\overline{z\frac{d}{dz}}\,\right),
\end{equation}
where, in general, an average value is defined as $\bar{A}=\langle A\rangle/\langle1\rangle$ and the differenciation acts on the $k^3(\varepsilon)$ function. For nonparabolic bands described by Eq.~(\ref{eq:lh_bands_conduction_approx}) the internal energy becomes
\begin{equation}
	\dfrac{U_{MB}}{n}
	=\frac{3}{2}k_0T
	\dfrac{^1D^{\nicefrac{1}{2}}_1(\beta)}{^0D^{\nicefrac{3}{2}}_0},
\end{equation}
where $^nD^m_l$ are the generalized Fermi integrals, see \textcite{Zawadzki1965}
\begin{equation}
	^nD^m_l(\beta)=
	\int\limits_0^{\infty} \exp(-z)z^n(z+\beta z^2)^m(1+2\beta z)^l\,dz.
\end{equation}
Here $\beta=k_0T/\varepsilon_g$ characterizes the degree of nonparabolicity. For $\beta=0$ one deals with a parabolic band and in this case D integrals can be expressed by the $\Gamma$ special function. It can then be easily verified that one obtains $U_{MB}/n=(3/2)k_0T$, which is a well known result for a nondegenerate electron gas. However, for a nondegenerate gas in which electrons have a nonparabolic $\varepsilon(k)$ dispersion this is not the case. Figure \ref{fig:spec-heat} shows the calculated mean internal energy as a function of $\beta$. With increasing $\beta$, i.e. growing nonparabolicity,  the mean internal energy increases from $(3/2)k_0T$ to $3k_0T$. The corresponding mean specific heat at a constant volume is $C_v=\partial U/\partial T$ and it also increases from $(3/2)k_0$ to $3k_0$.
It is known that the relativistic nondegenerate electron gas in vacuum has identical statistical properties to those presented above: its mean internal energy per electron at high temperatures is not $(3/2)k_0T$ but $3k_0T$ and its mean specific heat per electron at a constant volume is not $(3/2)k_0$ but $3k_0$, see \textcite{Synge1957}, \textcite{Kubo1965}. The reason for this similarity is the correspondence of  dispersion relations indicated above.

%% file: interband_tunelling.tex
\section{Interband Tunelling}
\begin{figure}
	\includegraphics[width=0.4\textwidth]{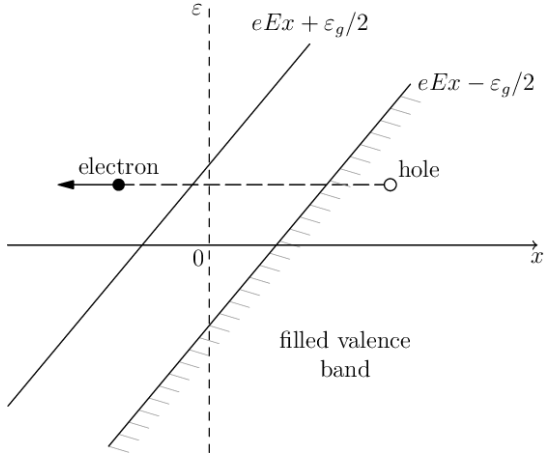}
	\caption{\label{fig:tunnel}Interband electron tunneling in a constant electric field within the two-band model of a semiconductor (schematically).}
\end{figure}

The phenomenon of tunneling between valence and conduction bands can also be interpreted in terms of the relativistic analogy. We begin by reproducing the simplest semiclassical calculation for a constant electric field within the framework of the two-band model of a semiconductor. Figure \ref{fig:tunnel} shows the space dependence of the potential energy $V(x)=eEx$ and the corresponding energy gap between $+(\varepsilon_g/2)+V(x)$ and $-(\varepsilon_g/2)+V(x)$. The electron-hole pair creation results from the electron tunneling from the filled valence band through the classically forbidden region. The probability of such a tunnel process is described semiclassically by
\begin{equation}
	P\simeq\exp\left(-\frac{2}{\hslash}\int_{x_-}^{x_+}q(x)\,dx\right),
\end{equation}
where the imaginary momentum is, cf.  Eq.~(\ref{eq:disp_rel_ul})
\begin{equation}
	q(x)=\frac{1}{u}\left[\left(\dfrac{\varepsilon_g}{2}\right)^2
	-\left(\varepsilon-eEx\right)^2\right]^\frac{1}{2},
\end{equation}
in which $\varepsilon$ is the energy and $x_\pm$ denote the classical turning points for which          $q(x_\pm)=0$. The transverse momentum components $p_y$ and $p_z$ have been omitted. Changing the variables $s=(\varepsilon-eEx)/(\varepsilon_g/2)$ and using $\varepsilon_g/2=m^\ast_0u^2$ one obtains
\begin{equation}
	P\simeq
		\exp\left(
			-\frac{2}{\hslash u}
			\frac{1}{\vert e\vert\,E}
			\left(\frac{\varepsilon_g}{2}\right)^2
			\int\limits_{-1}^{+1}\sqrt{1-s^2}\,ds
		\right)=
			\dfrac{\pi {m^\ast_0}^2u^3}{\vert e\vert\,E\hslash}.
\end{equation}
If we define the critical electric field $E_{cr}$ by the condition $P\simeq1$, we obtain $eE_{cr}\lambda_z\approx(\pi/2)\varepsilon_g$, where, as before, $\lambda_z=\hslash/m^\ast_0u$. The physical meaning of $E_{cr}$ is that  it gives the potential drop equal to the gap      over the distance $\lambda_z$.  Taking $\varepsilon_g=\SI{0.2}{eV}$ and $\lambda_z=\SI{20}{\angstrom}$ one obtains $E_{cr}\simeq\SI{e6}{V/cm}$. This is close to operating fields for semiconductor tunnel diodes \cite{Calawa1960}.

The above reasoning can be transposed to the tunneling in the frame of the Dirac equation for electrons and positrons, see e.g. \textcite{Akhiezer1981}, \textcite{Greiner1994}. Figure \ref{fig:tunnel} is still valid for this case. Using the relativistic analogy, i.e. replacing $\varepsilon_g/2$ by $m_0c^2$, $m^\ast_0$ by  $m_0$ and $u$ by $c$, the transition probability is
\begin{equation}
P\simeq\left(-\dfrac{\pi{m_0}^2c^3}{\vert e\vert\,E\hslash}\right).
\end{equation}
The condition for the critical field is now $eE_{cr}\lambda_c\simeq(\pi/2)2m_0c^2$, which means that $E_{cr}$ should cause the potential drop of $2m_0c^2$ over the distance $\lambda_c$. This gives $E_{cr}(\mathrm{relat})\simeq\SI{e16}{V/cm}$, which is not accessible  in terrestrial  conditions.  This shows the power of relativistic analogy: the physics is similar for relativistic electrons in vacuum and two-band electrons in semiconductors  but the parameters involved are much more “user friendly” in semiconductors.

%% file: e_in_mag_field.tex
\section{\label{sec:e_in_mag_field}Electrons in Magnetic Field}
We begin this section by considering properties of relativistic electrons in vacuum in the presence of a magnetic field and then proceed with the relativistic analogy describing magnetic behavior of electrons in semiconductors. In order to find eigenenergies of relativistic electrons in a uniform magnetic field $\mathbf{H}\parallel{z}$ one has to solve the Dirac equation for the problem. This can be done analytically taking the vector potential $\mathbf{A}=[-Hy,0,0]$, see \textcite{Rabi1928}, \textcite{Johnson1949}, \textcite{Strange1998}. The result is
\begin{gather}
	\begin{split}
		\label{eq:Dirac_eq1}
		&\mathscr{E}(n,p_z,\pm) =\\
		&\quad=\pm\left\{\left(m_0c^2\right)^2+2m_0c^2\left[D(n,p_z)
			\pm\frac{1}{2}\hslash \omega_c\right]\right\}^\frac{1}{2},
	\end{split}\\[1.2ex]
	\label{eq:Dirac_eq2}
	D(n,p_z) = \hslash\omega_c\left(n+\frac{1}{2}\right)
		+p_z^2/2m_0\quad,
\end{gather}
where $n$ is the Landau quantum number, $\omega_c=eH/m_0c$ is the cyclotron frequency, and $p_z$ is the momentum parallel to magnetic field. The signs $\pm$ are related to the electron spin orientations parallel and antiparallel to magnetic field.  Expression (\ref{eq:Dirac_eq1}) resembles Eq.~(\ref{eq:Dirac_rel_e}) with the term $p^2$ replaced by $D(n,p_z)$, reflecting the fact that the cyclotron and spin motions  are quantized by the magnetic field. For the magnetic energy small compared to the rest energy $m_0c^2$ one can expand the square root and obtain the standard expression for the uniform orbital and spin quantization
\begin{equation}
	\mathscr{E}=
		\pm\left[m_0c^2 +\hslash\omega_c\left(n+\frac{1}{2}\right)
		+\dfrac{p_z^2}{2m_0}\pm\frac{1}{2}\hslash\omega_c\right].
\end{equation}
The spin (Pauli) splitting is also equal to $\hslash\omega_c$. Introducing the spin Lande factor $g$
\begin{equation}
	\mathscr{E}(\pm)=\frac{1}{2}g\mu_BH\quad ,
\end{equation}
where $\mu_B=e\hslash/2m_0c$ is the Bohr magneton, it is seen that for free nonrelativistic electrons there is $g=2$. The lowest level in the system, corresponding to $n=0$ and the negative spin state, has always the energy $\mathscr{E}=0$ which does not depend on magnetic field.

\begin{figure}
	\includegraphics[width=0.4\textwidth]{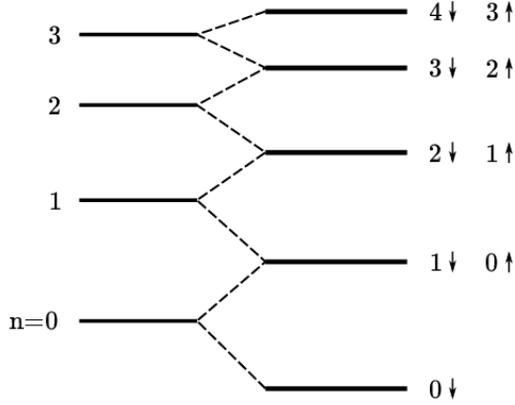}
	\caption{\label{fig:magn-sch}Quantized energy levels at $p_z=0$ for a relativistic electron in a magnetic field according to the Dirac equation (schematically). Orbital quantization is indicated on the left, the spin-split levels on the right.}
\end{figure}

The relativistic formula (\ref{eq:Dirac_eq1}) introduces two features, schematically illustrated in Fig.~\ref{fig:magn-sch}. First, the spacing between orbital levels is not constant, it diminishes with the increasing energy. This can be interpreted as a decrease of the cyclotron frequency $\omega_c=eH/m(\mathscr{E})c$ due to the relativistic increase of the mass according to relation (\ref{eq:rel_mass_inc}). We have
\begin{equation}
	\label{eq:rel_mass_inc}
	\dfrac{1}{m(\mathscr{E})}
		=\lim\limits_{H\rightarrow0}
		\dfrac{\mathscr{E}(n+1,p_z,\pm)-\mathscr{E}(n,p_z,\pm)}
			{\hslash eH/c}=\dfrac{c^2}{\mathscr{E}}\quad.
\end{equation}
This effect is commonly encountered in electron cyclotrons.

In addition, it follows from Eq.~(\ref{eq:Dirac_eq1}) and Fig.~\ref{fig:magn-sch} that, in the relativistic range of energies, also the spin splitting decreases with energy. This can be regarded as a decrease of the $g$-factor or of the Bohr magneton.  At this point one can ask a question: how does the spin magnetic moment of a relativistic electron behave when its energy  increases? One can find an explicit energy dependence of $\mu_B$ by keeping $g=2$, see \textcite{Zawadzki1971}. We have
\begin{equation}
	\label{eq:explicit_energy_dep}
	\mu_B(\mathscr{E})
		=\lim\limits_{H\rightarrow0}\dfrac{\mathscr{E}(n+1,p_z,+)-\mathscr{E}(n,p_z,-)}{2H}
		=\dfrac{e\hslash}{2m(\mathscr{E})c}
\end{equation}
where $m(\mathscr{E})$ is the above relativistic electron mass given in Eq.~(\ref{eq:rel_mass_inc}). Thus, the spin magnetic moment of a relativistic electron is given by the Bohr magneton in which the electron mass in the numerator is the relativistic mass. This means that the spin magnetic moment  goes to zero as the electron energy increases. If one does not want to employ the energy-dependent mass, one can  write $\mu_B(\mathscr{E})=e\hslash c/2\mathscr{E}$. The relativistic decrease of spin splitting with the energy is manifested also in the expansion of the relativistic Hamiltonian to $v^2/c^2$ terms \cite{Zawadzki2005a}. It should be mentioned that the same result for the spin magnetic moment of the Dirac electron was recently obtained by consideration of Zitterbewegung (trembling motion), see \textcite{Sasabe2014}.

In order to trace the relativistic analogy for magnetic properties of electrons in NGS, we need to outline a generalization of the $\mathbf{k}\cdot\mathbf{p}$ theory for the presence of an external magnetic field. We call it the $\mathbf{P}\cdot\mathbf{p}$ theory for reasons given below. To include properly the spin effects one needs to take into account the spin-orbit interaction related to the periodic potential. Thus,  the initial Hamiltonian for the problem reads
\begin{equation}
	\label{eq:init_hamiltonian}
	\hat{\mathscr{H}}
		=\dfrac{1}{2m_0}P^2
		+V_0(\mathbf{r})
		+\dfrac{\hslash}{4m_0^2c^2}
			\left(\boldsymbol{\sigma}\times\boldsymbol{\nabla}V_0\right)\cdot\mathbf{P}
		+\mu_B\mathbf{H}\cdot\boldsymbol{\sigma}
\end{equation}
where $\mathbf{P}=\mathbf{p}+(e/c)\mathbf{A}$ is the kinetic momentum, $\mathbf{A}$ is the vector potential of magnetic field $H$, $e$ is the absolute value of electron charge, $\boldsymbol{\sigma}$ are the Pauli spin operators and other quantities have been defined above. The vector potential is assumed to be slowly varying over the unit cell $\Omega$. The spin-orbit interaction and the Pauli term are written in the standard form. In the presence of $\mathbf{A}$, the Hamiltonian (\ref{eq:init_hamiltonian}) is not periodic and one may not look for its eigenstates in the form of Bloch functions. In a somewhat simplified derivation one tries to find for the eigenenergy problem
\begin{equation}
	\label{eq:eigenenergy_problem}
	\hat{\mathscr{H}}\Psi=\varepsilon\Psi ,
\end{equation}
a solution in the form
\begin{equation}
	\label{eq:eigenenergy_solution_form}
	\Psi(\mathbf{r})=\sum\limits_lf_l(\mathbf{r})u_{l0}(\mathbf{r}),
\end{equation}
where the summation is over the energy bands (Zawadzki, 1980). Functions $u_{l0}$ are the Luttinger-Kohn periodic amplitudes introduced in  Eq.~(\ref{eq:LK_fun}). They satisfy the eigenvalue problem of Eq.~(\ref{eq:eigenval_solution}) and are orthonormal within the unit cell. On the other hand, the envelope functions $f_l(\mathbf{r})$ are assumed to be slowly varying in the coordinate space, extending over many unit cells. Putting the form (\ref{eq:eigenenergy_solution_form}) into the problem (\ref{eq:eigenenergy_problem}) one obtains
\begin{widetext}
	\begin{multline}
		\label{eq:eigenenegy_solution}
		\sum\limits_l\left[
			\dfrac{1}{2m_0}(p^2u_{l0}) + \dfrac{1}{m_0}(\mathbf{p}u_{l0})\cdot\mathbf{p} +
			\dfrac{1}{2m_0}u_{l0}p^2 + \dfrac{e}{2m_0c}(\mathbf{p}u_{l0})\cdot\mathbf{A} +
			\dfrac{e}{2m_0c}u_{l0}(\mathbf{p}\cdot\mathbf{A}) +
			\dfrac{e}{2m_0c}\mathbf{A}\cdot(\mathbf{p}u_{l0}) +\right.\\
		\left.
			+\dfrac{e}{2m_0c}u_{l0}\mathbf{A}\cdot\mathbf{p} +
			\dfrac{e^2}{2m_0c^2}A^2u_{l0} + V_0u_{l0} +
			\dfrac{\hslash}{4m_0^2c^2}
				(\boldsymbol{\sigma}\times\boldsymbol{\nabla}V_0)
				\cdot(\mathbf{p}u_{l0}) +
			\dfrac{\hslash}{4m_0^2c^2}
				(\boldsymbol{\sigma}\times\boldsymbol{\nabla}V_0)u_{l0}
				\cdot\mathbf{p} +\right.\\[1.2ex]
		\left.+\dfrac{\hslash}{4m_0^2c^2}
				(\boldsymbol{\sigma}\times\boldsymbol{\nabla}V_0)u_{l0}
				\cdot\left(\frac{e}{c}\mathbf{A}\right) +
			\mu_B\mathbf{H}\cdot\boldsymbol{\sigma}u_{l0}
		\right]f_l(\mathbf{r})
		= \varepsilon\sum\limits_lf_l(\mathbf{r})u_{l0}
	\end{multline}
\end{widetext}
Next, both sides of the above equation are multiplied on the left by $(1/\Omega)u^\ast_{l'0}$ and integrated over the unit cell. Since it has been assumed that both $f_l(\mathbf{r})$ and $\mathbf{A}(\mathbf{r})$ are slowly varying, they can be considered constant within the unit cell and taken out of the integral sign. Making use of Eq.~(\ref{eq:eigenval_solution}) and of the orthonormality of $u_{l0}$, the eigenenergy problem (\ref{eq:eigenenegy_solution}) is obtained in the form
\begin{gather}
	\sum\limits_l\left\{
		\left(\dfrac{1}{2m_0}P^2
		+\varepsilon_{l0}-\varepsilon\right)\delta_{l\smash{'}l}
		+ \boldsymbol{\kappa}_{l\smash{'}l}\cdot\mathbf{P}
	+ \right.\notag\\
	\left.\label{eq:ee_prob_1}
		+\left[\dfrac{\hslash}{4m_0^2c^2}
			(\boldsymbol{\sigma}\times\boldsymbol{\nabla}V_0)
			\cdot\mathbf{P}\right]_{l\smash{'}l}
		+ \mu_B\mathbf{H}\cdot\boldsymbol{\sigma}_{l\smash{'}l}
	\right\}f_l(\mathbf{r}) = 0\\[1.2ex]
	\label{eq:ee_prob_2}
	\boldsymbol{\kappa}_{l\smash{'}l}
		= \dfrac{1}{m_0}\langle
			u_{l'0}\vert\mathbf{p}
			+\dfrac{\hslash}{4m_0c^2}
				(\boldsymbol{\sigma}\times\boldsymbol{\nabla}V_0)
		\vert u_{l0}\rangle.
\end{gather}
Similarly to the $\mathbf{k}\cdot\mathbf{p}$ theory discussed above, in case of narrow-gap or medium-gap materials one can  include a finite number of interacting bands and try to find analytical solutions. For carriers in the presence of a magnetic field the minimum number of levels at $k=0$ is three, because the spin-orbit interaction of valence levels determines the spin properties of conduction electrons. Following \textcite{Bowers1959} one solves the three-level model (eight states including spin). There are two s-like $\Gamma_6$ conduction levels, four p-like $\Gamma_8$ levels separated by the energy gap $\varepsilon_g$, and two p-like $\Gamma_7$ levels split-off by the spin-orbit energy $\Delta$. Neglecting the free electron terms in Eq.~(\ref{eq:ee_prob_1}) and taking the vector potential in the form $\mathbf{A}=[-Hy,0,0]$ one can solve the resulting $8\times8$ set of equations by a single column of harmonic oscillator functions $\phi_n[(y-y_0)/L]$, where $y_0= k_xL^2$ and $L=(\hslash c/eH)^{\nicefrac{1}{2}}$ \cite{Bowers1959, Zawadzki, Zawadzki1991}. Taking, as before, the zero of energy in the middle of the gap, the resulting energies are
\begin{multline}
	\left(\varepsilon-\dfrac{\varepsilon_g}{2}\right)
		\left(\varepsilon+\dfrac{\varepsilon_g}{2}\right)
		\left(\varepsilon+\dfrac{\varepsilon_g}{2}+\Delta\right) +\\
	\label{eq:eigenenergies}
	-\kappa^2\hslash^2\left[s(2n+1)+k_z^2\right]
		\left(\varepsilon+\dfrac{\varepsilon_g}{2}+\dfrac{2}{3}\Delta\right)
	\pm\dfrac{1}{3}\kappa^2\hslash^2\Delta s = 0
\end{multline}
where $n$ is the Landau quantum number, $s=eH/\hslash c$ and $\kappa=(-i/m_0)\langle S\vert p_z\vert Z\rangle$. Signs $\pm$ correspond to two effective spin states. Three roots of this equation give the energy levels in the conduction, light-hole and split-off energy bands. For $\varepsilon\ll\varepsilon_g+2\Delta/3$ the resulting quadratic equation gives for the conduction band
\begin{gather}
	\label{eq:conduction_band_1}
	\varepsilon(n,k_z,\pm)=\left[\left(\frac{\varepsilon_g}{2}\right)^2 +
		\varepsilon_gD(n,k_z,\pm)\right]^\frac{1}{2},\\
	\label{eq:conduction_band_2}
	D(n,k_z,\pm)=\hslash\omega_c\left(n+\frac{1}{2}\right) +
		\dfrac{\hslash^2k_z^2}{2m^\ast_0}
		\pm\frac{1}{2}\mu_B\,g^\ast_0\,H\quad,
\end{gather}
where $\omega_c=eH/m^\ast_0c$ is the effective cyclotron frequency and $g^\ast_0$ is the effective spin $g$-value at the band edge. For the InSb-type of materials one obtains
\begin{align}
	\label{eq:InSb_cond_band1}
	\dfrac{1}{m^\ast_0}
		&=\phantom{-}\dfrac{4\kappa^2}{3\varepsilon_g}
		\cdot\dfrac{\Delta+(3/2)\varepsilon_g}{\Delta+\varepsilon_g}\quad,\\
	\label{eq:InSb_cond_band2}
	g^\ast_0
		&=-\dfrac{m_0}{m^\ast_0}
		\cdot\dfrac{\Delta}{\Delta+(3/2)\varepsilon_g}\quad.
\end{align}
For HgTe-type of materials the conduction band energies in the region $\varepsilon\ll(2/3)\Delta$ are also given by Eq.~(\ref{eq:conduction_band_1}) with $\varepsilon_g$ replaced by $\varepsilon_0$ (the interaction energy). In this case there is: $1/m^\ast_0=4\kappa^2/3\varepsilon_0$ and $g^\ast_0=-m_0/m^\ast_0$.

We can now trace the relativistic analogy for electrons in quantizing magnetic fields. It can be seen that Eqs.~(\ref{eq:conduction_band_1}) and (\ref{eq:conduction_band_2}) for the electron energies in NGS have the same form as Eqs.~(\ref{eq:Dirac_eq1}) and (\ref{eq:Dirac_eq2}) for relativistic electrons in vacuum. The main difference is that, in the relativistic case, the spin splitting is equal to the orbital splitting and corresponds to the value of $g=+2$, whereas in semiconductors the spin splitting is usually (not always) smaller than the orbital splitting and corresponds to negative values of $g$. This is a consequence of the spin-orbit interaction expressed by the spin-orbit energy $\Delta$. However,  qualitatively, electrons in both systems behave quite similarly. In particular, it follows from Eqs.~(\ref{eq:conduction_band_1}) and (\ref{eq:conduction_band_2}) that both orbital and spin energy differences decrease with energy. The decrease of orbital splittings  can again be related to the increase of effective mass with energy, described above. As to the spin splitting, it is seen from Eq.~(\ref{eq:InSb_cond_band2}) that the $g$-factor is inversely proportional to the mass, so one can expect that, with the increase of the mass, the $g$-factor decreases. Calculations based on Eq.~(\ref{eq:eigenenergies}) were carried in the limit of small magnetic fields numerically by \textcite{Bowers1959} and analytically by \textcite{Zawadzki1963}. The analytical result for InSb-type materials is
\begin{equation}
	\label{eq:InSb_analytical_result}
	g^\ast(\varepsilon) = 2\left[1
		-\left(\dfrac{m_0}{m^\ast(\varepsilon)}-1\right)
		\dfrac{\Delta}{3(\varepsilon+\varepsilon_g/2)+2\Delta}\right]
\end{equation}
where $m^\ast(\varepsilon)$ is the energy-dependent effective mass. The unities in square brackets are due to the free electron terms. It is seen that, if the latter are neglected and one takes the limit of $\Delta\gg\varepsilon_g$, one obtains $g^\ast(\varepsilon)=m_0/m^\ast(\varepsilon)$, which is strictly analogous to the relativistic result (\ref{eq:explicit_energy_dep}). The only difference is that, for the reasons of tradition, in the relativistic formulation we have kept $g=2$ and put the energy dependence into the Bohr magneton, whereas for semiconductors we have put in Eq.~(\ref{eq:conduction_band_1}) the energy dependence into the $g$-factor.

\begin{figure}
	\includegraphics[width=0.5\textwidth]{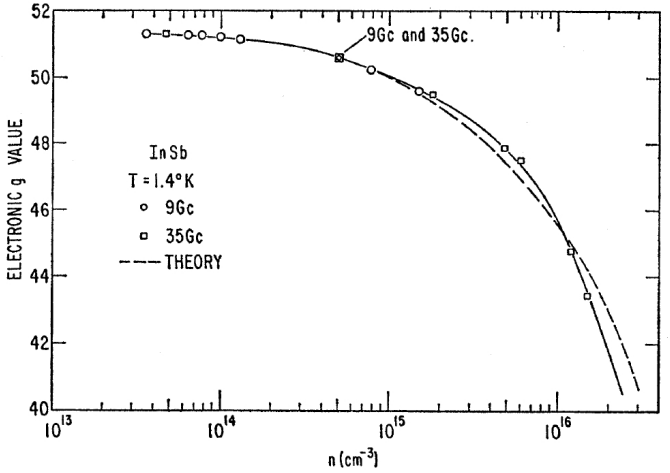}
	\caption{\label{fig:g-Isaac}Electron spin $g$-factor at the Fermi level versus electron concentration in InSb. The dashed line is calculated according to formula (\ref{eq:InSb_analytical_result}). After \textcite{Isaacson1968}.}
\end{figure}

The energy dependence of the spin $g$-value in InSb was first measured by \textcite{Bemski1960}. Figure \ref{fig:g-Isaac} shows more complete experimental results for $g^\ast$ measured by the spin resonance for electrons in InSb as a function of the free electron density $n$. The dashed line is theoretical, following Eq.~(\ref{eq:InSb_analytical_result}). As we already explained above when discussing the effective mass, an increase of the electron density raises the Fermi energy at which the $g$-value is measured. Thus Fig.~\ref{fig:g-Isaac} illustrates the decrease of $g^\ast(\varepsilon)$ with the increasing electron energy in the band. One should note that the electron $g$-value in InSb is negative as a result of the large spin-orbit energy, see Eq.~(\ref{eq:InSb_cond_band2}). A similar decrease of electron $g$-value with increasing electron density was observed  by means of quantum transport in HgSe having the “inverted” band structure \cite{Kacman1971}.

\begin{figure}
	\includegraphics[width=0.5\textwidth]{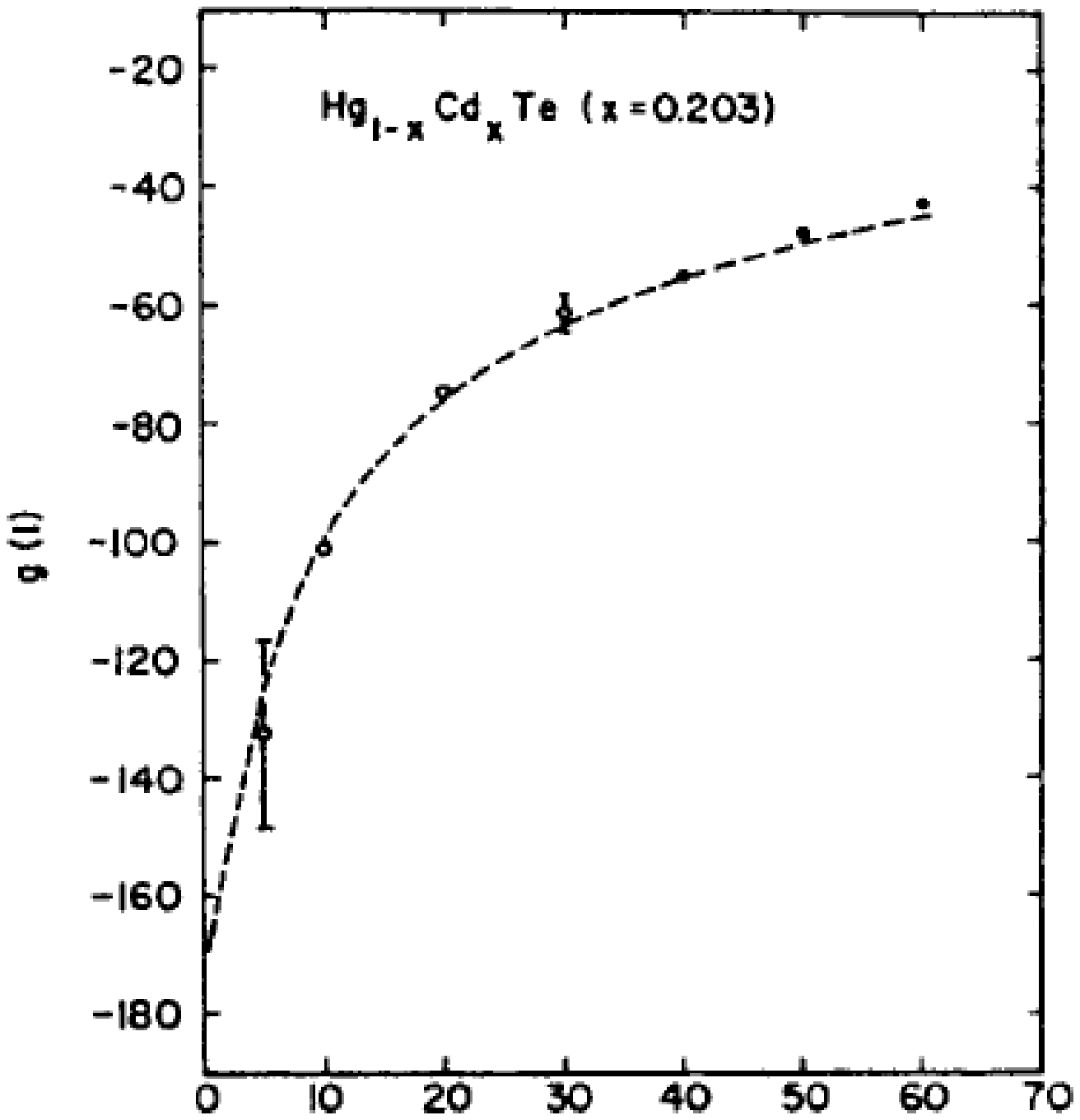}
	\caption{\label{fig:g-McCom}Spin g-factor of the first Landau level in $\mathrm{H\smash{g}}_{0.797}\mathrm{Cd}_{0.203}\mathrm{Te}$ versus magnetic field.  The dashed line is calculated according to Eq.~(\ref{eq:eigenenergies}). After \textcite{McCombe1970}.}
\end{figure}

The energy in the band can be raised not only by going to higher electron densities but also by increasing an external magnetic field. Figure \ref{fig:g-McCom} shows an experimental decrease of electron $g$-value in $\mathrm{Hg}_{0.797}\mathrm{Cd}_{0.203}\mathrm{Te}$ as the intensity of $H$ becomes larger. One should note the very large negative values of $g^\ast$ at low magnetic fields. This is due to the very small energy gap for the above chemical composition.

\begin{figure}
	\includegraphics[width=0.5\textwidth]{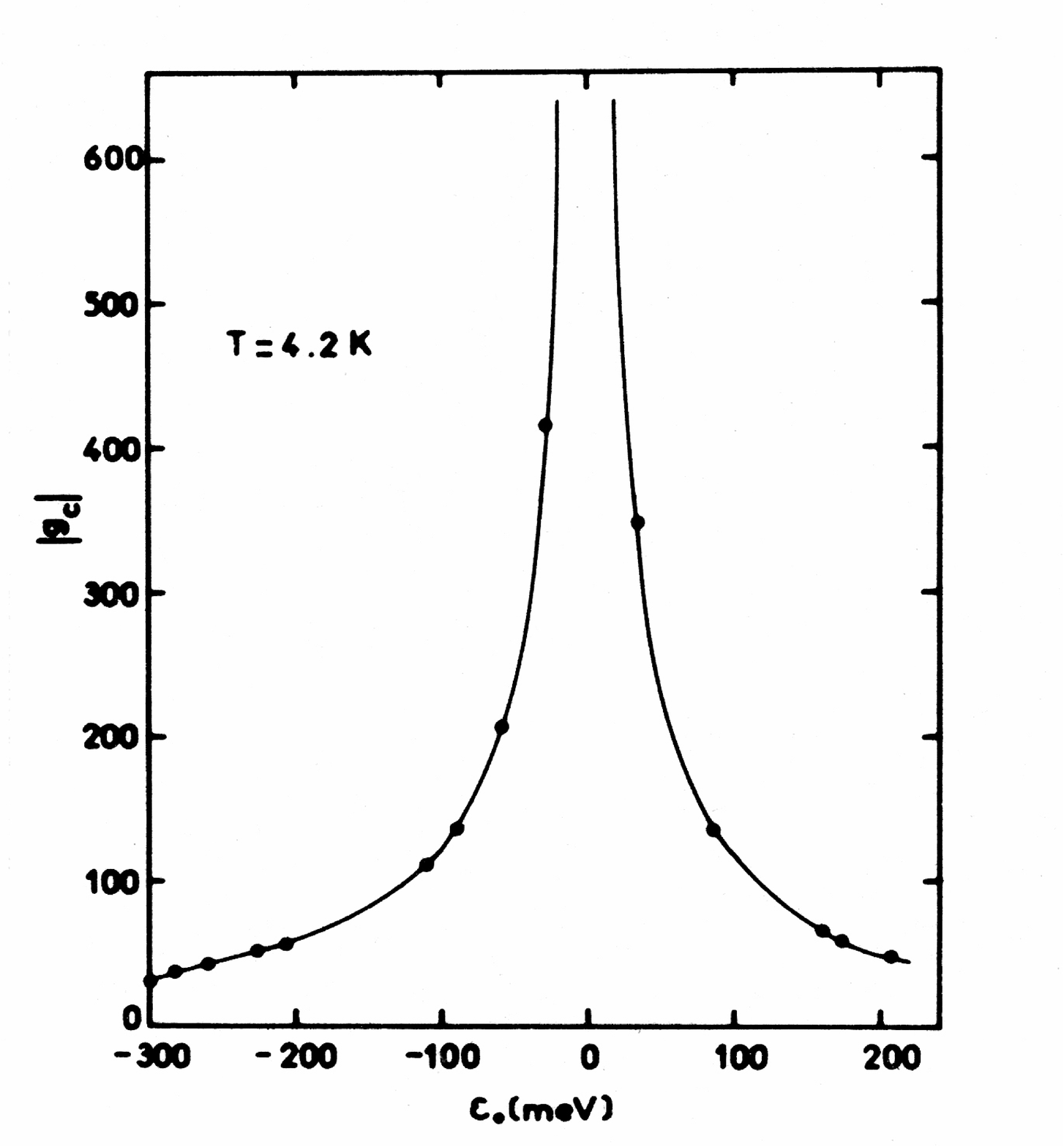}
	\caption{\label{fig:g-Rig}Spin $g$ factor at the band edge of $\mathrm{H\smash{g}}_{1-x}\mathrm{Cd}_{x}\mathrm{Te}$ versus interaction gap. As the effective mass vanishes the $g$-value becomes very high. After \textcite{Guldner1977}, see also \textcite{Rigaux}.}
\end{figure}

Spectacular results for spin $g$-values in $\mathrm{Hg}_{1-x}\mathrm{Cd\vphantom{g}}_{x}\mathrm{Te}$ are quoted in Fig.~\ref{fig:g-Rig}. They were obtained by interband magneto-absorption experiments on samples having different chemical compositions. The data for various interband transitions allow one to separate the orbital and spin quantizations in a magnetic field, the first related to the mass and  the second to the $g$-value. According to our previous analysis,  at the chemical composition $x=0.165$ the band-edge mass vanishes, see Fig.~\ref{fig:mHCTrig}. Since the spin $g$-factor is inversely proportional to the mass, it approaches infinity for the same value of $x$, as seen in Fig.~\ref{fig:g-Rig}. The highest measured value of $g^\ast$ is about \num{-400}. Similar data were reported by \textcite{Weiler1978}, who observed the spin $g$-value of about \num{-500}. A comparable behavior of the effective electron mass and $g$-value as functions of the interaction gap was demonstrated for $\mathrm{Pb}_{1-x}\mathrm{Sn}_{x}\mathrm{Te}$ solid solutions by \textcite{Gureev1978}.

As we indicated above, the two-band model works well for $\varepsilon_g=0$. The same may be said about the two-level model in the presence of a magnetic field. Cancelling in Eqs.~(\ref{eq:conduction_band_1}) and (\ref{eq:conduction_band_2}) $\varepsilon_g$ with $m^\ast_0$ in the orbital and spin terms one obtains an analytical expression for the energy. It is easy to see that, for $k_z=0$, there is $\varepsilon\sim H^{\nicefrac{1}{2}}$. This behavior is characteristic of massless fermions, see below. The $H^{\nicefrac{1}{2}}$ behavior of the conduction Landau levels was observed in interband magneto-optical transitions on $\mathrm{Hg}_{1-x}\mathrm{Cd\vphantom{g}}_{x}\mathrm{Te}$ in zero-gap situation by \textcite{Kim1976}.

We emphasize again that the behavior described above corresponds to the relativistic analogy since, according to Eqs.~(\ref{eq:explicit_energy_dep}) and (\ref{eq:InSb_analytical_result}), both spin magnetic moments of relativistic and NGS electrons are inversely proportional to the corresponding energy-dependent masses. According to our knowledge, the energy dependence of spin magnetic moment has not been measured experimentally for free relativistic electrons in vacuum.

%% file: e_in_x_fields.tex
\section{Electrons in Crossed Electric and Magnetic Fields}
We now consider semiconductor electrons in crossed electric and magnetic fields which provide a spectacular example of the relativistic analogy. We begin by using the one-band effective mass approximation. In other words we treat electrons in the conduction band of a semiconductor as free nonrelativistic electrons with the free electron mass $m_0$ replaced by the effective mass $m^\ast_c$, see  \textcite{Aronov1963}, \textcite{Zak1966}, \textcite{Zawadzki1966}. The eigenenergy  equation for the envelope function is
\begin{equation}
	\label{eq:eigenenergy_envelope}
	\left[
		\dfrac{1}{2m^\ast_c}\left(\mathbf{\hat{p}}
		+\dfrac{e}{c}\mathbf{A}\right)^2
		+e\mathbf{E}\cdot\mathbf{r}
	\right]f(\mathbf{r}) =
	\varepsilon f(\mathbf{r})\quad.
\end{equation}
We choose the electric field along the $y$ direction $\mathbf{E}=[0,E,0]$. It is then convenient to take for $\mathbf{H}\parallel z$ the vector potential $\mathbf{A}=[-Hy,0,0]$. With the above choice of $\mathbf{A}$ one looks for the envelope function in the form $f(\mathbf{r})= C(ik_xx+ik_zz)\phi(y)$. Solutions of Eq.~(\ref{eq:eigenenergy_envelope}) without the electric field term are given by the harmonic oscillation functions. The electric field, introducing the term linear in $y$, does not change the character of solutions. It simply shifts the potential well which results in the shift of the oscillator center. Eigenenergies of Eq.~(\ref{eq:eigenenergy_envelope}) are
\begin{equation}
	\label{eq:envelope_solution}
	\varepsilon=\hslash\omega^\ast_c\left(n+\frac{1}{2}\right)+
		\dfrac{\hslash^2k_z^2}{2m^\ast_c}
		-eEk_xL^2
		-\dfrac{m^\ast_cc^2}{2}\dfrac{E^2}{H^2}
	\quad,
\end{equation}
where $n=1,2,3,\ldots$ enumerate the Landau levels and $L=(\hslash c/eH)^{\nicefrac{1}{2}}$ is the magnetic radius. For $E=0$, Eq.~(\ref{eq:envelope_solution}) is reduced to the eigenenergies for the electron in a magnetic field alone.

\begin{figure}
	\includegraphics[width=0.5\textwidth]{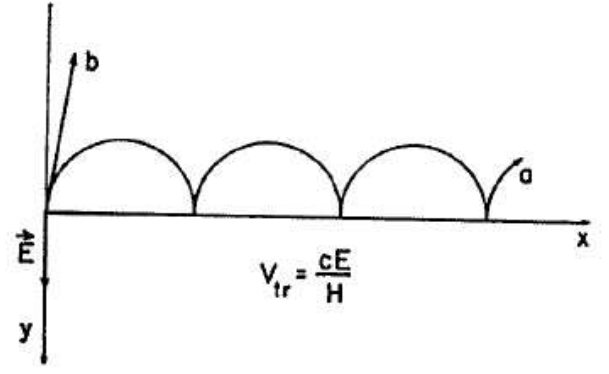}
	\caption{\label{fig:cros-class}Classical motion of an electron in crossed electric and magnetic fields according to the relativistic equation of motion. a) For $E/H \ll 1$ the motion is a superposition of the cyclotron motion and motion with the velocity $v_\mathrm{tr}= cE/H$ transverse to both fields. b) For $E/H > 1$ the motion is predominantly along the direction of electric field slightly deflected by magnetic field.}
\end{figure}

One can interpret the above result by considering the classical motion of a free electron in crossed fields. The nonrelativistic equation  of motion
\begin{equation}
	\label{eq:nonrelat_motion}
	m^\ast_c\ddot{\mathbf{r}}=-e\mathbf{E}-(e/c)(\mathbf{v}\times\mathbf{H})
\end{equation}
can be easily solved, see e.g. \textcite{Landau1959}. The resulting motion for the initial conditions $\mathbf{r}=\mathbf{v}=0$ is a cycloid, i.e. a superposition of the circular cyclotron motion in the $x$-$y$ plane with the frequency $\omega_c=eH/m^\ast_cc$ and the transverse motion along the $x$ direction with the velocity $v_{\mathrm{tr}}=cE/H$ transverse to both fields. This is illustrated with trace $a$ in Fig.~\ref{fig:cros-class}.  As follows from the energies (\ref{eq:envelope_solution}) and the classical motion, the described electron behavior is essentially of the magnetic type: the cyclotron motion is still the dominant feature and there is no net acceleration   along the electric field.

\begin{figure}
	\includegraphics[width=0.5\textwidth]{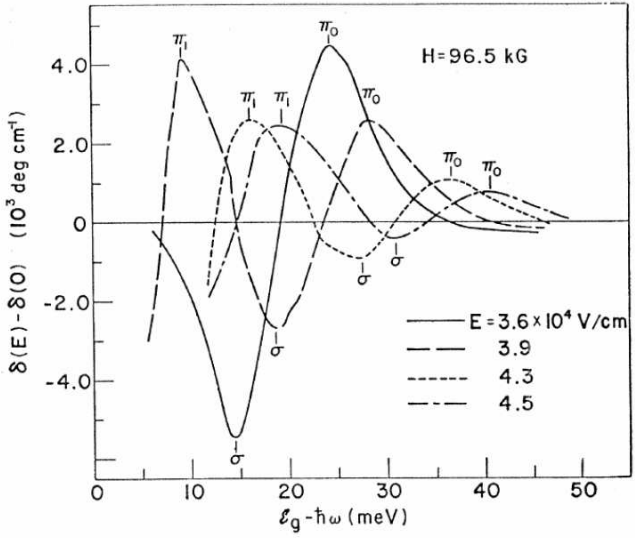}
	\caption{\label{fig:crosVoig}Resonant Voigt effect in crossed electric and magnetic fields for photon energies below the direct gap of germanium for a constant magnetic field $H$ and various electric fields $E$. The resonant behavior illustrates magnetic character of the motion and the interband transitions shift to lower energies as $E$ increases. After \textcite{Vrehen1967}.}
\end{figure}

\begin{figure}
	\includegraphics[width=0.5\textwidth]{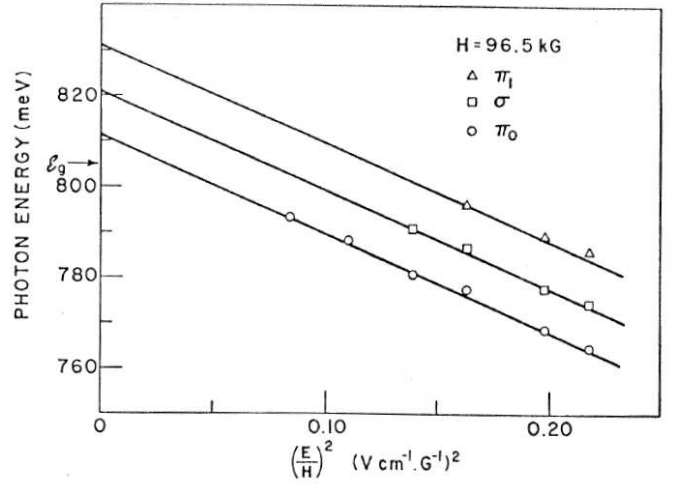}
	\caption{\label{fig:crosEnrg}Photon energies of Voigt-effect peaks in crossed fields, as determined from Fig.~\ref{fig:crosVoig}. All Landau level peaks undergo equal shifts to lower energies proportional to $(E/H)^2$. After \textcite{Vrehen1967}.}
\end{figure}

Experiments on interband optical absorption and dispersion on germanium in crossed electric and magnetic fields confirm these conclusions. As follows from Eq.~(\ref{eq:envelope_solution}), the main effect of electric field is to shift all conduction Landau levels downwards by the amount $m^\ast_cc^2E^2/2H^2$. The valence Landau levels  shift upwards since their mass $m^\ast_v$ has the negative sign. As a consequence, the effect should be observable in interband optical transitions whose energies decrease at the rate $(m^\ast_c+m^\ast_v)c^2E^2/2H^2$, as first remarked by \textcite{Aronov1963}. The experimental results are shown in Fig.~\ref{fig:crosVoig} and the decrease of interband energies is illustrated in Fig.~\ref{fig:crosEnrg}. Thus the one-band effective mass approximation describes the data very well. However, this cannot be the full story since it is not clear what to do with the term $(E/H)^2$ when $H$ goes to zero. On the other hand, once one accepts the initial Eq.~(\ref{eq:eigenenergy_envelope}) the energies (\ref{eq:envelope_solution}) follow as an exact result. In other words,  according to Eq.~(\ref{eq:eigenenergy_envelope}) and Fig.~\ref{fig:cros-class}, as soon as $H\neq0$ the solutions are of the magnetic type. In still other words, even a very small magnetic field seems to have a dramatic effect on the character of the motion. This does not make much physical sense because for large electric and small magnetic fields an electric type of motion should prevail.

\begin{figure}
	\includegraphics[width=0.5\textwidth]{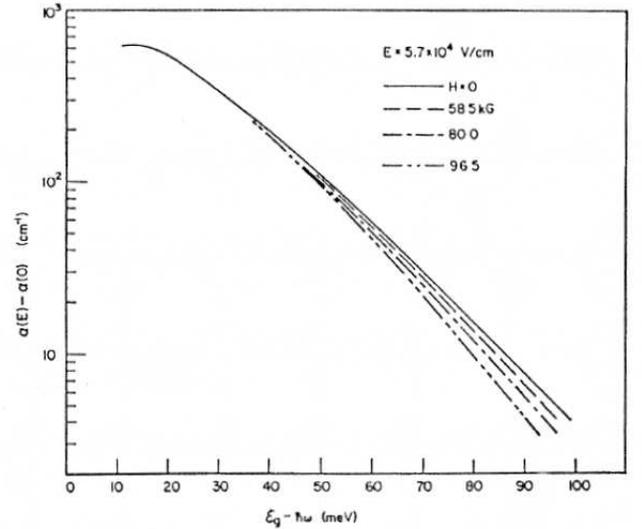}
	\caption{\label{fig:cros-expon}. Exponential optical absorption below the direct gap of germanium due to electric field $E=\SI{5.7e4}{V/cm}$ for various transverse magnetic fields illustrating the electric-type of motion. The  magnetic field diminishes  photon-assisted tunneling. After \textcite{Reine1967}.}
\end{figure}

In fact, experiments performed for high electric and low magnetic field intensities clearly indicate that one deals with electric-type solutions. Figure \ref{fig:cros-expon} illustrates results for the optical absorption of germanium (direct transitions across the gap of about \SI{0.804}{eV}) in the presence of a constant high electric field and different values of transverse magnetic field. It is seen that for $H=0$ one observes the usual Franz-Keldysh effect, i.e. the photon-assisted tunneling, see \textcite{Franz1958}, \textcite{Keldysh1958}. The absorption has an exponential nonresonant behavior, which is characteristic of the electric-type of motion. Low magnetic fields do not affect dramatically the electric behavior, diminishing slightly the absorption below the gap, see \textcite{Zawadzki1969}. These results indicate that the theory should treat electric and magnetic fields symmetrically, whereas the above treatment, as presented in Eqs.~(\ref{eq:eigenenergy_envelope}) and (\ref{eq:nonrelat_motion}), does not possess this property.

The relativistic analogy helps to find the key to the problem. According to the nonrelativistic motion in crossed fields, the troublesome high values of $E/H$ ratio correspond to high transverse velocities $v_{\mathrm{tr}}=cE/H$. It is clear that for sufficiently large values of $E/H$ this description should not be valid, as  it would lead to a possibility of producing $v_{\mathrm{tr}}>c$, which cannot be. Thus, one may not use the nonrelativistic equation of motion and the corresponding quantum treatment of Eq.~(\ref{eq:eigenenergy_envelope}) for large values of $E/H$. For electrons in vacuum  it is then necessary to use the relativistic equation of motion
\begin{equation}
	\dfrac{d}{dt}\left[m_0\left(1-\dfrac{v^2}{c^2}\right)^{-\frac{1}{2}}\mathbf{v}\right]
	=-e\mathbf{E}-\dfrac{e}{c}(\mathbf{v}\times\mathbf{H})\quad.
\end{equation}
This problem is treated in textbooks on electrodynamics, see e.g. \textcite{Landau1959}, \textcite{Jackson1962}. For small values of $E/H$ the treatment reduces to the nonrelativistic limit and the motion is described, as before, by the cycloid  of  Fig.~\ref{fig:cros-class}, trace $a$. However, for $E/H>1$ the motion becomes of the electric type: there are no oscillations and there is a net acceleration in the direction $y$ of electric field. In other words, for $E/H <1$ one can perform the Lorentz transformation eliminating $E$, while for $E/H >1$ one can perform the transformation eliminating $H$. In Fig.~\ref{fig:cros-class} the electric regime corresponds to the trace $b$: there are no oscillations, the electron does not come back to the line $y=0$ but accelerates along the $y$ direction with some deflection caused by the magnetic field.

To obtain the corresponding results in quantum theory it is necessary to solve the Dirac equation for a free electron in crossed fields. However, since we are interested in general features of the motion and not in details of spin effects, it is enough to consider the simpler Klein-Gordon equation for a spinless particle. Then the problem reads
\begin{equation}
	\left[\left(\mathbf{\hat{p}}+\dfrac{e}{c}\mathbf{A}\right)^2
		-\left(\dfrac{\varepsilon}{c}+\dfrac{eV}{c}\right)^2
		+m_0^2c^2\right]\Psi
	=0 \quad.
\end{equation}
Taking as before $V=eEy$ and $A= [-Hy, 0, 0]$, and separating  $x$ and $z$ variables one obtains
\begin{equation}
	\label{eq:KG_develop}
	\left[
		-\dfrac{\hslash^2}{2m_0}\dfrac{\partial^2}{\partial y^2}
		+\alpha y+
		\dfrac{e^2}{2m_0c^2}(H^2-E^2)y^2
	\right]\varphi(y)
	=\lambda\varphi(y)
\end{equation}
where $\alpha$ and $\lambda$ are $c$-numbers. Examination of Eq.~(\ref{eq:KG_develop}) is instructive. For $E=0$ one has an eigenvalue problem for the harmonic oscillator with the resulting Landau levels quantized in terms of $\hslash\omega_c$. In other words, one deals with the magnetic-type of motion. As long as $E<H$, i.e. as long as the magnetic term in the parentheses is larger than the electric one, one still has a parabolic well, so that the eigenenergies are quantized and the motion remains of the  magnetic type. However, for $E>H$ the coefficient in front of $y^2$ term becomes negative, there is no potential well anymore, i.e. no quantization, and one deals with electric-type solutions  (the Weber functions). For free electrons the transition between the two cases occurs for $E/H=1$, in agreement with the classical relativistic result. Thus relativity gives a symmetric description of the crossed-field situation with respect to magnetic and electric fields.

Going back to the relativistic analogy one should expect a similar description from the two-level model of the $\mathbf{P}\cdot\mathbf{p}$ theory supplemented by the presence of electric field. Thus, one should include in Eq.~(\ref{eq:ee_prob_1}) the electric potential $e\mathbf{E}\cdot\mathbf{r}$. Since the operator $\mathbf{r}$ is rigorously diagonal in the band index, see \textcite{Zak1966}, the complete formulation is
\begin{multline}
	\label{eq:ee_prob_w_EF}
	\sum\limits_l\left[
		\left(\dfrac{1}{2m_0}P^2
			+e\mathbf{E}\cdot\mathbf{r}+
			\varepsilon_{l0}-\varepsilon
		\right)\delta_{l\smash{'}l}
	+\right.\\
	\left.+\dfrac{1}{m_0}\mathbf{p}_{l\smash{'}l}\cdot\mathbf{P}\right]
	f_l(\mathbf{r})=0
\end{multline}
where, as before, $\mathbf{P}=\mathbf{p}+(e/c)\mathbf{A}$, and  $l'=1,2,3,\ldots$ runs over all bands. In Eq.~(\ref{eq:ee_prob_w_EF}) we have neglected the spin-orbit interaction. We now consider the two-band model neglecting free electron terms and obtain set of two equations
\begin{equation}
	\label{eq:2band_model}
	\left[
		\begin{array}{cc}
			-\varepsilon+\dfrac{\varepsilon_g}{2}+e\mathbf{E}\cdot\mathbf{r}&
			\dfrac{1}{m_0}\mathbf{p}_{\mathrm{cv}}\cdot\mathbf{\hat{P}}\\[1.5ex]
			\dfrac{1}{m_0}\mathbf{p}_{\mathrm{vc}}\cdot\mathbf{\hat{P}}&
			-\varepsilon-\dfrac{\varepsilon_g}{2}+e\mathbf{E}\cdot\mathbf{r}
		\end{array}
	\right]\left[
		\begin{array}{c}
			\vphantom{\dfrac{1}{m_0}}f_1\\[1.5ex]\vphantom{\dfrac{1}{m_0}}f_2
		\end{array}
	\right]=0
\end{equation}
Set (\ref{eq:2band_model}) has manifestly the structure of the Dirac equation with the band-edge energies and  potential on the diagonal and kinetic momenta off the diagonal. With the previous choice of $\mathbf{E}$ and $\mathbf{A}$ one can separate $x$ and $z$ variables and solve the set by substitution. The final equation is, see \textcite{Zawadzki1966}
\begin{multline}
	\left[
		\vphantom{\left(\dfrac{eE}{eEy-\varepsilon-\varepsilon_g/2}\right)^2}
		-\dfrac{\hslash^2}{2m^\ast_0}\dfrac{\partial^2}{\partial y^2}
		-\alpha y
		+\dfrac{m^\ast_0}{2}\left(\dfrac{e^2H^2}{{m^\ast_0}^2c^2}
		-\dfrac{2e^2E^2}{m^\ast_0\varepsilon_g}\right)y^2+\right.\\
	\left.
		+\frac{3}{8}\dfrac{\hslash^2}{m^\ast_0}
			\left(\dfrac{eE}{eEy-\varepsilon-\varepsilon_g/2}\right)^2
	\right]\varphi(y) = \lambda\varphi(y)
	\label{eq:ee_final_form}
\end{multline}
where $\alpha$ and $\lambda$ are $c$-numbers. The last term in the square bracket is small and can be neglected, see \textcite{Weiler1967}. Comparing Eq.~(\ref{eq:ee_final_form}) with the one-band equation (\ref{eq:eigenenergy_envelope}) one can see the basic difference between the two descriptions. The one-band description is linear in the electric potential, whereas in the two-band description the electric potential appears also squared. It is the negative quadratic term  that makes the difference. The relative importance of magnetic and electric quadratic terms  gives the possibility of having two types of motion, as observed experimentally. It follows from Eq.~(\ref{eq:ee_final_form}) that the transition between the magnetic and electric type of motion  occurs for $e^2H^2/{m^\ast_0}^2c^2=2e^2E^2/m^\ast\varepsilon_g$, i.e. for
\begin{equation}
	cE/H=u
\end{equation}
in analogy to the relativistic case (\ref{eq:KG_develop}), with $c$ replaced by $u$.

One can summarize the above analysis by saying that the nonrelativistic theory does not give a physically sound description of electrons in crossed magnetic and electric fields because it always predicts the magnetic type of motion, even for weak magnetic and strong electric fields. It is the relativistic theory that gives the physically sound result predicting the magnetic motion for strong magnetic and weak electric fields and the electric motion in the opposite case. This is in fact observed experimentally in semiconductors and correctly described with the use of the two-band model equivalent to semirelativity. In other words, the two-band description for electrons in crossed electric and magnetic fields leads not only to quantitative differences, as compared to the one-band description, but to the qualitatively different picture.

\begin{figure}
	\includegraphics[width=0.5\textwidth]{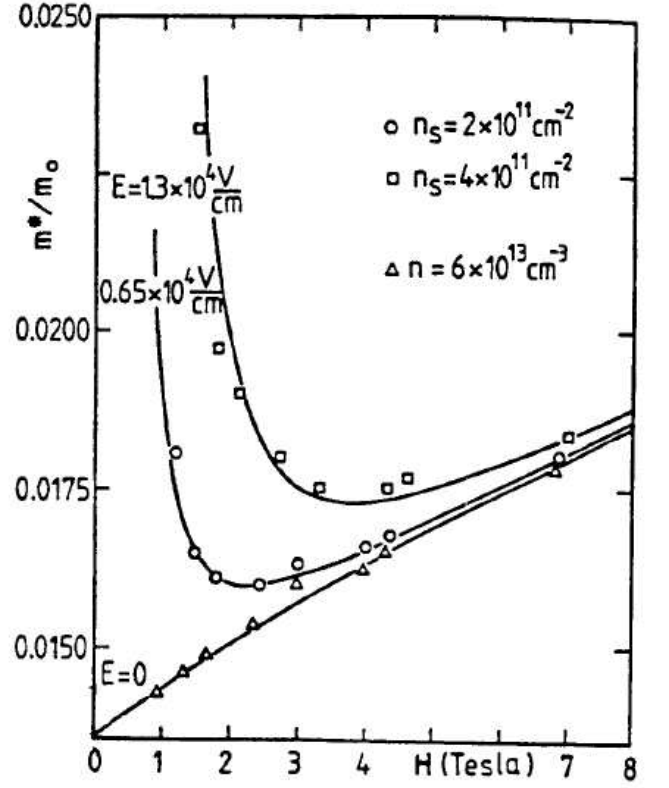}
	\caption{\label{fig:time-dil}Electron cyclotron masses in InSb in crossed magnetic and electric fields for two values of $E$ versus $H$. Electron masses for $E=0$ are included for comparison. Solid lines are calculated. A steep increase of values at low magnetic fields for $E\neq0$ is due to semirelativistic enhancement of the mass when the transverse velocity $v_\mathrm{tr}=cE/H$ becomes comparable to the maximum  velocity $u$. After \textcite{Zawadzki1985}.}
\end{figure}

Another spectacular demonstration of the semirelativistic behavior of NGS electrons in crossed fields is provided by experiments on the cyclotron resonance (CR). With the advancement of metal-oxide-semiconductor structures  it became much easier to apply strong electric fields without the danger of burning  semiconducting material. Figure \ref{fig:time-dil} shows experimental cyclotron mass of conduction electrons in InSb measured  in crossed fields as a function of $H$ for $E=0$ and two nonvanishing electric fields. For $E=0$ the CR mass  increases with $H$ because of band’s nonparabolicity, but begins at the band-edge value at $H=0$. On the other hand, for $E\neq0$ the CR mass is strongly enhanced as $H\rightarrow0$, which can be interpreted as the semirelativistic growth when $v_{\mathrm{tr}}= cE/H$ increases and approaches $u$.

To treat the problem theoretically one uses again the $\mathbf{P}\cdot\mathbf{p}$ theory supplemented by the electric potential $e\mathbf{E}\cdot\mathbf{r}$, see \textcite{Zawadzki1985, Zawadzki1986}. One needs here the three-level model since the electron spin $g$-factor is not negligible in InSb. One also assumes $\Delta\gg\varepsilon_g$ which is well satisfied for this material. Assuming the usual configuration, taking the same electric and magnetic potentials and separating the variables $x$ and $z$, one reduces the $2\times2$ set of equations into the following effective equation for the conduction band
\begin{equation}
	\label{eq:red_eff_conduction_band}
	\left[
		-\dfrac{\hslash^2}{2m^\ast_0}\dfrac{\partial^2}{\partial y^2}
		-\alpha y
		+\dfrac{m^\ast_0}{2}\omega_{\mathrm{eff}}^2y^2
	\right]\phi_\pm=\lambda_\pm\phi_\pm
\end{equation}
where $\omega_{\mathrm{eff}}^2=\omega_c^2-2e^2E^2/m^\ast_0\varepsilon_g$, $\alpha=\hslash\omega_c-2eE\varepsilon/\varepsilon_g$ and
\begin{equation}
	\lambda_\pm
		=\dfrac{\varepsilon^2}{\varepsilon_g}
		-\dfrac{1}{4}\varepsilon_g
		\mp\dfrac{1}{2}g^\ast_0\mu_BH
		-\dfrac{\hslash^2k_z^2}{2m^\ast_0}
		-\dfrac{\hslash^2k_x^2}{2m^\ast_0}
\end{equation}
in the standard notation. To arrive at the above equation one neglects the small term resulting from the noncommutation of $p_y$ and $eEy$ operators, see Eq.~(\ref{eq:ee_final_form}). The CR experiments are done in the magnetic regime of crossed fields, i.e. for $\omega_{\mathrm{eff}}>0$ which determines quantized energies. After completing the square and carrying the harmonic-oscillator-type quantization in Eq.~(\ref{eq:red_eff_conduction_band}), one solves the resulting quadratic equation for the energy and obtains for $k_z=0$
\begin{equation}
	\label{eq:solution_for_energy}
	\varepsilon(l,k_x,\pm)=
		\dfrac{eE}{H}\hslash k_x
		+(1-\delta^2)^\frac{1}{2}\left[\left(\dfrac{\varepsilon_g}{2}\right)^2
		+\varepsilon_gD_l^\pm\right]^\frac{1}{2}
\end{equation}
where
\begin{equation}
	D_l^\pm
		=\hslash\omega_c(1-\delta^2)^\frac{1}{2}(l+\dfrac{1}{2})
		\pm\frac{1}{2}g^\ast_0\mu_BH
\end{equation}
and $\delta^2=v_{\mathrm{tr}}^2/u^2$. Figure \ref{fig:time-dil} shows the theoretical   results for the cyclotron mass defined as $m^\ast=e\hslash H/(\varepsilon^+_{l+1}-\varepsilon^+_{l})c$ and calculated for $l=1$.

It is remarkable that the difference $\varepsilon(l+1)-\varepsilon(l)$ can be interpreted as the transverse Doppler shift (TDS) of the radiation frequency emitted by a moving source. This is related to the Voigt geometry of the experiment. In the special theory of relativity TDS is described by $\omega=\omega_0(1-v^2/c^2)^{\nicefrac{1}{2}}$, which corresponds to the $(1-\delta^2)^{\nicefrac{1}{2}}$ in front of the square root in Eq.~(\ref{eq:solution_for_energy}). Physically it means that the electron oscillates with the effective cyclotron frequency $\omega_{\mathrm{eff}}$ and simultaneously moves with the drift velocity $v_{\mathrm{tr}}=cE/H$. Because of the drift, the immobile observer sees the frequency reduced by the factor $(1-\delta^2)^{\nicefrac{1}{2}}$. In the theory of relativity TDS is considered to be a direct manifestation of the time dilatation.

It should be mentioned that \textcite{Aronov1991}, in their review of optical effects in semiconductors in the presence of crossed electric and magnetic fields also referred to the relativistic analogy.

%% file: zwitterbewegung.tex
\section{\label{sec:ZB}Zitterbewegung (Trembling Motion)}
Now we consider a somewhat mysterious phenomenon for electrons in vacuum and in semiconductors called the Zitterbewegung (ZB, trembling motion). In addition to the relativistic analogy, this phenomenon  gives us an opportunity to analyze an important difference between electron’s momentum and its quasimomentum in semiconductors.

The phenomenon of Zitterbewegung and its name were conceived by \textcite{Schrodinger1930} who observed that, in the Dirac equation, the $4\times4$ velocity matrices do not commute with the free-electron Hamiltonian. As a consequence, the electron velocity is not a constant of the motion also in absence of external fields. Such an effect must be of a quantum nature as it does not obey Newton’s first law of the classical motion. Schr\"{o}dinger’s idea stimulated numerous theoretical investigations but no experiments since the predicted ZB frequency is $\hslash\omega\simeq2m_0c^2\simeq\SI{1}{MeV}$ and its amplitude is about $\lambda_c=\hslash/m_0c\simeq\SI{3.86e-3}{\angstrom}$. These values are not accessible to current detection techniques. It was recognized that ZB is due to an interference of states corresponding to positive and negative electron energies resulting from the Dirac equation, see \textcite{Bjorken1964}, \textcite{Sakurai1967}, \textcite{Greiner1990}. \textcite{Lock1979} showed that, if the electron is represented by a wave packet, its ZB has a transient character, i.e. it decays with time.

\begin{figure}
	\includegraphics[width=0.5\textwidth]{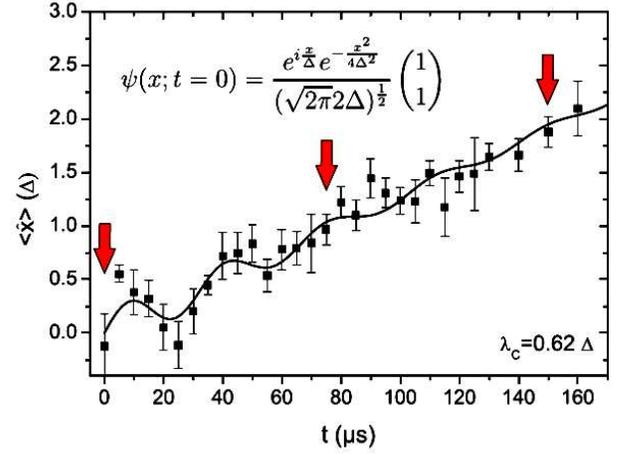}
	\caption{\label{fig:GerrZB}Simulation of the $(1+1)$ Dirac equation with the resulting  Zitterbewegung  of velocity by cold atomic ions interacting with laser radiation. After \textcite{Gerritsma2010}.}
\end{figure}

In Fig.~\ref{fig:GerrZB} we show the ZB phenomenon resulting from the one-dimensional Dirac equation, as simulated by \textcite{Gerritsma2010} with the use of laser beams interacting with cold atomic ions. An important property of this simulation is that one can modify the parameters of DE, so that both the frequency  and  amplitude of ZB acquire measurable values. It is seen that, since the electron is represented by a Gaussian wave packet, the Zitterbewegung has indeed a decaying character.

It was conceived years after the pioneering proposition of Schr\"{o}dinger’s that the trembling electron motion should occur also in crystalline solids if their band structure is represented by the two-band model reminiscent of the Dirac equation, see \textcite{Lurie1970}, \textcite{Ferrari1990}, \textcite{Vonsovskii1990}, \textcite{Zawadzki1997}. Thus we are back to the relativistic analogy. The simple reason for this result is that, for the two-band description, the velocity operator does not commute with the Hamiltonian. Let us consider the band Hamiltonian for an InSb-type narrow-gap material within the model including $\Gamma_6$ (conduction), $\Gamma_8$ (light and heavy hole), and $\Gamma_7$ (split-off) bands. It represents an $8\times8$ operator matrix \cite{Bowers1959, Zawadzki}. Assuming $\Delta\gg\varepsilon_g$ and neglecting the free-electron terms since they are negligible for NGS, one obtains a $6\times6$ Hamiltonian having $\pm\varepsilon_g/2$ terms on the diagonal and linear momenta $\mathbf{\hat{p}}$ off the diagonal, similarly to DE for free electrons. However, the three $6\times6$ matrices multiplying the momentum components do not have the properties of $4\times4$ Dirac matrices, which considerably complicates subsequent calculations. For this reason, with only a slight loss of generality, one takes $p_z\neq0$ and $p_x=p_y=0$. In the remaining matrix two rows  and columns corresponding to the heavy holes contain only zeros and they can be omitted. The remaining Hamiltonian for the conduction and light hole bands reads
\begin{equation}
	\label{eq:H_conduction_light_hole_band}
	\hat{H}=u\hat{\alpha}_3\hat{p}_z+\frac{1}{2}\varepsilon_g\hat{\beta}
\end{equation}
where $\hat{\alpha}_3$ and $\hat{\beta}$ are the well known $4\times4$ Dirac matrices. The Hamiltonian (\ref{eq:H_conduction_light_hole_band}) has the form appearing in DE, with $c$ replaced by $u$ and $m_0c^2$ by $\varepsilon_g/2$, so that it is possible to calculate the Zitterbewegung  following the procedure used by Schr\"{o}dinger \cite{Zawadzki2005, Zawadzki2011}.

\begin{figure}
	\includegraphics[width=0.5\textwidth]{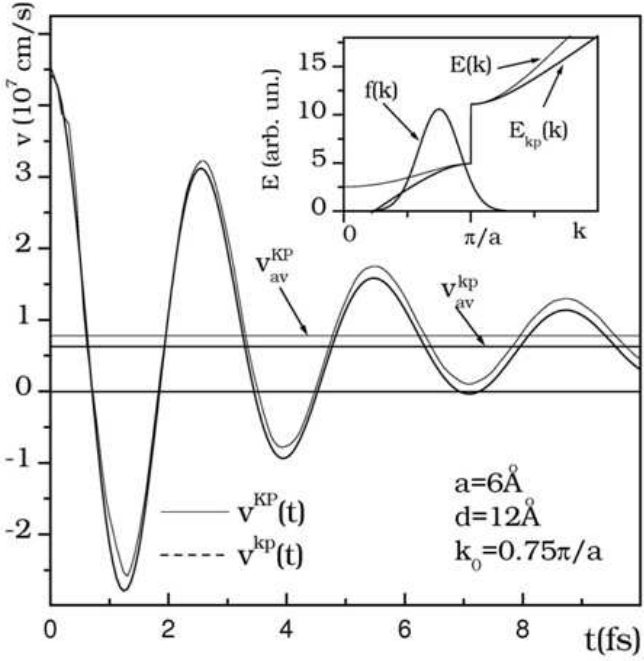}
	\caption{\label{fig:ZB-nat}Zitterbewegung (trembling motion) of velocity for electron  moving in the Kronig-Penney 1D periodic potential, as calculated from the two-band model  (shown in the inset) or directly from the potential. After \textcite{Zawadzki2010}.}
\end{figure}

The electron velocity is $\dot{z}=(1/i\hslash)[z,\hat{H}]=u\hat{\alpha}_3$. In order to determine $\hat{\alpha}_3(t)$ one calculates $\dot{\hat{\alpha}}_3(t)$ by commuting $\hat{\alpha}_3$ with $\hat{H}$. A resulting simple linear differential equation for $\hat{\alpha}_3(t)$ is solved, which gives $\dot{z}(t)$, and one calculates $z(t)$ integrating with respect to time. The final result is
\begin{equation}
	\label{eq:ZB_H_solution}
	z(t)=z(0)
		+\dfrac{u^2p_z}{\hat{H}}
		+\dfrac{i\hslash u}{2\hat{H}}\hat{A}_0
			\left[\exp\left(\dfrac{-2i\hat{H}t}{\hslash}\right)-1\right]\quad,
\end{equation}
where $\hat{A}_0=\hat{\alpha}_3(0)-up_z/\hat{H}$. There is $1/\hat{H}=\hat{H}/\varepsilon^2$, in which $\varepsilon$ is the electron energy. The first two terms of Eq.~(\ref{eq:ZB_H_solution}) represent the classical electron motion. The third term describes time dependent oscillations with a frequency of $\omega_z=\varepsilon_g/\hslash$. Since $\hat{A}_0\simeq1$, the amplitude of oscillations is $\hslash u/2m^\ast_0u^2=\lambda_z/2$. In Fig.~\ref{fig:ZB-nat} we show calculated velocity oscillations  for an electron moving in a one-dimensional Kronig-Penney periodic potential. Inset illustrates the two-band model resulting from the potential. The decreasing amplitude of oscillations is due to the fact that the electron is represented by a Gaussian wave packet, similarly to the situation showed in Fig.~\ref{fig:GerrZB}.

As mentioned above, in relativistic quantum  mechanics the analogous oscillations are called Zitterbewegung. We remark that the result (\ref{eq:ZB_H_solution}) is given in terms of operators. This means that, in order to get physically observable quantities, one must average this result over a state. The same remark applies to the original treatment of Schr\"{o}dinger’s. It should be noted that, while one obtains similar description of ZB for free relativistic electrons in vacuum and those in NGS due to the formal analogy between the Dirac equation and the two-band equation, physical reasons for the ZB phenomena in both cases  are different. In solids, the electron has an oscillating component of the velocity due to the motion in a periodic potential (see below), while free relativistic electrons do not move in a periodic potential and their “two-band” description results from the nature of Dirac formalism.

Finally, we emphasize that the phenomenon of ZB in solids contradicts the common conviction that electrons in a periodic potential behave like free particles with the electron mass replaced by an effective mass. In order to keep the total energy constant when moving in a periodic potential, an electron accelerates and slows down, see \textcite{Smith1961}, \textcite{Zawadzki2010}, \textcite{Zawadzki2013}. It is this energy conservation that is responsible for the Zitterbewegung in crystalline solids. The instantaneous velocity is related to electron’s momentum: $\mathbf{v}=\mathbf{p}/m_0$, which is not a constant of the motion in the presence of a periodic potential. It is the quasimomentum $\hslash\mathbf{k}$ which is a constant of the motion and the corresponding constant velocity $\mathbf{v}=\hslash\mathbf{k}/m^\ast$ is an \emph{average velocity} of the electron in a crystal, see \textcite{Kireev1978}, \textcite{Zawadzki2013}.

\textcite{Wilam2010} investigated
experimentally and theoretically the spin resonance in asymmetric silicon
structures in a magnetic field. The findings were analyzed in terms of the
Rashba spin splitting that causes non-commutativity of the velocity and
Hamiltonian operators. The precession of electron spin with the Larmor
frequency results via the spin-orbit interaction in an ac current. The
latter is a source of the Joule heat which is manifested in additional
effects in the spin resonance.

Very recently, a coherent Zitterbewegung of electrons was observed experimentally in n-type InGaAs in the presence of a magnetic field by \textcite{Stepanov2016}. The Zitterbewegung of electron velocity originates in the interference between two spin states split by the magnetic field, while non-commutativity of the Hamiltonian and velocity operators is related to the spin-orbit interaction manifested in the Bychkov-Rashba and Dresselhaus interactions. Many electrons tremble with the same phase being all excited across the InGaAs gap with laser pulses to the same spin state. The ZB motion is measured as an AC current. The amplitude of ZB oscillations is estimated to be about \SI{20}{nm} and their frequency at the field $\mathbf{B} = \SI{3}{T}$ is \SI{26}{GHz}.

%% file: graphene.tex
\section{Graphene}
Now we describe monolayer graphene discovered by \textcite{Novoselov2004}. As mentioned in the Introduction, graphene, together with carbon nanotubes and topological insulators, are considered separately from the 3D materials because they are characterized by reduced dimensionalities, have different crystal symmetries from other NGS and possess rather special properties.

\begin{figure}
	\includegraphics[width=0.4\textwidth]{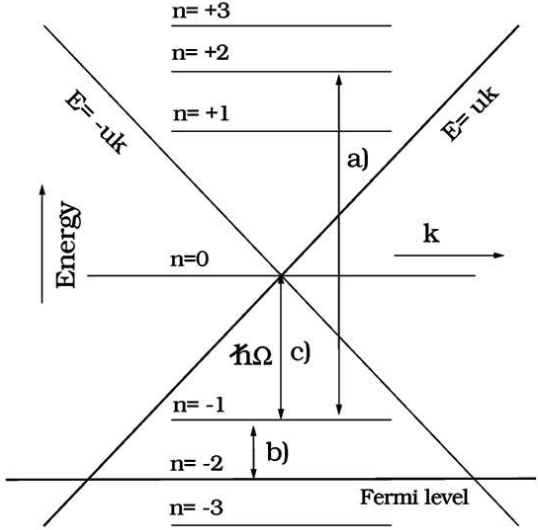}
	\caption{\label{fig:Gr-linear}Linear energy dispersion $\varepsilon(k)$ and the Landau levels for monolayer graphene in a magnetic field (schematically). The basic energy is $\hslash\Omega = \sqrt{2}\hslash u/L$.}
\end{figure}

Graphene is a two-dimensional (2D) monolayer of carbon atoms with the honeycomb atomic array. Graphene’s band structure, at each of the two K points of the Brillouin zone, is described by the 2D $\mathbf{k}\cdot\mathbf{p}$ Hamiltonian (omitting spin). It can be considered as a special case of the two-band model with a vanishing energy gap, see \textcite{Wallace1947}, \textcite{Slonczewski1958}, \textcite{Semenoff1984},
\begin{equation}
	\label{eq:graphene_H}
	\hat{H}=u\left[
		\begin{array}{cc}
			0&
			\hat{p}_x-i\hat{p}_y\\
			\hat{p}_x+i\hat{p}_y&
			0
		\end{array}
	\right]\quad ,
\end{equation}
where $u$ has dimensions of velocity. In the absence of external fields, solutions of the eigenenergy equation are given by 2D exponentials $\exp(i\mathbf{k}\cdot\mathbf{p})$ and it is easy to see that the resulting energy dispersion is linear in quasimomentum
\begin{equation}
\varepsilon=\pm u\hslash k,
\end{equation}
where $k^2=k_x^2+k_y^2$. This dispersion is shown in Fig.~\ref{fig:Gr-linear}. In view of the relativistic analogy, this situation may be regarded as “extreme relativistic limit” because the linear $\varepsilon(k)$ relation in the semirelativistic and relativistic dispersions corresponds to high values of momentum and high velocities. Since the velocity is $v=d\varepsilon/d\hslash k=u$, its absolute value is always equal to $u$ independently of the value of $k$.  Remarkably, $u\simeq\SI{1e8}{cm/s}$ is again very close to the “universal” value for many materials, although the symmetry of bands is different, see \textcite{Novoselov2005}, \textcite{Zhang2005}.

The velocity effective mass can be defined as before: $m^\ast v=\hslash k$, which leads to: $1/m^\ast=(1/\hslash^2k)d\varepsilon/dk$. Thus
\begin{equation}
	\label{eq:gr_vel_eff_mass}
	m^\ast=\dfrac{\hslash k}{u}=\dfrac{\varepsilon}{u^2}\quad .
\end{equation}
This means that at $k=0$, i.e. at the band edge (called “the Dirac point”), the effective mass vanishes and one deals with 2D “massless fermions”. However, as the energy increases the mass increases as well and is not zero anymore. It follows
\begin{equation}
	\varepsilon=m^\ast u^2\quad.
\end{equation}
Thus the analogy to the relativistic relation holds. If one applies an external electric force to the electron, the change of its energy cannot change the velocity, so it goes entirely into the change of the effective mass according to Eq.~(\ref{eq:gr_vel_eff_mass}).

\begin{figure}
	\includegraphics[width=0.4\textwidth]{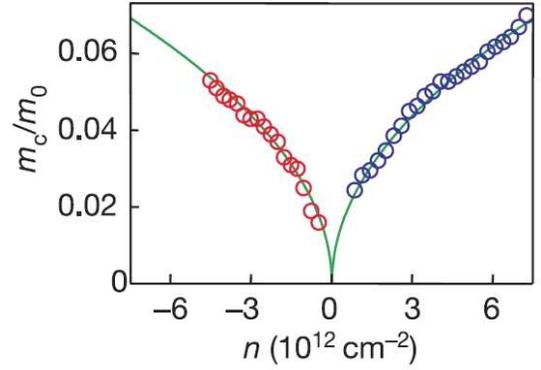}
	\caption{\label{fig:Gr-Nvs}Velocity effective mass at the Fermi energy versus electron density in monolayer graphene determined by the Shubnikov-deHass effect. Red circles -- valence band, blue circles -- conduction band. The dependence of the mass on $n^{\nicefrac{1}{2}}$ corresponds to the linear band dispersion. After \textcite{Novoselov2005}.}
\end{figure}

The proportionality of the mass to quasimomentum $\hslash k$, which is a signature of the linear $\varepsilon(k)$ dispersion, can be verified experimentally. For the degenerate 2D electron gas the density is $n=k^2/\pi$ (including spin and valley degeneracy). This gives the mass at the Fermi level
\begin{equation}
m^\ast=\dfrac{\hslash(\pi n)^\frac{1}{2}}{u}\quad.
\end{equation}
Figure~\ref{fig:Gr-Nvs} shows dependence of the velocity mass in graphene on $n^{\nicefrac{1}{2}}$ as measured by \textcite{Novoselov2005} with the use of Shubnikov-deHass effect for both positive and negative electron energies.

\begin{figure}
	\includegraphics[width=0.4\textwidth]{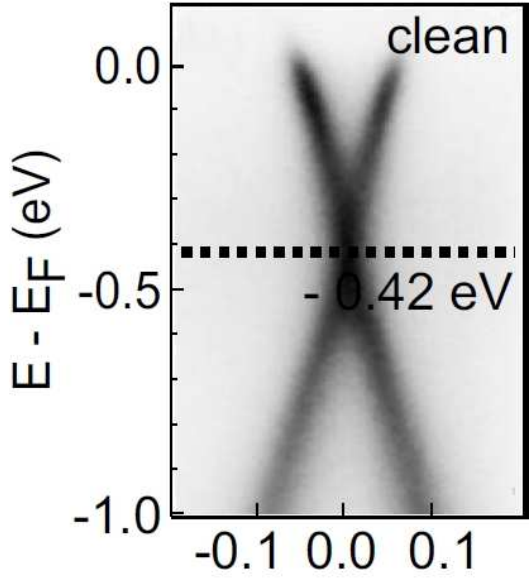}
	\caption{\label{fig:GrARP}Linear energy dispersions of conduction and valence bands in monolayer graphene determined by angle resolved photoemission spectroscopy (ARPES). The linear $\varepsilon(k)$ dispersions correspond to the extreme semirelativistic regime. After \textcite{Coletti2010}.}
\end{figure}

The linear $\varepsilon(k)$ band dispersions in monolayer graphene have been spectacularly illustrated in angle-resolved photoemission spectroscopy (ARPES), a powerful technique which is able to trace energy bands in thin layers of solids. In Fig.~\ref{fig:GrARP} we quote an example of such studies of graphene energy bands.

\begin{figure}
	\includegraphics[width=0.5\textwidth]{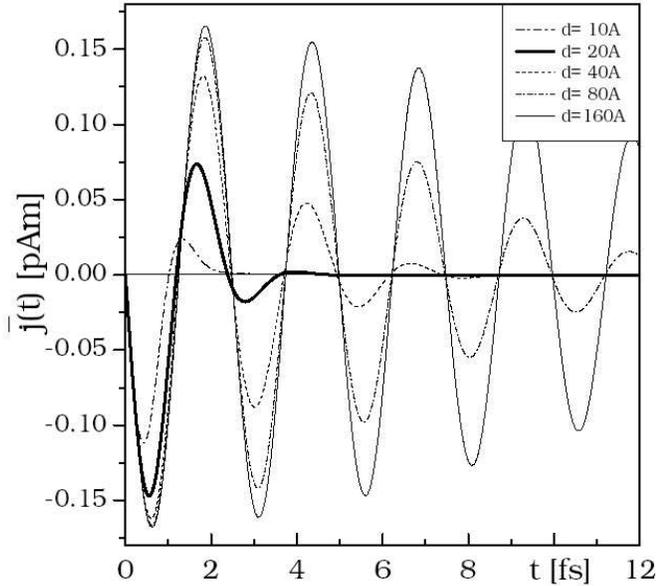}
	\caption{\label{fig:Gr-ZB}Oscillatory electric current caused by the Zitterbewegung (trembling motion) in monolayer graphene versus time, as calculated for a Gaussian wave packet with $k_{y0} = \SI{1.2e9}{m^{-1}}$ and various packet widths.  The transient character of ZB due to the packet representation is seen. After \textcite{Rusin2007}.}
\end{figure}

Next we consider the phenomenon of electron Zitterbewegung  in monolayer graphene, i.e. in the “extreme relativistic limit” of zero energy gap, see \textcite{Rusin2007}. Using the Hamiltonian (\ref{eq:graphene_H}) one calculates the quantum velocity: $v_i=\partial\hat{H}/\partial p_i$. The latter does not commute with the Hamiltonian, so that the velocity is not a constant of the motion. In the Heisenberg picture there is $\mathbf{\hat{v}}(t)=\exp(i\hat{H}t/\hslash)\mathbf{\hat{v}}\exp(-i\hat{H}t/\hslash)$. Using Eq.~(\ref{eq:graphene_H}) one obtains
\begin{equation}
	v_x^{(11)}=u\dfrac{k_x}{k}\sin(2ukt)\quad.
\end{equation}
The above equation describes the trembling motion with the frequency $\omega_z=2uk$, determined by the energy difference between upper and lower energy branches for a given value of $k$. One calculates an average velocity of a charge carrier represented by a 2D Gaussian wave packet with the nonzero momentum $k_0$. The results for an electric current $j_x=ev_x$ are plotted in Fig.~\ref{fig:Gr-ZB} for different packet widths. It is seen that the ZB frequency does not depend on the width, while the amplitude and the decay time do. One can compare Figs.~\ref{fig:GerrZB} and \ref{fig:Gr-ZB} noting that in Fig.~\ref{fig:GerrZB} one plots  electron’s position while in Fig.~\ref{fig:Gr-ZB} the current is proportional to velocity. It can be concluded that, once again, free relativistic electrons and semiconductor electrons described  by the two-band model behave quite similarly.

\begin{figure}
	\includegraphics[width=0.5\textwidth]{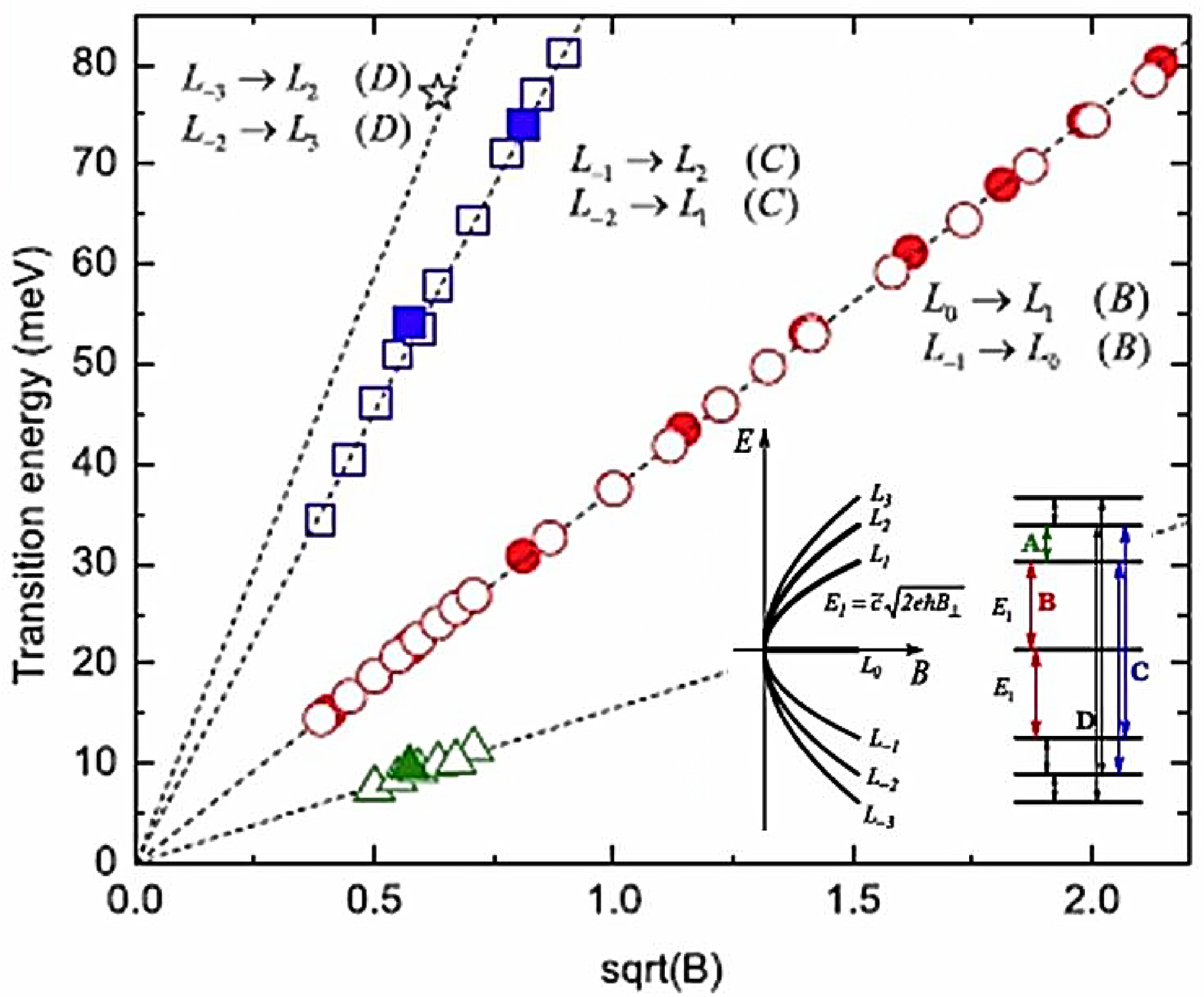}
	\caption{\label{fig:Gr-Sad}Energies of experimental magneto-optical transitions in ultrathin graphite layers (indicated in the inset) plotted versus $H^{\nicefrac{1}{2}}$. Black symbols are obtained for tilted magnetic field keeping the perpendicular field component constant. The dashed lines are calculated with the same velocity $u = \SI{1.03e8}{cm/s}$. After \textcite{Sadowski2006}.}
\end{figure}

The “ultrarelativistic” nature of electrons in graphene is also reflected  in the  presence of a magnetic field. The Hamiltonian for this situation is (spin is omitted)
\begin{equation}
	\hat{H}=\left[
		\begin{array}{cc}
			0&
			\hat{P}_x-i\hat{P}_y\\
			{P}_x+i{P}_y&
			0
		\end{array}
	\right]\quad,
\end{equation}
where $\hat{P}_i=\hat{p}_i+eA_i$. Using the asymmetric gauge: $\mathbf{\hat{A}}=[-Hy,0,0]$ for $\mathbf{H}\parallel z$ transverse to the monolayer plane, one separates $x$ variable by taking $\Psi=\exp(ik_xx)\phi(y)$. Using the magnetic radius $L$, the variable $\xi=y/L-k_xL$, and defining the standard raising and lowering operators for the harmonic oscillator $\hat{a}=(\xi+\partial/\partial\xi)$ and $a^{\dagger}=(\xi-\partial/\partial\xi)$ the Hamiltonian is rewritten in the form
\begin{equation}
	\hat{H}=-\hslash\Omega\left[
		\begin{array}{cc}
			0&\hat{a}\\\hat{a}^{\dagger}&0
		\end{array}
	\right]
\end{equation}
where the frequency is $\Omega=\sqrt{2}u/L$. The solutions are given in terms of the oscillator functions $\phi_n(\xi)$ and the eigenenergies are
\begin{equation}
	\varepsilon_{ns}=s\hslash\Omega\sqrt{n}\quad ,
\end{equation}
where $n=0,1,2,3,\ldots$ and plus and minus signs of $s$ correspond to the conduction and valence bands, respectively. Thus, importantly, the spectrum of orbital magnetic quantization contains the zero of energy. This is illustrated in  Fig.~\ref{fig:Gr-linear}.

The above treatment does not include electron spin. The spin g-factor for electrons in graphene is almost exactly +2 because the spin-orbit interaction in carbon is very weak. As a consequence, the spin splitting is much smaller than the orbital splitting  and does not behave in the semirelativistic manner.

The magnetic quantization described above was confirmed in intraband and interband magneto-optical experiments performed on very thin graphite layers, see Fig.~\ref{fig:Gr-Sad}. The involved transitions are indicated in the inset. The experiment  verifies  main conclusions of the theory: energy dependence on $(nH)^{\nicefrac{1}{2}}$, existence of the zero-energy level and the velocity value  $u= \SI{1.03e8}{cm/s}$. Interestingly, as we indicated above, the magnetic levels resulting from the Dirac equation follow the same scheme as the one shown in Fig.~\ref{fig:Gr-linear}. Because the spin splitting “compensates”  the orbital splitting, the level $\mathscr{E}=0$ appears. Thus the relativistic analogy still holds, even though in this case it occurs  somewhat accidentally. One should remark on this occasion that the name  “Dirac fermions” given to electrons in graphene does not apply to their spin.

%% file: nanotubes.tex
\section{Carbon Nanotubes}
Carbon nanontubes (CNT) are graphene sheets rolled into tubes. They represent one-dimensional (1D) systems in which charge carriers can move only along the tubes (in y direction). Omitting spin, the $\mathbf{k}\cdot\mathbf{p}$ Hamiltonian at the K point of the Brillouin zone is a $2\times2$ operator, see \textcite{Ajiki1993}
\begin{equation}
	\hat{H}=u\left[
		\begin{array}{cc}
			0	&	a_n-i\hat{p}\\
			a_n+i\hat{p}	&	0
		\end{array}
	\right]\quad,
\end{equation}
where $u$ is a velocity coefficient, $\hat{p}$ is the quasimomentum in the $y$ direction. This Hamiltonian is similar to that given for graphene in Eq.~(\ref{eq:graphene_H}) except that, because of the periodic boundary conditions around the tube’s circumference $L_c$, the quasimomentum $p_x$ is quantized and takes discrete values $a_n=\hslash k_x(n)=\hslash(2\pi/L_c)(n-\nu/3)$, where $n=0,\pm 1,\pm 2\ldots$, and $\nu=\pm 1$ for semiconducting CNT. The resulting subband energies are  
\begin{equation}
	\label{eq:NT_subband_energies}
	\varepsilon(k) =\pm u\left(a_n^2+\hslash^2k^2\right)^\frac{1}{2}\quad,
\end{equation}
where $\hslash k$ replaces $p$. The upper signs are for the conduction and the lower for the valence subbands. The above relation is the 1D analog of the dispersion for free relativistic electrons in vacuum  and  can be cast into the standard form of Eq.~(\ref{eq:lh_bands_conduction_approx}), see \textcite{Zawadzki2006}. The geometry of CNT has important consequences. There exist energy gaps $\varepsilon_g=2ua_n$ and effective masses at the subband edges: $m^\ast_0 =a_n/u$, which have different values for various subbands.

The electron velocity is $v=d\varepsilon/d\hslash k=u^2\hslash k/\varepsilon$. For large $k$ the velocity reaches saturation value $u=(\varepsilon_g/2m^\ast_0)^{\nicefrac{1}{2}}$, the same for all subbands. As for other NGS, the maximum velocity $u$ plays for electrons in CNTs the role of the light velocity $c$ in relativity. We define an energy-dependent effective mass $m^\ast$ relating velocity to quasimomentum: $m^\ast v = \hslash k$, and calculate $m^\ast = \hslash k/v = \varepsilon/u^2$, which gives the 1D analogue of the relativistic formula. We can also express the mass $m^\ast$ by the velocity. Beginning with the relation $\hslash^2k^2u^2=v^2\varepsilon^2/u^2$, using $\varepsilon^2=(m^\ast_0u^2)^2+u^2\hslash^2k^2$ and solving for quasimomentum, one obtains
\begin{equation}
	\hslash k=m^\ast_0\gamma v\quad,
\end{equation}
where $\gamma=(1-v^2/u^2)^{-\nicefrac{1}{2}}$. Using the above definition $\hslash k =m^\ast v$ we have $m^\ast=m_0\gamma$ which has the familiar relativistic form with $u$ replacing $c$. In this notation the semirelativistic formula reads $\varepsilon=m^\ast_0\gamma u^2$.

Next we assume, similarly to the special relativity: $d(\hslash k)/dt = F$, where $F$ is the force. One can now define another effective mass $M^\ast$, relating force to acceleration
\begin{equation}
	M^\ast a=F\quad.
\end{equation}
Since $a=dv/dt=(dv/d(\hslash k)(d\hslash k/dt)=d^2\varepsilon/d(\hslash k)^2F$, one obtains $1/M^\ast=d^2\varepsilon/d(\hslash k)^2$. With the use of the dispersion (\ref{eq:NT_subband_energies}) one has finally
\begin{equation}
	M^\ast=\dfrac{\varepsilon^2}{{m^\ast_0}^2u^6}=m^\ast_0\gamma^3\quad.
\end{equation}
This again has the corresponding relation in special relativity when the acceleration is parallel to the force.

It is instructive to estimate the introduced quantities. One obtains $u = \SI{0.98e8}{cm/s}$, see \textcite{Ajiki1993}. This is, again, very similar to the value of $u$ obtained for other narrow gap materials. The lowest energy gap is $\varepsilon_g=2ua_0$, where $a_0=\hslash 2\pi/3L_c$. For the circumference $L_c=\SI{60}{\angstrom}$ one gets $\varepsilon_g(0)=\SI{0.45}{eV}$. The effective mass is $m^\ast_0/m_0=a_0/um_0= 0.041$ for the same conditions. The above parameters are similar to those of typical narrow-gap InAs. However, CNTs of higher diameters have smaller $\varepsilon_g$ and $m^\ast_0$.

When considering quantum effects in CNT it is again useful to introduce the quantity 
\begin{equation}
\lambda_z=\dfrac{\hslash}{m^\ast_0u}=\dfrac{\hslash}{a_n}\quad.
\end{equation}
This length is analogous to the Compton wavelength $\lambda_c$ for relativistic electrons. Again, $\lambda_z$ is a few orders of magnitude larger than $\lambda_c$. Taking $m^\ast_0 = 0.041 m_0$ and $u = \SI{0.98e8}{cm/s}$  one obtains $\lambda_z=\SI{28.6}{\angstrom}$. Using the procedure presented above one can calculate the 1D phenomenon of Zitterbewegung in analogy to the idea described by \textcite{Schrodinger1930}. In addition to the classical motion one obtains an oscillatory component with the frequency $\hslash\omega_z\simeq\varepsilon_g$ and the amplitude $\lambda_z$, see \textcite{Rusin2007}.

There exist also metallic carbon nanotubes with the vanishing energy gap \cite{Saito1998}. The $\varepsilon(k)$ dispersion in the lowest 1D subband of such systems is linear in quasimomentum . In terms of the relativistic analogy this situation represents the “ultrarelativistic” regime and the properties of charge carriers are similar to those in graphene, as discussed above. All in all, the above examples of graphene and carbon nanotubes demonstrate that the relativistic analogy extends to two-dimensional and one-dimensional semiconductor systems.           

%% file: insulators.tex
\begin{figure}
	\includegraphics[width=0.5\textwidth]{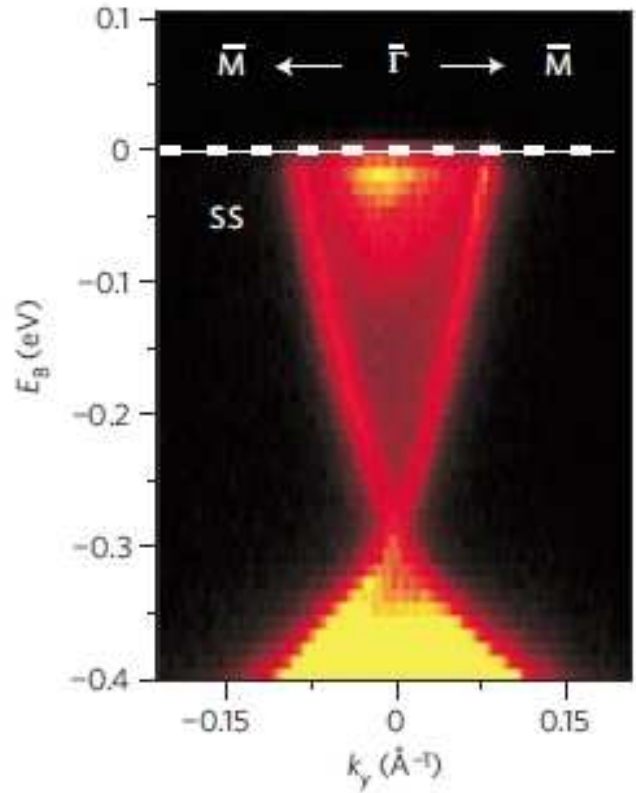}
	\caption{\label{fig:TI-Xia}Energy dispersions $\varepsilon(k)$ of topological surface states in $\mathrm{Bi}_2\mathrm{Se}_3$ observed by ARPES studies. The linear $\varepsilon(k)$ conduction band corresponds to the extreme semirelativistic regime. After \textcite{Xia2009}.}
\end{figure}

\section{Topological Insulators}

\begin{figure}
	\includegraphics[width=0.5\textwidth]{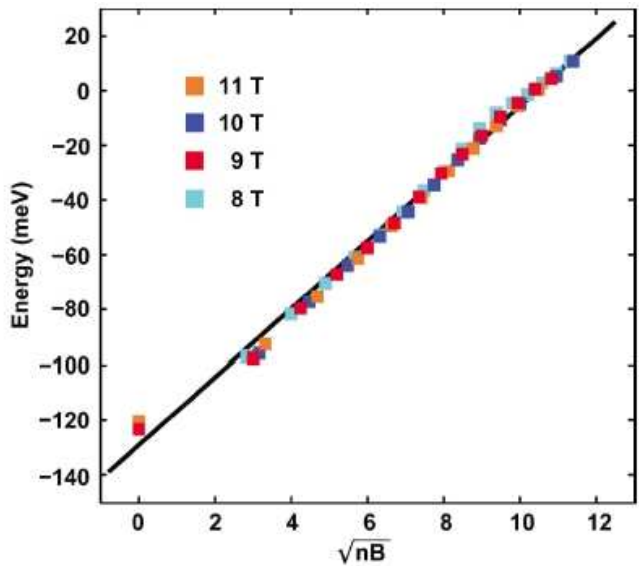}
	\caption{\label{fig:TI-magnCh}Energies of STM peaks observed on $\mathrm{Bi}_2\mathrm{Se}_3$ topological surface states quantized into Landau levels $n$ by magnetic field, plotted versus  $(nH)^{\nicefrac{1}{2}}$. After \textcite{Cheng2010}.}
\end{figure}

Finally, we briefly consider very intensively studied subject of the so called topological insulators (TIs). We do not go into the origin of TIs and are not in a position to quote important papers on the subject since there are very many of them. We concentrate, as before, on the semirelativistic aspect of these systems. In the simplest description, a 3D TI is a 2D metallic-like state on the surface of a 3D insulator. It appears that the first theoretical possibility of such  states had been predicted by \textcite{Volkov1985} who observed that, if one puts into contact two pieces of $\mathrm{Pb}_{1-x}\mathrm{Sn}_{x}\mathrm{Te}$ having opposite signs of energy gap, one finds at the contact 2D states having linear $\varepsilon(k)$ dispersion (massless fermions). \textcite{Kane2005} came up with the “modern” theoretical version of similar states which has started a real surge of theoretical and experimental investigations continuing until present.

\begin{figure*}
	\includegraphics[width=0.9\textwidth]{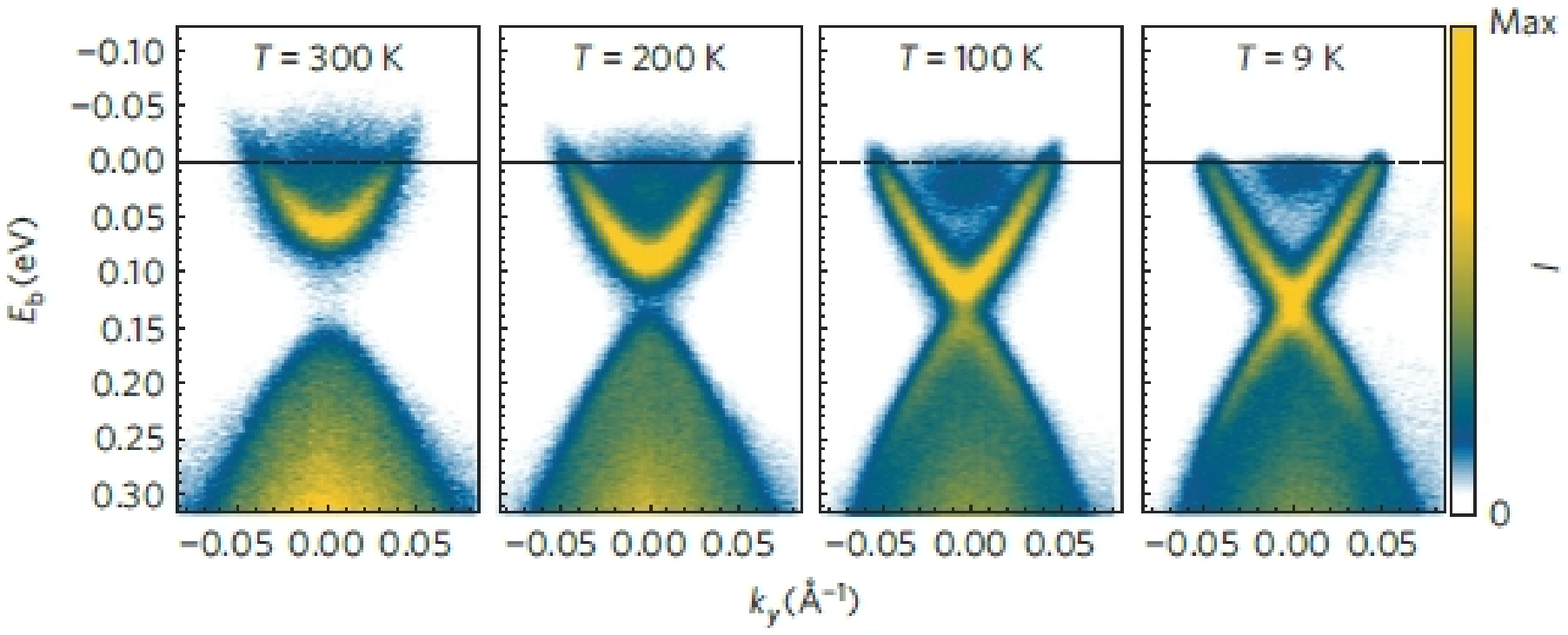}
	\caption{\label{fig:TIdziaw}Energy dispersions of surface states in $\mathrm{Pb}_{0.77}\mathrm{Sn}_{0.23}\mathrm{Se}$ for four temperatures, as observed by ARPES studies. Panels at \SI{300}{K} and \SI{200}{K} correspond to the “trivial” order of bulk energy bands, panels at $T\leq\SI{100}{K}$ correspond to the “nontrivial” order of bulk bands at which metallic crystalline surface states with zero gap and linear energy dispersions are formed at the surface. The linear $\varepsilon(k)$ bands correspond to the extreme semirelativistic regime. After \textcite{Dziawa2012}.}
\end{figure*}

\begin{figure}
	\includegraphics[width=0.36\textwidth]{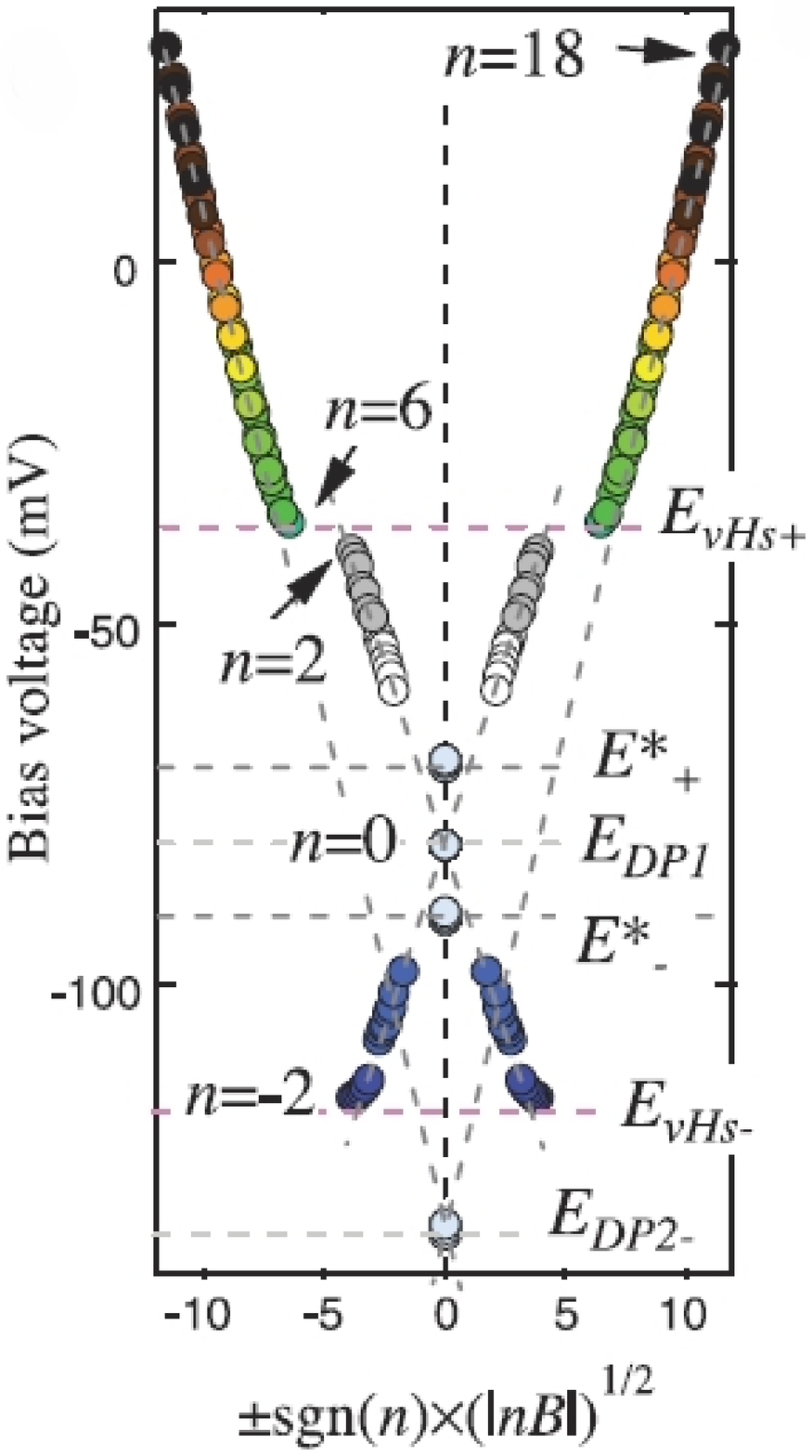}
	\caption{\label{fig:TI-magOk}Energies of STM peaks observed on $\mathrm{Pb}_{0.66}\mathrm{Sn}_{0.34}\mathrm{Se}$ crystalline surface states quantized into Landau levels $n$ by magnetic field, plotted versus $(nH)^{\nicefrac{1}{2}}$. After \textcite{Okada2013}.}
\end{figure}

Figure~\ref{fig:TI-Xia} shows $\varepsilon(k)$ dispersion of a 2D surface state observed by means of angle resolved  photoemission spectroscopy (ARPES) on the surface of $\mathrm{Bi}_2\mathrm{Se}_3$. It is seen that one deals with an almost linear dispersion, but the $\varepsilon(k)$ branches below and above the Dirac point are not symmetric, unlike the dispersion in graphene. Still, in terms of the relativistic analogy this dispersion represents for the upper branch the extreme semirelativistic regime of massless fermions. As we showed above for graphene, in the presence of a magnetic field the characteristic $2\times2$ Hamiltonian of Eq.~(\ref{eq:graphene_H}) gives the orbital quantized energies: $\varepsilon(n) = \varepsilon_D + u(2eHn/c)^{\nicefrac{1}{2}}$ with $n=0, 1, 2,\ldots$. Figure~\ref{fig:TI-magnCh} illustrates the $(nH)^{\nicefrac{1}{2}}$ dependence of magnetic quantization of surface states in $\mathrm{Bi}_2\mathrm{Se}_3$ observed with the use of scanning tunneling microscopy (STM). It should be noted that the spin splitting of the energies  exists in TI in the presence of a magnetic field but this is not observed in STM spectra because the spin is coupled to quasimomentum, so that spin-up and spin-down states belong to different dispersion cones.

An interesting situation occurs in $\mathrm{Pb}_{1-x}\mathrm{Sn}_{x}\mathrm{Se}$ mixed crystals in which the so called topological crystalline insulator (TCI) states are formed. The TCIs are special kinds of TIs, related not to the time and space inversion symmetry of their “host” crystals, but to the symmetry of energy bands in their “host” crystals, see \textcite{Fu2011}, \textcite{Hsieh2012}. And so, some band orderings are called “trivial” because they do not allow for TCI formation, whereas others are “nontrivial” since they allow for TCIs. As we mentioned above, see section \ref{sec:band_struct}, the band structure of lead chalcogenides is very sensitive to temperature and can reach, as $T$ decreases, zero of energy gap and change the band ordering from trivial to nontrivial in the above sense. In PbSnSe system the trivial ordering corresponds to the $L_6^-$ band above the $L_6^+$ band and the nontrivial one to the reversed order. Critical temperatures $T_c$ vary for different chemical compositions $x$. In $\mathrm{Pb}_{0.77}\mathrm{Sn}_{0.23}\mathrm{Se}$ the critical temperature is below \SI{100}{K}.

In Fig.~\ref{fig:TIdziaw} we quote ARPES spectra of surface states in $\mathrm{Pb}_{0.77}\mathrm{Sn}_{0.23}\mathrm{Se}$ for four decreasing temperatures in the range of $\SI{300}{K}<T<\SI{9}{K}$. As the temperature is lowered, the spectra show an evolution of the energy gap and the corresponding curvatures of conduction and valence bands to the situation  with zero gap and the corresponding linear energy bands. At the critical temperature the gap in the bulk vanishes and the transition from trivial to nontrivial band ordering occurs. When the  temperature is lowered further, the zero gap and linear energy branches of the surface states remain since one continues to be in the nontrivial band ordering of the bulk crystal.

As to the relativistic analogy, the first three panels of Fig.~\ref{fig:TIdziaw} can be regarded as a textbook illustration of relativistic-like energy bands for decreasing gaps and the resulting diminishing band-edge effective masses. They also illustrate that, as the gap decreases, the gradual transition from parabolic to linear $\varepsilon(k)$ band dispersion occurs at smaller $k$ values. In addition, in PbSnSe the conduction and valence branches are almost mirror images of each other, which is often not the case in other materials, cf. Fig.~\ref{fig:TI-Xia}. All in all, the first three panels follow almost literally the scheme presented in our Fig.~\ref{fig:Dir-NGS}.

Again, the linear $\varepsilon(k)$ energies in the presence of a magnetic field are quantized into $\varepsilon\sim(nH)^{\nicefrac{1}{2}}$, where $n=0,1,2,\ldots$. This “extreme relativistic” quantization is verified on TCI in PbSnSe by STM studies, as illustrated in Fig.~\ref{fig:TI-magOk}. One should, however, remark that the $(nH)^{\nicefrac{1}{2}}$ quantization is not rigorously observed in all TI, see e.g. \textcite{Hanaguri2010}.

One concludes from our brief review of graphene, carbon nanotubes and topological insulators that electrons in these 2D and 1D systems fit very well into the frame of relativistic analogy, representing in fact the “extreme semirelativistic” conditions. In other words, their properties extend the relativistic analogy to systems of reduced dimensionality which do not exist in vacuum.

%% file: discussion.tex
\section{Discussion}
As we mentioned above, the analogy between behavior of free relativistic electrons in vacuum and that of electrons in narrow-gap semiconductors, although far reaching, is subject to approximations and limitations. The approximations, enumerated in section \ref{sec:band_struct}, are not essential as they are well satisfied, especially in NGS. On the other hand, as shown at the end of section \ref{sec:band_struct}, first nonparabolic corrections to any spherical energy band are always of the relativistic type. The restrictions present a more serious limitation. The relativistic analogy extends to not too high energies in the band, approximately to the inflection points on the $\varepsilon(k)$ dispersion curves. At higher energies the slopes of $\varepsilon(k)$ dependences begin to decrease and the analogy fails. Still, in practice the relativistic-like dispersions usually work very well for achievable charge carrier densities in III-V, II-VI and IV-VI NGS compounds, as illustrated in Figs.~\ref{fig:mInSb}, \ref{fig:mHgSe} and \ref{fig:mPbSe}, respectively. For spheroidal energy bands the analogy holds after a simple transformation of the $k$-space. On the other hand, it works only approximately for light holes in warped valence bands. The analogy involves the momentum of electrons in vacuum  and quasimomentum of electrons in semiconductors, as explicitly seen in the phenomenon of Zitterbewegung discussed in Section \ref{sec:ZB}. We emphasize this point, as it is rarely treated in textbooks.

We discussed above quite a few phenomena and properties for illustrating the relativistic analogy, but this catalogue is by no means complete. It is well known from the relativistic quantum mechanics that solutions of the Dirac equation are four-component spinors. The two-band $\mathbf{k}\cdot\mathbf{p}$ theory gives similar multi-component wave functions, as seen for example in Section \ref{sec:e_in_mag_field}. But multi-component  functions give also rise to scattering effects for both relativistic and NGS electrons which were not discussed above \cite{Zawadzki1982}. This aspect is related to the so called Foldy-Wouthuysen transformation which allows one to separate four-component spinors into two-component functions corresponding to positive and negative energies in DE \cite{Foldy1950}. An analogous transformation exists also for the two-band model in $\mathbf{k}\cdot\mathbf{p}$ theory \cite{Zawadzki2005, Zawadzki2006} which was not discussed above. A striking similarity between the Dirac equation for vacuum and the two-band $\mathbf{k}\cdot\mathbf{p}$ model for NGS is provided also by the fact that, according to the Dirac interpretation, the negative energies resulting from DE are completely filled with electrons (the Fermi sea), so that an electron-positron pair creation by a photon may be  interpreted as an excitation of an electron across the gap $2m_0c^2$ leaving a hole in the Fermi sea. An identical interpretation is given in semiconductors to an electron excitation by a photon from the full valence to the empty conduction band leaving a hole in the valence band. Tracing the relativistic analogy for many-body phenomena would require an application of the quantum field theory which we did not attempt here.

It is known by now that the so called Dirac cones, i.e. linear $\varepsilon(k)$ energy dispersions of cylindrical symmetry, exist also in organic conductors \cite{Tajima2007} and d-wave superconductors \cite{Balatsky2006}. More such systems will certainly appear in the future which will extend the relativistic analogy beyond narrow gap semiconductors. As we showed above, the analogy is not limited to three dimensions, but is valid also for 2D (graphene, some topological  insulators)  and 1D (carbon nanotubes) physical objects.

With regard to graphene and topological insulators we want to stress again that the terms “Dirac fermions”, “Dirac point”, “Dirac cone” et cetera, are somewhat misleading because they refer in the recent usage only to the linear dispersion relations, whereas the Dirac equation gives for relativistic electrons and positrons not the linear dispersion but the square-root dispersion given in Eq.~(\ref{eq:Dirac_rel_e}). Thus, it would be more appropriate to call “the Dirac fermions” electrons in all narrow-gap materials for which the description by the two-band model applies. Further, it follows from our considerations that the orbital properties of the “Dirac fermions” in the solid state resemble those of the relativistic Dirac electrons, while their spin properties are usually different. This is particularly true of the topological insulators whose spin properties are quite unusual.

We mentioned above some semirelativistic properties and effects which have been observed for charge carriers in semiconductors but not observed for relativistic electrons in vacuum. As a consequence, the relativistic analogy has been used in the past and can be used in the future not only from the relativity to semirelativity but also in the opposite direction.

Last but not least, we should  briefly mention why the semirelativity in semiconductors is not the “true” relativity. There are two main reasons for that. First, in true relativity the highest possible velocity of particles is equal to the light velocity $c$, whereas in semiconductors the highest possible electron or hole velocity $u$ is roughly hundred times smaller than the light velocity in the crystal. This makes the standard relativity considerations of signal propagation and simultaneity of events in vacuum not valid in solids. Second, relativity is based on the equivalence of different reference frames. However, in a crystal, different frames of reference are not equivalent because the frame attached to the lattice is singled out. For this reason we did not try to apply the Lorentz transformations to the time dilatation, elimination of electric or magnetic fields, etc.  Still, some semirelativistic effects in NGS behave so similarly to the corresponding relativistic effects in vacuum that one can suspect also a similarity of more fundamental laws underlying these two seemingly different physical realities.

%% file: article.bbl
\begin{thebibliography}{117}
\expandafter\ifx\csname natexlab\endcsname\relax\def\natexlab#1{#1}\fi
\expandafter\ifx\csname bibnamefont\endcsname\relax
  \def\bibnamefont#1{#1}\fi
\expandafter\ifx\csname bibfnamefont\endcsname\relax
  \def\bibfnamefont#1{#1}\fi
\expandafter\ifx\csname citenamefont\endcsname\relax
  \def\citenamefont#1{#1}\fi
\expandafter\ifx\csname url\endcsname\relax
  \def\url#1{\texttt{#1}}\fi
\expandafter\ifx\csname urlprefix\endcsname\relax\def\urlprefix{URL }\fi
\providecommand{\bibinfo}[2]{#2}
\providecommand{\eprint}[2][]{\url{#2}}

\bibitem[{\citenamefont{Ajiki and Ando}(1993)}]{Ajiki1993}
\bibinfo{author}{\bibnamefont{Ajiki}, \bibfnamefont{H.}}, and
  \bibinfo{author}{\bibfnamefont{T.}~\bibnamefont{Ando}}, \bibinfo{year}{1993},
  \bibinfo{journal}{{{J. Phys. Soc. Jap.}}}
  \textbf{\bibinfo{volume}{62}}, \bibinfo{pages}{2470}.

\bibitem[{\citenamefont{Akhiezer and Berestetskii}(1981)}]{Akhiezer1981}
\bibinfo{author}{\bibnamefont{Akhiezer}, \bibfnamefont{A.~I.}}, and
  \bibinfo{author}{\bibfnamefont{V.~B.} \bibnamefont{Berestetskii}},
  \bibinfo{year}{1981}, \emph{\bibinfo{title}{{Quantum electrodynamics}}}
  (\bibinfo{publisher}{Nauka}, \bibinfo{address}{Moscow}), \bibinfo{edition}{4}
  edition.

\bibitem[{\citenamefont{Aronov}(1963)}]{Aronov1963}
\bibinfo{author}{\bibnamefont{Aronov}, \bibfnamefont{A.}},
  \bibinfo{year}{1963}, \bibinfo{journal}{{{Fiz. Tverd. Tela}}}
  \textbf{\bibinfo{volume}{5}}, \bibinfo{pages}{552} \bibinfo{note}{[Sov.
  Phys.--Solid State \textbf{5}, 402 (1963)]}.

\bibitem[{\citenamefont{Aronov and Pikus}(1966)}]{Aronov1967}
\bibinfo{author}{\bibnamefont{Aronov}, \bibfnamefont{A.}}, and
  \bibinfo{author}{\bibfnamefont{G.}~\bibnamefont{Pikus}},
  \bibinfo{year}{1966} \bibinfo{journal}{{{Zh.
  Eksp. Teor. Fiz.}}} \textbf{\bibinfo{volume}{51}}, \bibinfo{pages}{505}; \textbf{51}, 281
  \bibinfo{note}{[Sov. Phys.--JETP \textbf{24}, 339;
  \textbf{24}, 188 (1966)]}.

\bibitem[{\citenamefont{Aronov and Pikus}(1991)}]{Aronov1991}
\bibinfo{author}{\bibnamefont{Aronov}, \bibfnamefont{A.}}, and
  \bibinfo{author}{\bibfnamefont{G.}~\bibnamefont{Pikus}},
  \bibinfo{year}{1991}, in \emph{\bibinfo{booktitle}{Landau Level
  Spectroscopy}}, edited by
  \bibinfo{editor}{\bibfnamefont{G.}~\bibnamefont{Landwehr}} and
  \bibinfo{editor}{\bibfnamefont{E.~I.} \bibnamefont{Rashba}}
  (\bibinfo{publisher}{North-Holland}, \bibinfo{address}{Amsterdam}),
  p. \bibinfo{pages}{513}.

\bibitem[{\citenamefont{Arzeli\`{e}s}(1968)}]{Arzelies1968}
\bibinfo{author}{\bibnamefont{Arzeli\`{e}s}, \bibfnamefont{H.}},
  \bibinfo{year}{1968}, \emph{\bibinfo{title}{{Thermodynamique relativiste et
  quantique}}} (\bibinfo{publisher}{Gauthier-Villars}, \bibinfo{address}{Paris}).

\bibitem[{\citenamefont{Balatsky} \emph{et~al.}(2006)\citenamefont{Balatsky,
  Vekhter, and Zhu}}]{Balatsky2006}
\bibinfo{author}{\bibnamefont{Balatsky}, \bibfnamefont{A.~V.}},
  \bibinfo{author}{\bibfnamefont{I.}~\bibnamefont{Vekhter}}, and
  \bibinfo{author}{\bibfnamefont{J.-X.} \bibnamefont{Zhu}},
  \bibinfo{year}{2006}, \bibinfo{journal}{{{Rev. Mod.
  Phys.}}} \textbf{\bibinfo{volume}{78}}, \bibinfo{pages}{373}.

\bibitem[{\citenamefont{Barut and Bracken}(1981)}]{Barut1981}
\bibinfo{author}{\bibnamefont{Barut}, \bibfnamefont{A.~O.}}, and
  \bibinfo{author}{\bibfnamefont{A.~J.} \bibnamefont{Bracken}},
  \bibinfo{year}{1981}, \bibinfo{journal}{{{Phys. Rev. D}}}
  \textbf{\bibinfo{volume}{23}}, \bibinfo{pages}{2454}.

\bibitem[{\citenamefont{Bemski}(1960)}]{Bemski1960}
\bibinfo{author}{\bibnamefont{Bemski}, \bibfnamefont{G.}},
  \bibinfo{year}{1960}, \bibinfo{journal}{{{Phys. Rev.
  Lett.}}} \textbf{\bibinfo{volume}{4}}, \bibinfo{pages}{62}.

\bibitem[{\citenamefont{Bjorken and Drell}(1964)}]{Bjorken1964}
\bibinfo{author}{\bibnamefont{Bjorken}, \bibfnamefont{J.~D.}}, and
  \bibinfo{author}{\bibfnamefont{S.~D.} \bibnamefont{Drell}},
  \bibinfo{year}{1964}, \emph{\bibinfo{title}{{Relativistic Quantum
  Mechanics}}} (\bibinfo{publisher}{McGraw-Hill}).

\bibitem[{\citenamefont{Bowers and Yafet}(1959)}]{Bowers1959}
\bibinfo{author}{\bibnamefont{Bowers}, \bibfnamefont{R.}}, and
  \bibinfo{author}{\bibfnamefont{Y.}~\bibnamefont{Yafet}},
  \bibinfo{year}{1959}, \bibinfo{journal}{{{Phys. Rev.}}}
  \textbf{\bibinfo{volume}{115}}, \bibinfo{pages}{1165}.

\bibitem[{\citenamefont{Calawa} \emph{et~al.}(1960)\citenamefont{Calawa,
  Rediker, Lax, and McWhorter}}]{Calawa1960}
\bibinfo{author}{\bibnamefont{Calawa}, \bibfnamefont{A.~R.}},
  \bibinfo{author}{\bibfnamefont{R.~H.} \bibnamefont{Rediker}},
  \bibinfo{author}{\bibfnamefont{B.}~\bibnamefont{Lax}}, and
  \bibinfo{author}{\bibfnamefont{A.~L.} \bibnamefont{McWhorter}},
  \bibinfo{year}{1960}, \bibinfo{journal}{{{Phys. Rev.
  Lett.}}} \textbf{\bibinfo{volume}{5}}, \bibinfo{pages}{55}.

\bibitem[{\citenamefont{Cheng} \emph{et~al.}(2010)\citenamefont{Cheng, Song,
  Zhang, Zhang, Wang, Jia, Wang, Wang, Zhu, Chen, Ma, He}
  \emph{et~al.}}]{Cheng2010}
\bibinfo{author}{\bibnamefont{Cheng}, \bibfnamefont{P.}},
  \bibinfo{author}{\bibfnamefont{C.}~\bibnamefont{Song}},
  \bibinfo{author}{\bibfnamefont{T.}~\bibnamefont{Zhang}},
  \bibinfo{author}{\bibfnamefont{Y.}~\bibnamefont{Zhang}},
  \bibinfo{author}{\bibfnamefont{Y.}~\bibnamefont{Wang}},
  \bibinfo{author}{\bibfnamefont{J.-F.} \bibnamefont{Jia}},
  \bibinfo{author}{\bibfnamefont{J.}~\bibnamefont{Wang}},
  \bibinfo{author}{\bibfnamefont{Y.}~\bibnamefont{Wang}},
  \bibinfo{author}{\bibfnamefont{B.-F.} \bibnamefont{Zhu}},
  \bibinfo{author}{\bibfnamefont{X.}~\bibnamefont{Chen}},
  \bibinfo{author}{\bibfnamefont{X.}~\bibnamefont{Ma}},
  \bibinfo{author}{\bibfnamefont{K.}~\bibnamefont{He}}, \emph{et~al.},
  \bibinfo{year}{2010}, \bibinfo{journal}{{{Phys. Rev.
  Lett.}}} \textbf{\bibinfo{volume}{105}}, \bibinfo{pages}{076801}.

\bibitem[{\citenamefont{Coletti} \emph{et~al.}(2010)\citenamefont{Coletti,
  Riedl, Lee, Krauss, Patthey, von Klitzing, Smet, and Starke}}]{Coletti2010}
\bibinfo{author}{\bibnamefont{Coletti}, \bibfnamefont{C.}},
  \bibinfo{author}{\bibfnamefont{C.}~\bibnamefont{Riedl}},
  \bibinfo{author}{\bibfnamefont{D.~S.} \bibnamefont{Lee}},
  \bibinfo{author}{\bibfnamefont{B.}~\bibnamefont{Krauss}},
  \bibinfo{author}{\bibfnamefont{L.}~\bibnamefont{Patthey}},
  \bibinfo{author}{\bibfnamefont{K.}~\bibnamefont{von Klitzing}},
  \bibinfo{author}{\bibfnamefont{J.~H.} \bibnamefont{Smet}}, and
  \bibinfo{author}{\bibfnamefont{U.}~\bibnamefont{Starke}},
  \bibinfo{year}{2010}, \bibinfo{journal}{{{Phys. Rev. B}}}
  \textbf{\bibinfo{volume}{81}}, \bibinfo{pages}{235401}.

\bibitem[{\citenamefont{Conley and Mahan}(1967)}]{Conley1967}
\bibinfo{author}{\bibnamefont{Conley}, \bibfnamefont{J.~W.}}, and
  \bibinfo{author}{\bibfnamefont{G.~D.} \bibnamefont{Mahan}},
  \bibinfo{year}{1967}, \bibinfo{journal}{{{Phys. Rev.}}}
  \textbf{\bibinfo{volume}{161}}, \bibinfo{pages}{681}.

\bibitem[{\citenamefont{Dimmock and Wright}(1964)}]{Dimmock1964}
\bibinfo{author}{\bibnamefont{Dimmock}, \bibfnamefont{J.~O.}}, and
  \bibinfo{author}{\bibfnamefont{G.~B.} \bibnamefont{Wright}},
  \bibinfo{year}{1964}, \bibinfo{journal}{{{Phys. Rev.}}}
  \textbf{\bibinfo{volume}{135}}, \bibinfo{pages}{A 821}.

\bibitem[{\citenamefont{Dirac}(1958)}]{Dirac1958}
\bibinfo{author}{\bibnamefont{Dirac}, \bibfnamefont{P.~A.~M.}},
  \bibinfo{year}{1958}, \emph{\bibinfo{title}{{The Principles of Quantum
  Mechanics}}} (\bibinfo{publisher}{Oxford University Press},
  \bibinfo{address}{Oxford}), \bibinfo{edition}{4} edition.

\bibitem[{\citenamefont{Dornhaus and Nimtz}(1976)}]{Dornhaus1976}
\bibinfo{author}{\bibnamefont{Dornhaus}, \bibfnamefont{R.}}, and
  \bibinfo{author}{\bibfnamefont{G.}~\bibnamefont{Nimtz}},
  \bibinfo{year}{1976}, in \emph{\bibinfo{booktitle}{Solid State Physics}}
  (Springer Tracts in Modern Physics \textbf{\bibinfo{volume}{78}})
  (\bibinfo{publisher}{Springer}, \bibinfo{address}{Berlin}), p. \bibinfo{pages}{1}.


\bibitem[{\citenamefont{Dziawa} \emph{et~al.}(2012)\citenamefont{Dziawa,
  Kowalski, Dybko, Buczko, Szczerbakow, Szot, {\L}usakowska, Balasubramanian,
  Wojek, Berntsen, Tjernberg, and Story}}]{Dziawa2012}
\bibinfo{author}{\bibnamefont{Dziawa}, \bibfnamefont{P.}},
  \bibinfo{author}{\bibfnamefont{B.~J.} \bibnamefont{Kowalski}},
  \bibinfo{author}{\bibfnamefont{K.}~\bibnamefont{Dybko}},
  \bibinfo{author}{\bibfnamefont{R.}~\bibnamefont{Buczko}},
  \bibinfo{author}{\bibfnamefont{A.}~\bibnamefont{Szczerbakow}},
  \bibinfo{author}{\bibfnamefont{M.}~\bibnamefont{Szot}},
  \bibinfo{author}{\bibfnamefont{E.}~\bibnamefont{{\L}usakowska}},
  \bibinfo{author}{\bibfnamefont{T.}~\bibnamefont{Balasubramanian}},
  \bibinfo{author}{\bibfnamefont{B.~M.} \bibnamefont{Wojek}},
  \bibinfo{author}{\bibfnamefont{M.~H.} \bibnamefont{Berntsen}},
  \bibinfo{author}{\bibfnamefont{O.}~\bibnamefont{Tjernberg}}, and
  \bibinfo{author}{\bibfnamefont{T.}~\bibnamefont{Story}},
  \bibinfo{year}{2012}, \bibinfo{journal}{{{Nat. Mater.}}}
  \textbf{\bibinfo{volume}{11}}, \bibinfo{pages}{1023}.

\bibitem[{\citenamefont{Ferrari and Russo}(1990)}]{Ferrari1990}
\bibinfo{author}{\bibnamefont{Ferrari}, \bibfnamefont{L.}}, and
  \bibinfo{author}{\bibfnamefont{G.}~\bibnamefont{Russo}},
  \bibinfo{year}{1990}, \bibinfo{journal}{{{Phys. Rev. B}}}
  \textbf{\bibinfo{volume}{42}}, \bibinfo{pages}{7454}.

\bibitem[{\citenamefont{Foldy and Wouthuysen}(1950)}]{Foldy1950}
\bibinfo{author}{\bibnamefont{Foldy}, \bibfnamefont{L.~L.}}, and
  \bibinfo{author}{\bibfnamefont{S.~A.} \bibnamefont{Wouthuysen}},
  \bibinfo{year}{1950}, \bibinfo{journal}{{{Phys. Rev.}}}
  \textbf{\bibinfo{volume}{78}}, \bibinfo{pages}{29}.

\bibitem[{\citenamefont{Franz}(1958)}]{Franz1958}
\bibinfo{author}{\bibnamefont{Franz}, \bibfnamefont{W.}}, \bibinfo{year}{1958},
  \bibinfo{journal}{{{Z. Naturforsch.}}}
  \textbf{\bibinfo{volume}{13a}}, \bibinfo{pages}{484}.

\bibitem[{\citenamefont{Fu}(2011)}]{Fu2011}
\bibinfo{author}{\bibnamefont{Fu}, \bibfnamefont{L.}}, \bibinfo{year}{2011},
  \bibinfo{journal}{{{Phys. Rev. Lett.}}}
  \textbf{\bibinfo{volume}{106}}, \bibinfo{pages}{106802}.

\bibitem[{\citenamefont{Galazka and Sosnowski}(1967)}]{Galazka1967}
\bibinfo{author}{\bibnamefont{Galazka}, \bibfnamefont{R.~R.}}, and
  \bibinfo{author}{\bibfnamefont{L.}~\bibnamefont{Sosnowski}},
  \bibinfo{year}{1967}, \bibinfo{journal}{{{Phys. Status Solidi
  (b)}}} \textbf{\bibinfo{volume}{20}}, \bibinfo{pages}{113}.

\bibitem[{\citenamefont{Gerritsma} \emph{et~al.}(2010)\citenamefont{Gerritsma,
  Kirchmair, Z{\"{a}}hringer, Solano, Blatt, and Roos}}]{Gerritsma2010}
\bibinfo{author}{\bibnamefont{Gerritsma}, \bibfnamefont{R.}},
  \bibinfo{author}{\bibfnamefont{G.}~\bibnamefont{Kirchmair}},
  \bibinfo{author}{\bibfnamefont{F.}~\bibnamefont{Z{\"{a}}hringer}},
  \bibinfo{author}{\bibfnamefont{E.}~\bibnamefont{Solano}},
  \bibinfo{author}{\bibfnamefont{R.}~\bibnamefont{Blatt}}, and
  \bibinfo{author}{\bibfnamefont{C.~F.} \bibnamefont{Roos}},
  \bibinfo{year}{2010}, \bibinfo{journal}{{Nature}}
  \textbf{\bibinfo{volume}{463}}, \bibinfo{pages}{68}.

\bibitem[{\citenamefont{Ginzburg}(1979)}]{Ginzburg1979}
\bibinfo{author}{\bibnamefont{Ginzburg}, \bibfnamefont{V.~L.}},
  \bibinfo{year}{1979}, supplement in \emph{\bibinfo{booktitle}{Special Theory of
  Relativity}} by \bibinfo{editor}{\bibfnamefont{V.~A.}
  \bibnamefont{Ugarov}} (\bibinfo{publisher}{Mir}, \bibinfo{address}{Moscow}),
  p. \bibinfo{pages}{317}.

\bibitem[{\citenamefont{Greiner}(1990)}]{Greiner1990}
\bibinfo{author}{\bibnamefont{Greiner}, \bibfnamefont{W.}},
  \bibinfo{year}{1990}, \emph{\bibinfo{title}{{Relativistic Quantum Mechanics.
  Wave Equations}}} (\bibinfo{publisher}{Springer},
  \bibinfo{address}{Berlin}).

\bibitem[{\citenamefont{Greiner and Reinhardt}(1994)}]{Greiner1994}
\bibinfo{author}{\bibnamefont{Greiner}, \bibfnamefont{W.}}, and
  \bibinfo{author}{\bibfnamefont{J.}~\bibnamefont{Reinhardt}},
  \bibinfo{year}{1994}, \emph{\bibinfo{title}{{Quantum Electrodynamics}}}
  (\bibinfo{publisher}{Springer}, \bibinfo{address}{Berlin}),
  \bibinfo{edition}{2} edition.

\bibitem[{\citenamefont{Groves and Paul}(1963)}]{Groves1963}
\bibinfo{author}{\bibnamefont{Groves}, \bibfnamefont{S.}}, and
  \bibinfo{author}{\bibfnamefont{W.}~\bibnamefont{Paul}}, \bibinfo{year}{1963},
  \bibinfo{journal}{{{Phys. Rev. Lett.}}}
  \textbf{\bibinfo{volume}{11}}, \bibinfo{pages}{194}.

\bibitem[{\citenamefont{Guldner} \emph{et~al.}(1977)\citenamefont{Guldner,
  Rigaux, Mycielski, and Couder}}]{Guldner1977}
\bibinfo{author}{\bibnamefont{Guldner}, \bibfnamefont{Y.}},
  \bibinfo{author}{\bibfnamefont{C.}~\bibnamefont{Rigaux}},
  \bibinfo{author}{\bibfnamefont{A.}~\bibnamefont{Mycielski}}, and
  \bibinfo{author}{\bibfnamefont{Y.}~\bibnamefont{Couder}},
  \bibinfo{year}{1977}, \bibinfo{journal}{{{Phys. Status Solidi
  (b)}}} \textbf{\bibinfo{volume}{82}}, \bibinfo{pages}{149}.

\bibitem[{\citenamefont{Gureev} \emph{et~al.}(1978)\citenamefont{Gureev,
  Zasavitsky, Matsonashvili, and Shotov}}]{Gureev1978}
\bibinfo{author}{\bibnamefont{Gureev}, \bibfnamefont{D.}},
  \bibinfo{author}{\bibfnamefont{I.}~\bibnamefont{Zasavitsky}},
  \bibinfo{author}{\bibfnamefont{B.}~\bibnamefont{Matsonashvili}}, and
  \bibinfo{author}{\bibfnamefont{A.}~\bibnamefont{Shotov}},
  \bibinfo{year}{1978}, in \emph{\bibinfo{booktitle}{Proc. III Int. Conf. on
  Physics of Narrow Gap Semicond.}}, edited by
  \bibinfo{editor}{\bibfnamefont{J.}~\bibnamefont{Rauluszkiewicz}},
  \bibinfo{editor}{\bibfnamefont{M.}~\bibnamefont{G{\'{o}}rska}}, and
  \bibinfo{editor}{\bibfnamefont{E.}~\bibnamefont{Kaczmarek}}
  (\bibinfo{publisher}{PWN Polish Scientific Publishers},
  \bibinfo{address}{Warsaw}), p. \bibinfo{pages}{109}.

\bibitem[{\citenamefont{Guye and Lavanchy}(1915)}]{Guye1915}
\bibinfo{author}{\bibnamefont{Guye}, \bibfnamefont{C.~E.}}, and
  \bibinfo{author}{\bibfnamefont{C.}~\bibnamefont{Lavanchy}},
  \bibinfo{year}{1915}, \bibinfo{journal}{{{Compt. Rend. Acad. Sci.}}}
  \textbf{\bibinfo{volume}{161}}, \bibinfo{pages}{52}.

\bibitem[{\citenamefont{Hanaguri} \emph{et~al.}(2010)\citenamefont{Hanaguri,
  Igarashi, Kawamura, Takagi, and Sasagawa}}]{Hanaguri2010}
\bibinfo{author}{\bibnamefont{Hanaguri}, \bibfnamefont{T.}},
  \bibinfo{author}{\bibfnamefont{K.}~\bibnamefont{Igarashi}},
  \bibinfo{author}{\bibfnamefont{M.}~\bibnamefont{Kawamura}},
  \bibinfo{author}{\bibfnamefont{H.}~\bibnamefont{Takagi}}, and
  \bibinfo{author}{\bibfnamefont{T.}~\bibnamefont{Sasagawa}},
  \bibinfo{year}{2010}, \bibinfo{journal}{{{Phys. Rev. B}}}
  \textbf{\bibinfo{volume}{82}}, \bibinfo{pages}{081305}.

\bibitem[{\citenamefont{Hsieh} \emph{et~al.}(2012)\citenamefont{Hsieh, Lin,
  Liu, Duan, Bansi, and Fu}}]{Hsieh2012}
\bibinfo{author}{\bibnamefont{Hsieh}, \bibfnamefont{T.~H.}},
  \bibinfo{author}{\bibfnamefont{H.}~\bibnamefont{Lin}},
  \bibinfo{author}{\bibfnamefont{J.}~\bibnamefont{Liu}},
  \bibinfo{author}{\bibfnamefont{W.}~\bibnamefont{Duan}},
  \bibinfo{author}{\bibfnamefont{A.}~\bibnamefont{Bansi}}, and
  \bibinfo{author}{\bibfnamefont{L.}~\bibnamefont{Fu}}, \bibinfo{year}{2012},
  \bibinfo{journal}{{{Nat. Comm.}}}
  \textbf{\bibinfo{volume}{3}}, \bibinfo{pages}{982}.

\bibitem[{\citenamefont{Isaacson}(1968)}]{Isaacson1968}
\bibinfo{author}{\bibnamefont{Isaacson}, \bibfnamefont{R.~A.}},
  \bibinfo{year}{1968}, \bibinfo{journal}{{{Phys. Rev.}}}
  \textbf{\bibinfo{volume}{169}}, \bibinfo{pages}{312}.

\bibitem[{\citenamefont{Jackson}(1962)}]{Jackson1962}
\bibinfo{author}{\bibnamefont{Jackson}, \bibfnamefont{J.~D.}},
  \bibinfo{year}{1962}, \emph{\bibinfo{title}{{Classical Electrodynamics}}}
  (\bibinfo{publisher}{John Wiley {\&} Sons Ltd.}, \bibinfo{address}{New
  York}), \bibinfo{edition}{2} edition.

\bibitem[{\citenamefont{Johnson and Lippmann}(1949)}]{Johnson1949}
\bibinfo{author}{\bibnamefont{Johnson}, \bibfnamefont{M.~H.}}, and
  \bibinfo{author}{\bibfnamefont{B.~A.} \bibnamefont{Lippmann}},
  \bibinfo{year}{1949}, \bibinfo{journal}{{{Phys. Rev.}}}
  \textbf{\bibinfo{volume}{76}}, \bibinfo{pages}{828}.

\bibitem[{\citenamefont{J{\"{u}}ttner}(1911)}]{Juttner1911}
\bibinfo{author}{\bibnamefont{J{\"{u}}ttner}, \bibfnamefont{F.}},
  \bibinfo{year}{1911}, \bibinfo{journal}{{{Ann. Phys.
  (Leipzig)}}} \textbf{\bibinfo{volume}{339}}, \bibinfo{pages}{856}.

\bibitem[{\citenamefont{Kacman and Zawadzki}(1971)}]{Kacman1971}
\bibinfo{author}{\bibnamefont{Kacman}, \bibfnamefont{P.}}, and
  \bibinfo{author}{\bibfnamefont{W.}~\bibnamefont{Zawadzki}},
  \bibinfo{year}{1971}, \bibinfo{journal}{{{Phys. Status Solidi
  (b)}}} \textbf{\bibinfo{volume}{47}}, \bibinfo{pages}{629}.

\bibitem[{\citenamefont{Kane and Mele}(2005)}]{Kane2005}
\bibinfo{author}{\bibnamefont{Kane}, \bibfnamefont{C.~L.}}, and
  \bibinfo{author}{\bibfnamefont{E.~J.} \bibnamefont{Mele}},
  \bibinfo{year}{2005}, \bibinfo{journal}{{{Phys. Rev.
  Lett.}}} \textbf{\bibinfo{volume}{95}}, \bibinfo{pages}{146802}.

\bibitem[{\citenamefont{Kane}(1957)}]{Kane1957}
\bibinfo{author}{\bibnamefont{Kane}, \bibfnamefont{E.~O.}},
  \bibinfo{year}{1957}, \bibinfo{journal}{{{J. Phys.
  Chem. Solids}}} \textbf{\bibinfo{volume}{1}}, \bibinfo{pages}{249}.

\bibitem[{\citenamefont{Kaplan}(1955)}]{Kaplan1955}
\bibinfo{author}{\bibnamefont{Kaplan}, \bibfnamefont{I.}},
  \bibinfo{year}{1955}, \emph{\bibinfo{title}{{Nuclear Physics}}}
  (\bibinfo{publisher}{Addison-Wesley}, \bibinfo{address}{Reading, Mass.}).

\bibitem[{\citenamefont{Keldysh}(1958)}]{Keldysh1958}
\bibinfo{author}{\bibnamefont{Keldysh}, \bibfnamefont{L.~V.}},
  \bibinfo{year}{1958}, \bibinfo{journal}{{{Zh. Eksp. Teor.
  Fiz.}}} \textbf{\bibinfo{volume}{34}}, \bibinfo{pages}{1138}
  \bibinfo{note}{[Sov. Phys.--JETP, \textbf{7}, 788, (1958)]}.

\bibitem[{\citenamefont{Keldysh}(1963)}]{Keldysh1964}
\bibinfo{author}{\bibnamefont{Keldysh}, \bibfnamefont{L.~V.}},
  \bibinfo{year}{1963}, \bibinfo{journal}{{{Zh. Eksp. Teor.
  Fiz.}}} \textbf{\bibinfo{volume}{45}}, \bibinfo{pages}{364}
  \bibinfo{note}{[Sov. Phys.--JETP, \textbf{18}, 253, (1963)]}.

\bibitem[{\citenamefont{Kim and Narita}(1976)}]{Kim1976}
\bibinfo{author}{\bibnamefont{Kim}, \bibfnamefont{R.~S.}}, and
  \bibinfo{author}{\bibfnamefont{S.}~\bibnamefont{Narita}},
  \bibinfo{year}{1976}, \bibinfo{journal}{{{Phys. Status Solidi
  (b)}}} \textbf{\bibinfo{volume}{73}}, \bibinfo{pages}{741}.

\bibitem[{\citenamefont{Kireev}(1978)}]{Kireev1978}
\bibinfo{author}{\bibnamefont{Kireev}, \bibfnamefont{P.~S.}},
  \bibinfo{year}{1978}, \emph{\bibinfo{title}{{Semiconductor Physics}}}
  (\bibinfo{publisher}{Mir}, \bibinfo{address}{Moscow}).

\bibitem[{\citenamefont{Konczykowski}(1974)}]{Konczykowski}
\bibinfo{author}{\bibnamefont{Konczykowski}, \bibfnamefont{M.}}, 1974,
  \bibinfo{title}{{unpublished}}.

\bibitem[{\citenamefont{Kubo}(1965)}]{Kubo1965}
\bibinfo{author}{\bibnamefont{Kubo}, \bibfnamefont{R.}}, \bibinfo{year}{1965},
  \emph{\bibinfo{title}{{Statistical Mechanics}}} (\bibinfo{publisher}{North-Holland},
  \bibinfo{address}{Amsterdam}).

\bibitem[{\citenamefont{Landau and Lifshits}(1959)}]{Landau1959}
\bibinfo{author}{\bibnamefont{Landau}, \bibfnamefont{E.~M.}}, and
  \bibinfo{author}{\bibfnamefont{L.~D.} \bibnamefont{Lifshits}},
  \bibinfo{year}{1959}, \emph{\bibinfo{title}{{Classical Theory of Fields}}}
  (\bibinfo{publisher}{Addison-Wesley}, \bibinfo{address}{Reading, Mass.}).

\bibitem[{\citenamefont{Lock}(1979)}]{Lock1979}
\bibinfo{author}{\bibnamefont{Lock}, \bibfnamefont{J.~A.}},
  \bibinfo{year}{1979}, \bibinfo{journal}{{{Am. J.
  Phys.}}} \textbf{\bibinfo{volume}{47}}, \bibinfo{pages}{797}.

\bibitem[{\citenamefont{Luri{\'{e}} and Cremer}(1970)}]{Lurie1970}
\bibinfo{author}{\bibnamefont{Luri{\'{e}}}, \bibfnamefont{D.}}, and
  \bibinfo{author}{\bibfnamefont{S.}~\bibnamefont{Cremer}},
  \bibinfo{year}{1970}, \bibinfo{journal}{{Physica}}
  \textbf{\bibinfo{volume}{50}}, \bibinfo{pages}{224}.

\bibitem[{\citenamefont{Luttinger and Kohn}(1955)}]{Luttinger1955}
\bibinfo{author}{\bibnamefont{Luttinger}, \bibfnamefont{J.~M.}}, and
  \bibinfo{author}{\bibfnamefont{W.}~\bibnamefont{Kohn}}, \bibinfo{year}{1955},
  \bibinfo{journal}{{{Phys. Rev.}}}
  \textbf{\bibinfo{volume}{97}}, \bibinfo{pages}{869}.

\bibitem[{\citenamefont{McCombe} \emph{et~al.}(1970)\citenamefont{McCombe,
  Wagner, and Prinz}}]{McCombe1970}
\bibinfo{author}{\bibnamefont{McCombe}, \bibfnamefont{B.}},
  \bibinfo{author}{\bibfnamefont{R.}~\bibnamefont{Wagner}}, and
  \bibinfo{author}{\bibfnamefont{G.}~\bibnamefont{Prinz}},
  \bibinfo{year}{1970}, \bibinfo{journal}{{{Solid State
  Commun.}}} \textbf{\bibinfo{volume}{8}}, \bibinfo{pages}{1687}.

\bibitem[{\citenamefont{Nimtz and Schlicht}(1983)}]{Nimtz}
\bibinfo{author}{\bibnamefont{Nimtz}, \bibfnamefont{G.}}, and
  \bibinfo{author}{\bibfnamefont{B.}~\bibnamefont{Schlicht}},
  \bibinfo{year}{1983}, in \emph{\bibinfo{booktitle}{Narrow-Gap
  Semiconductors}} (Springer Tracts in Modern Physics
  \textbf{\bibinfo{volume}{98}}), edited by
  \bibinfo{editor}{\bibfnamefont{G.}~\bibnamefont{Hoehler}}
  (\bibinfo{publisher}{Springer}, \bibinfo{address}{Berlin}),
  p. \bibinfo{pages}{1}.

\bibitem[{\citenamefont{Novoselov} \emph{et~al.}(2004)\citenamefont{Novoselov,
  Geim, Morozov, Jiang, Zhang, Dubonos, Grigorieva, and
  Firsov}}]{Novoselov2004}
\bibinfo{author}{\bibnamefont{Novoselov}, \bibfnamefont{K.~S.}},
  \bibinfo{author}{\bibfnamefont{A.~K.} \bibnamefont{Geim}},
  \bibinfo{author}{\bibfnamefont{S.~V.} \bibnamefont{Morozov}},
  \bibinfo{author}{\bibfnamefont{D.}~\bibnamefont{Jiang}},
  \bibinfo{author}{\bibfnamefont{Y.} \bibnamefont{Zhang}},
  \bibinfo{author}{\bibfnamefont{S.~V.} \bibnamefont{Dubonos}},
  \bibinfo{author}{\bibfnamefont{I.~V.} \bibnamefont{Grigorieva}}, and
  \bibinfo{author}{\bibfnamefont{A.~A.} \bibnamefont{Firsov}},
  \bibinfo{year}{2004}, \bibinfo{journal}{{Science}}
  \textbf{\bibinfo{volume}{306}}, \bibinfo{pages}{666}.

\bibitem[{\citenamefont{Novoselov} \emph{et~al.}(2005)\citenamefont{Novoselov,
  Geim, Morozov, Jiang, Katsnelson, Grigorieva, Dubonos, and
  Firsov}}]{Novoselov2005}
\bibinfo{author}{\bibnamefont{Novoselov}, \bibfnamefont{K.~S.}},
  \bibinfo{author}{\bibfnamefont{A.~K.} \bibnamefont{Geim}},
  \bibinfo{author}{\bibfnamefont{S.~V.} \bibnamefont{Morozov}},
  \bibinfo{author}{\bibfnamefont{D.}~\bibnamefont{Jiang}},
  \bibinfo{author}{\bibfnamefont{M.~I.} \bibnamefont{Katsnelson}},
  \bibinfo{author}{\bibfnamefont{I.~V.} \bibnamefont{Grigorieva}},
  \bibinfo{author}{\bibfnamefont{S.~V.} \bibnamefont{Dubonos}}, and
  \bibinfo{author}{\bibfnamefont{A.~A.} \bibnamefont{Firsov}},
  \bibinfo{year}{2005}, \bibinfo{journal}{{Nature}}
  \textbf{\bibinfo{volume}{438}}, \bibinfo{pages}{197}.

\bibitem[{\citenamefont{Okada} \emph{et~al.}(2013)\citenamefont{Okada, Serbyn,
  Lin, Walkup, Zhou, Dhital, Neupane, Xu, Wang, Sankar, Chou, Bansil}
  \emph{et~al.}}]{Okada2013}
\bibinfo{author}{\bibnamefont{Okada}, \bibfnamefont{Y.}},
  \bibinfo{author}{\bibfnamefont{M.}~\bibnamefont{Serbyn}},
  \bibinfo{author}{\bibfnamefont{H.}~\bibnamefont{Lin}},
  \bibinfo{author}{\bibfnamefont{D.}~\bibnamefont{Walkup}},
  \bibinfo{author}{\bibfnamefont{W.}~\bibnamefont{Zhou}},
  \bibinfo{author}{\bibfnamefont{C.}~\bibnamefont{Dhital}},
  \bibinfo{author}{\bibfnamefont{M.}~\bibnamefont{Neupane}},
  \bibinfo{author}{\bibfnamefont{S.}~\bibnamefont{Xu}},
  \bibinfo{author}{\bibfnamefont{Y.~J.} \bibnamefont{Wang}},
  \bibinfo{author}{\bibfnamefont{R.}~\bibnamefont{Sankar}},
  \bibinfo{author}{\bibfnamefont{F.}~\bibnamefont{Chou}},
  \bibinfo{author}{\bibfnamefont{A.}~\bibnamefont{Bansil}}, \emph{et~al.},
  \bibinfo{year}{2013}, \bibinfo{journal}{{Science}}
  \textbf{\bibinfo{volume}{341}}, \bibinfo{pages}{1496}.

\bibitem[{\citenamefont{Padovani and Stratton}(1966)}]{Padovani1966}
\bibinfo{author}{\bibnamefont{Padovani}, \bibfnamefont{F.~A.}}, and
  \bibinfo{author}{\bibfnamefont{R.}~\bibnamefont{Stratton}},
  \bibinfo{year}{1966}, \bibinfo{journal}{{{Phys. Rev.
  Lett.}}} \textbf{\bibinfo{volume}{16}}, \bibinfo{pages}{1202}.

\bibitem[{\citenamefont{Parker and Mead}(1968)}]{Parker1968}
\bibinfo{author}{\bibnamefont{Parker}, \bibfnamefont{G.~H.}}, and
  \bibinfo{author}{\bibfnamefont{C.~A.} \bibnamefont{Mead}},
  \bibinfo{year}{1968}, \bibinfo{journal}{{{Phys. Rev.
  Lett.}}} \textbf{\bibinfo{volume}{21}}, \bibinfo{pages}{605}.

\bibitem[{\citenamefont{Pfeffer and Zawadzki}(1990)}]{Pfeffer1990}
\bibinfo{author}{\bibnamefont{Pfeffer}, \bibfnamefont{P.}}, and
  \bibinfo{author}{\bibfnamefont{W.}~\bibnamefont{Zawadzki}},
  \bibinfo{year}{1990}, \bibinfo{journal}{{{Phys. Rev. B}}}
  \textbf{\bibinfo{volume}{41}}, \bibinfo{pages}{1561}.

\bibitem[{\citenamefont{Rabi}(1928)}]{Rabi1928}
\bibinfo{author}{\bibnamefont{Rabi}, \bibfnamefont{I.~I.}},
  \bibinfo{year}{1928}, \bibinfo{journal}{{{Z. Phys.}}}
  \textbf{\bibinfo{volume}{49}}, \bibinfo{pages}{507}.

\bibitem[{\citenamefont{Reine} \emph{et~al.}(1967)\citenamefont{Reine, Vrehen,
  and Lax}}]{Reine1967}
\bibinfo{author}{\bibnamefont{Reine}, \bibfnamefont{M.}},
  \bibinfo{author}{\bibfnamefont{Q.~H.~F.} \bibnamefont{Vrehen}}, and
  \bibinfo{author}{\bibfnamefont{B.}~\bibnamefont{Lax}}, \bibinfo{year}{1967},
  \bibinfo{journal}{{{Phys. Rev.}}}
  \textbf{\bibinfo{volume}{163}}, \bibinfo{pages}{726}.

\bibitem[{\citenamefont{Rigaux}(1980)}]{Rigaux}
\bibinfo{author}{\bibnamefont{Rigaux}, \bibfnamefont{C.}},
  \bibinfo{year}{1980}, in \emph{\bibinfo{booktitle}{Narrow Gap Semiconductors,
  Physics and Applications}}, edited by
  \bibinfo{editor}{\bibfnamefont{W.}~\bibnamefont{Zawadzki}}
  (\bibinfo{publisher}{Springer}, \bibinfo{address}{Berlin}),
  p. \bibinfo{pages}{110}.

\bibitem[{\citenamefont{Rogers} \emph{et~al.}(1940)\citenamefont{Rogers,
  McReynolds, and Rogers}}]{Rogers1940}
\bibinfo{author}{\bibnamefont{Rogers}, \bibfnamefont{M.~M.}},
  \bibinfo{author}{\bibfnamefont{A.~W.} \bibnamefont{McReynolds}}, and
  \bibinfo{author}{\bibfnamefont{F.~T.} \bibnamefont{Rogers}},
  \bibinfo{year}{1940}, \bibinfo{journal}{{{Phys. Rev.}}}
  \textbf{\bibinfo{volume}{57}}, \bibinfo{pages}{379}.

\bibitem[{\citenamefont{Roth} \emph{et~al.}(1959)\citenamefont{Roth, Lax, and
  Zwerdling}}]{Roth1959}
\bibinfo{author}{\bibnamefont{Roth}, \bibfnamefont{L.~M.}},
  \bibinfo{author}{\bibfnamefont{B.}~\bibnamefont{Lax}}, and
  \bibinfo{author}{\bibfnamefont{S.}~\bibnamefont{Zwerdling}},
  \bibinfo{year}{1959}, \bibinfo{journal}{{{Phys. Rev.}}}
  \textbf{\bibinfo{volume}{114}}, \bibinfo{pages}{90}.

\bibitem[{\citenamefont{Rusin and Zawadzki}(2007)}]{Rusin2007}
\bibinfo{author}{\bibnamefont{Rusin}, \bibfnamefont{T.~M.}}, and
  \bibinfo{author}{\bibfnamefont{W.}~\bibnamefont{Zawadzki}},
  \bibinfo{year}{2007}, \bibinfo{journal}{{{Phys. Rev. B}}}
  \textbf{\bibinfo{volume}{76}}, \bibinfo{pages}{195439}.


\bibitem[{\citenamefont{Sadowski} \emph{et~al.}(2006)\citenamefont{Sadowski,
  Martinez, Potemski, Berger, and de~Heer}}]{Sadowski2006}
\bibinfo{author}{\bibnamefont{Sadowski}, \bibfnamefont{M.~L.}},
  \bibinfo{author}{\bibfnamefont{G.}~\bibnamefont{Martinez}},
  \bibinfo{author}{\bibfnamefont{M.}~\bibnamefont{Potemski}},
  \bibinfo{author}{\bibfnamefont{C.}~\bibnamefont{Berger}}, and
  \bibinfo{author}{\bibfnamefont{W.~A.} \bibnamefont{de~Heer}},
  \bibinfo{year}{2006}, \bibinfo{journal}{{{Phys. Rev.
  Lett.}}} \textbf{\bibinfo{volume}{97}}, \bibinfo{pages}{266405}.

\bibitem[{\citenamefont{Saito} \emph{et~al.}(1998)\citenamefont{Saito,
  Dresselhaus, and Dresselhaus}}]{Saito1998}
\bibinfo{author}{\bibnamefont{Saito}, \bibfnamefont{R.}},
  \bibinfo{author}{\bibfnamefont{G.}~\bibnamefont{Dresselhaus}}, and
  \bibinfo{author}{\bibfnamefont{M.~S.} \bibnamefont{Dresselhaus}},
  \bibinfo{year}{1998}, \emph{\bibinfo{title}{{Physical Properties of Carbon
  Nanotubes}}} (\bibinfo{publisher}{Imperial College Press},
  \bibinfo{address}{London}).

\bibitem[{\citenamefont{Sakurai}(1967)}]{Sakurai1967}
\bibinfo{author}{\bibnamefont{Sakurai}, \bibfnamefont{J.~J.}},
  \bibinfo{year}{1967}, \emph{\bibinfo{title}{{Advanced Quantum Mechanics}}}
  (\bibinfo{publisher}{Addison-Wesley}, \bibinfo{address}{Reading, Mass.}).

\bibitem[{\citenamefont{Sasabe}(2014)}]{Sasabe2014}
\bibinfo{author}{\bibnamefont{Sasabe}, \bibfnamefont{S.}},
  \bibinfo{year}{2014}, \bibinfo{journal}{{{J. Mod.
  Phys.}}} \textbf{\bibinfo{volume}{05}}, \bibinfo{pages}{534}.

\bibitem[{\citenamefont{Schr{\"{o}}dinger}(1930)}]{Schrodinger1930}
\bibinfo{author}{\bibnamefont{Schr{\"{o}}dinger}, \bibfnamefont{E.}},
  \bibinfo{year}{1930}, \bibinfo{journal}{{{Sitzber. Preuss.
  Akad.}}} \textbf{\bibinfo{volume}{24}}, \bibinfo{pages}{418}.
  The derivation of Zitterbewegung is reproduced by \textcite{Barut1981}.

\bibitem[{\citenamefont{Semenoff}(1984)}]{Semenoff1984}
\bibinfo{author}{\bibnamefont{Semenoff}, \bibfnamefont{G.~W.}},
  \bibinfo{year}{1984}, \bibinfo{journal}{{{Phys. Rev.
  Lett.}}} \textbf{\bibinfo{volume}{53}}, \bibinfo{pages}{2449}.

\bibitem[{\citenamefont{Slonczewski and Weiss}(1958)}]{Slonczewski1958}
\bibinfo{author}{\bibnamefont{Slonczewski}, \bibfnamefont{J.~C.}}, and
  \bibinfo{author}{\bibfnamefont{P.~R.} \bibnamefont{Weiss}},
  \bibinfo{year}{1958}, \bibinfo{journal}{{{Phys. Rev.}}}
  \textbf{\bibinfo{volume}{109}}, \bibinfo{pages}{272}.

\bibitem[{\citenamefont{Smith}(1961)}]{Smith1961}
\bibinfo{author}{\bibnamefont{Smith}, \bibfnamefont{R.~A.}},
  \bibinfo{year}{1961}, \emph{\bibinfo{title}{{Wave Mechanics of Crystalline
  Solids}}} (\bibinfo{publisher}{Chapman {\&} Hall},
  \bibinfo{address}{London}).

\bibitem[{\citenamefont{Stepanov} \emph{et~al.}(2016)\citenamefont{Stepanov,
  Ersfeld, Poshakinskiy, Lepsa, Ivchenko, Tarasenko, and
  Beschoten}}]{Stepanov2016}
\bibinfo{author}{\bibnamefont{Stepanov}, \bibfnamefont{I.}},
  \bibinfo{author}{\bibfnamefont{M.}~\bibnamefont{Ersfeld}},
  \bibinfo{author}{\bibfnamefont{A.~V.} \bibnamefont{Poshakinskiy}},
  \bibinfo{author}{\bibfnamefont{M.}~\bibnamefont{Lepsa}},
  \bibinfo{author}{\bibfnamefont{E.~L.} \bibnamefont{Ivchenko}},
  \bibinfo{author}{\bibfnamefont{S.~A.} \bibnamefont{Tarasenko}}, and
  \bibinfo{author}{\bibfnamefont{B.}~\bibnamefont{Beschoten}},
  \bibinfo{year}{2016}, \emph{\bibinfo{title}{{Coherent Electron
  Zitterbewgung}}}, \bibinfo{journal}{unpublished} .

\bibitem[{\citenamefont{Strange}(1998)}]{Strange1998}
\bibinfo{author}{\bibnamefont{Strange}, \bibfnamefont{P.}},
  \bibinfo{year}{1998}, \emph{\bibinfo{title}{{Relativistic Quantum
  Mechanics}}} (\bibinfo{publisher}{Cambridge University Press},
  \bibinfo{address}{Cambridge}).

\bibitem[{\citenamefont{Strauss}(1967)}]{Strauss1967}
\bibinfo{author}{\bibnamefont{Strauss}, \bibfnamefont{A.~J.}},
  \bibinfo{year}{1967}, \bibinfo{journal}{{{Phys. Rev.}}}
  \textbf{\bibinfo{volume}{157}}, \bibinfo{pages}{608}.

\bibitem[{\citenamefont{Synge}(1957)}]{Synge1957}
\bibinfo{author}{\bibnamefont{Synge}, \bibfnamefont{J.~L.}},
  \bibinfo{year}{1957}, \emph{\bibinfo{title}{{The Relativistic Gas}}}
  (\bibinfo{publisher}{North-Holland}, \bibinfo{address}{Amsterdam}).

\bibitem[{\citenamefont{Tajima} \emph{et~al.}(2007)\citenamefont{Tajima,
  Sugawara, Tamura, Kato, Nishio, and Kajita}}]{Tajima2007}
\bibinfo{author}{\bibnamefont{Tajima}, \bibfnamefont{N.}},
  \bibinfo{author}{\bibfnamefont{S.}~\bibnamefont{Sugawara}},
  \bibinfo{author}{\bibfnamefont{M.}~\bibnamefont{Tamura}},
  \bibinfo{author}{\bibfnamefont{R.}~\bibnamefont{Kato}},
  \bibinfo{author}{\bibfnamefont{Y.}~\bibnamefont{Nishio}}, and
  \bibinfo{author}{\bibfnamefont{K.}~\bibnamefont{Kajita}},
  \bibinfo{year}{2007}, \bibinfo{journal}{{{EuroPhys. Lett.
  (EPL)}}} \textbf{\bibinfo{volume}{80}}, \bibinfo{pages}{47002}.

\bibitem[{\citenamefont{Volkov and Pankratov}(1985)}]{Volkov1985}
\bibinfo{author}{\bibnamefont{Volkov}, \bibfnamefont{B.~A.}}, and
  \bibinfo{author}{\bibfnamefont{O.~A.} \bibnamefont{Pankratov}},
  \bibinfo{year}{1985}, \bibinfo{journal}{{{Pis'ma Zh. Eksp. Teor.
  Fiz.}}} \textbf{\bibinfo{volume}{42}}, \bibinfo{pages}{145}
  \bibinfo{note}{[JETP Letters \textbf{42}, 178 (1985)]}.

\bibitem[{\citenamefont{Vonsovskii}
  \emph{et~al.}(1990)\citenamefont{Vonsovskii, Svirskii, and
  Svirskaya}}]{Vonsovskii1990}
\bibinfo{author}{\bibnamefont{Vonsovskii}, \bibfnamefont{S.~V.}},
  \bibinfo{author}{\bibfnamefont{M.~S.} \bibnamefont{Svirskii}}, and
  \bibinfo{author}{\bibfnamefont{L.~M.} \bibnamefont{Svirskaya}},
  \bibinfo{year}{1990}, \bibinfo{journal}{{{Teor. Mat. Fiz.}}}
  \textbf{\bibinfo{volume}{85}}, \bibinfo{pages}{211}
  \bibinfo{note}{[Theor. Math. Phys. \textbf{85}, 1159 (1990)]}.

\bibitem[{\citenamefont{Vrehen} \emph{et~al.}(1967)\citenamefont{Vrehen,
  Zawadzki, and Reine}}]{Vrehen1967}
\bibinfo{author}{\bibnamefont{Vrehen}, \bibfnamefont{Q.~H.~F.}},
  \bibinfo{author}{\bibfnamefont{W.}~\bibnamefont{Zawadzki}}, and
  \bibinfo{author}{\bibfnamefont{M.}~\bibnamefont{Reine}},
  \bibinfo{year}{1967}, \bibinfo{journal}{{{Phys. Rev.}}}
  \textbf{\bibinfo{volume}{158}}, \bibinfo{pages}{702}.

\bibitem[{\citenamefont{Wallace}(1947)}]{Wallace1947}
\bibinfo{author}{\bibnamefont{Wallace}, \bibfnamefont{P.~R.}},
  \bibinfo{year}{1947}, \bibinfo{journal}{{{Phys. Rev.}}}
  \textbf{\bibinfo{volume}{71}}, \bibinfo{pages}{622}.

\bibitem[{\citenamefont{Weiler} \emph{et~al.}(1978)\citenamefont{Weiler,
  Aggarwal, and Lax}}]{Weiler1978}
\bibinfo{author}{\bibnamefont{Weiler}, \bibfnamefont{M.~H.}},
  \bibinfo{author}{\bibfnamefont{R.~L.} \bibnamefont{Aggarwal}}, and
  \bibinfo{author}{\bibfnamefont{B.}~\bibnamefont{Lax}}, \bibinfo{year}{1978},
  in \emph{\bibinfo{booktitle}{Proc. III Int. Conf. on Physics of Narrow Gap
  Semicond.}}, edited by
  \bibinfo{editor}{\bibfnamefont{J.}~\bibnamefont{Rauluszkiewicz}},
  \bibinfo{editor}{\bibfnamefont{M.}~\bibnamefont{G{\'{o}}rska}}, and
  \bibinfo{editor}{\bibfnamefont{E.}~\bibnamefont{Kaczmarek}}
  (\bibinfo{publisher}{PWN Polish Scientific Publishers},
  \bibinfo{address}{Warsaw}), p. \bibinfo{pages}{137}.

\bibitem[{\citenamefont{Weiler} \emph{et~al.}(1967)\citenamefont{Weiler,
  Zawadzki, and Lax}}]{Weiler1967}
\bibinfo{author}{\bibnamefont{Weiler}, \bibfnamefont{M.~H.}},
  \bibinfo{author}{\bibfnamefont{W.}~\bibnamefont{Zawadzki}}, and
  \bibinfo{author}{\bibfnamefont{B.}~\bibnamefont{Lax}}, \bibinfo{year}{1967},
  \bibinfo{journal}{{{Phys. Rev.}}}
  \textbf{\bibinfo{volume}{163}}, \bibinfo{pages}{733}.

\bibitem[{\citenamefont{Wilamowski} \emph{et~al.}(2010)\citenamefont{Wilamowski}}]{Wilam2010}
\bibinfo{author}{\bibnamefont{Wilamowski}, \bibfnamefont{Z.}},
  \bibinfo{author}{\bibfnamefont{W.}~\bibnamefont{Ungier}},
  \bibinfo{author}{\bibfnamefont{M.}~\bibnamefont{Havlicek}}, and
    \bibinfo{author}{\bibfnamefont{W.}~\bibnamefont{Jantsch}},
  \bibinfo{year}{2010}, \bibinfo{title}{arXiv:1001.3746v2}.

\bibitem[{\citenamefont{Wolff}(1964)}]{Wolff1964}
\bibinfo{author}{\bibnamefont{Wolff}, \bibfnamefont{P.~A.}},
  \bibinfo{year}{1964}, \bibinfo{journal}{{{Journal of Physics and
  Chemistry of Solids}}} \textbf{\bibinfo{volume}{25}}, \bibinfo{pages}{1057}.

\bibitem[{\citenamefont{Xia} \emph{et~al.}(2009)\citenamefont{Xia, Qian, Hsieh,
  Wray, Pal, Lin, Bansil, Grauer, Hor, Cava, and Hasan}}]{Xia2009}
\bibinfo{author}{\bibnamefont{Xia}, \bibfnamefont{Y.}},
  \bibinfo{author}{\bibfnamefont{D.}~\bibnamefont{Qian}},
  \bibinfo{author}{\bibfnamefont{D.}~\bibnamefont{Hsieh}},
  \bibinfo{author}{\bibfnamefont{L.}~\bibnamefont{Wray}},
  \bibinfo{author}{\bibfnamefont{A.}~\bibnamefont{Pal}},
  \bibinfo{author}{\bibfnamefont{H.}~\bibnamefont{Lin}},
  \bibinfo{author}{\bibfnamefont{A.}~\bibnamefont{Bansil}},
  \bibinfo{author}{\bibfnamefont{D.}~\bibnamefont{Grauer}},
  \bibinfo{author}{\bibfnamefont{Y.~S.} \bibnamefont{Hor}},
  \bibinfo{author}{\bibfnamefont{R.~J.} \bibnamefont{Cava}}, and
  \bibinfo{author}{\bibfnamefont{M.~Z.} \bibnamefont{Hasan}},
  \bibinfo{year}{2009}, \bibinfo{journal}{{{Nat. Phys.}}}
  \textbf{\bibinfo{volume}{5}}, \bibinfo{pages}{398}.


\bibitem[{\citenamefont{Zak and Zawadzki}(1966)}]{Zak1966}
\bibinfo{author}{\bibnamefont{Zak}, \bibfnamefont{J.}}, and
  \bibinfo{author}{\bibfnamefont{W.}~\bibnamefont{Zawadzki}},
  \bibinfo{year}{1966}, \bibinfo{journal}{{{Phys. Rev.}}}
  \textbf{\bibinfo{volume}{145}}, \bibinfo{pages}{536}.

\bibitem[{\citenamefont{Zawadzki}(1963)}]{Zawadzki1963}
\bibinfo{author}{\bibnamefont{Zawadzki}, \bibfnamefont{W.}},
  \bibinfo{year}{1963}, \bibinfo{journal}{{{Phys. Lett.}}}
  \textbf{\bibinfo{volume}{4}}, \bibinfo{pages}{190}.

\bibitem[{\citenamefont{Zawadzki}(1969)}]{Zawadzki1969}
\bibinfo{author}{\bibnamefont{Zawadzki}, \bibfnamefont{W.}},
  \bibinfo{year}{1969}, in \emph{\bibinfo{booktitle}{Physics of
  Solids in Intense Magnetic Fields}}, edited by
  \bibinfo{editor}{\bibfnamefont{E.}~\bibnamefont{Haidemenakis}}
  (\bibinfo{publisher}{Plenum Press},
  \bibinfo{address}{New York}), p. \bibinfo{pages}{311}.

\bibitem[{\citenamefont{Zawadzki}(1970)}]{Zawadzki1970}
\bibinfo{author}{\bibnamefont{Zawadzki}, \bibfnamefont{W.}},
  \bibinfo{year}{1970}, in \emph{\bibinfo{booktitle}{Optical Properties of
  Solids}}, edited by
  \bibinfo{editor}{\bibfnamefont{E.}~\bibnamefont{Haidemenakis}}
  (\bibinfo{publisher}{Gordon {\&} Breach}, \bibinfo{address}{New York}), p.
  \bibinfo{pages}{179}.

\bibitem[{\citenamefont{Zawadzki}(1971)}]{Zawadzki1971}
\bibinfo{author}{\bibnamefont{Zawadzki}, \bibfnamefont{W.}},
  \bibinfo{year}{1971}, \bibinfo{journal}{{{Phys. Rev. D}}}
  \textbf{\bibinfo{volume}{3}}, \bibinfo{pages}{1728}.

\bibitem[{\citenamefont{Zawadzki}(1974)}]{Zawadzki1974}
\bibinfo{author}{\bibnamefont{Zawadzki}, \bibfnamefont{W.}},
  \bibinfo{year}{1974}, \bibinfo{journal}{{{Adv. Phys.}}}
  \textbf{\bibinfo{volume}{23}}, \bibinfo{pages}{435}.

\bibitem[{\citenamefont{Zawadzki}(1980)}]{Zawadzki}
\bibinfo{author}{\bibnamefont{Zawadzki}, \bibfnamefont{W.}},
  \bibinfo{year}{1980}, in \emph{\bibinfo{booktitle}{Narrow Gap Semiconductors,
  Physics and Applications}}, edited by
  \bibinfo{editor}{\bibfnamefont{W.}~\bibnamefont{Zawadzki}}
  (\bibinfo{publisher}{Springer}, \bibinfo{address}{Berlin}),
  p. \bibinfo{pages}{85}.

\bibitem[{\citenamefont{Zawadzki}(1982)}]{Zawadzki1982}
\bibinfo{author}{\bibnamefont{Zawadzki}, \bibfnamefont{W.}},
  \bibinfo{year}{1982}, in \emph{\bibinfo{booktitle}{Handbook on
  Semiconductors, Vol 1: Band Theory and Transport Properties}}, edited by
  \bibinfo{editor}{\bibfnamefont{W.}~\bibnamefont{Paul}}
  (\bibinfo{publisher}{North-Holland}, \bibinfo{address}{Amsterdam}), p.
  \bibinfo{pages}{713}.

\bibitem[{\citenamefont{Zawadzki}(1991)}]{Zawadzki1991}
\bibinfo{author}{\bibnamefont{Zawadzki}, \bibfnamefont{W.}},
  \bibinfo{year}{1991}, in \emph{\bibinfo{booktitle}{Landau Level
  Spectroscopy}}, edited by
  \bibinfo{editor}{\bibfnamefont{G.}~\bibnamefont{Landwehr}} and
  \bibinfo{editor}{\bibfnamefont{E.~I.} \bibnamefont{Rashba}}
  (\bibinfo{publisher}{North-Holland}, \bibinfo{address}{Amsterdam}),
  p. \bibinfo{pages}{483}.

\bibitem[{\citenamefont{Zawadzki}(1997)}]{Zawadzki1997}
\bibinfo{author}{\bibnamefont{Zawadzki}, \bibfnamefont{W.}},
  \bibinfo{year}{1997}, in \emph{\bibinfo{booktitle}{High Magnetic Fields in
  the Physics of Semiconductors II}}, edited by
  \bibinfo{editor}{\bibfnamefont{G.}~\bibnamefont{Landwehr}} and
  \bibinfo{editor}{\bibfnamefont{W.}~\bibnamefont{Ossau}}
  (\bibinfo{publisher}{World Scientific}, \bibinfo{address}{Singapore}), p.
  \bibinfo{pages}{755}.

\bibitem[{\citenamefont{Zawadzki}(2005{\natexlab{a}})}]{Zawadzki2005a}
\bibinfo{author}{\bibnamefont{Zawadzki}, \bibfnamefont{W.}},
  \bibinfo{year}{2005}{\natexlab{a}}, \bibinfo{journal}{{{Am.
  J. Phys.}}} \textbf{\bibinfo{volume}{73}}, \bibinfo{pages}{756}.

\bibitem[{\citenamefont{Zawadzki}(2005{\natexlab{b}})}]{Zawadzki2005}
\bibinfo{author}{\bibnamefont{Zawadzki}, \bibfnamefont{W.}},
  \bibinfo{year}{2005}{\natexlab{b}}, \bibinfo{journal}{{{Phys.
  Rev. B}}} \textbf{\bibinfo{volume}{72}}, \bibinfo{pages}{085217}.

\bibitem[{\citenamefont{Zawadzki}(2006)}]{Zawadzki2006}
\bibinfo{author}{\bibnamefont{Zawadzki}, \bibfnamefont{W.}},
  \bibinfo{year}{2006}, \bibinfo{journal}{{{Phys. Rev. B}}}
  \textbf{\bibinfo{volume}{74}}, \bibinfo{pages}{205439}.

\bibitem[{\citenamefont{Zawadzki}(2013)}]{Zawadzki2013}
\bibinfo{author}{\bibnamefont{Zawadzki}, \bibfnamefont{W.}},
  \bibinfo{year}{2013}, \bibinfo{journal}{{{Acta Phys. Pol. A}}}
  \textbf{\bibinfo{volume}{123}}, \bibinfo{pages}{132}.

\bibitem[{\citenamefont{Zawadzki} \emph{et~al.}(1985)\citenamefont{Zawadzki,
  Klahn, and Merkt}}]{Zawadzki1985}
\bibinfo{author}{\bibnamefont{Zawadzki}, \bibfnamefont{W.}},
  \bibinfo{author}{\bibfnamefont{S.}~\bibnamefont{Klahn}}, and
  \bibinfo{author}{\bibfnamefont{U.}~\bibnamefont{Merkt}},
  \bibinfo{year}{1985}, \bibinfo{journal}{{{Phys. Rev.
  Lett.}}} \textbf{\bibinfo{volume}{55}}, \bibinfo{pages}{983}.

\bibitem[{\citenamefont{Zawadzki} \emph{et~al.}(1986)\citenamefont{Zawadzki,
  Klahn, and Merkt}}]{Zawadzki1986}
\bibinfo{author}{\bibnamefont{Zawadzki}, \bibfnamefont{W.}},
  \bibinfo{author}{\bibfnamefont{S.}~\bibnamefont{Klahn}}, and
  \bibinfo{author}{\bibfnamefont{U.}~\bibnamefont{Merkt}},
  \bibinfo{year}{1986}, \bibinfo{journal}{{{Phys. Rev. B}}}
  \textbf{\bibinfo{volume}{33}}, \bibinfo{pages}{6916}.

\bibitem[{\citenamefont{Zawadzki and Ko{\l}odziejczak}(1964)}]{Zawadzki1964}
\bibinfo{author}{\bibnamefont{Zawadzki}, \bibfnamefont{W.}}, and
  \bibinfo{author}{\bibfnamefont{J.}~\bibnamefont{Ko{\l}odziejczak}},
  \bibinfo{year}{1964}, \bibinfo{journal}{{{Phys. Status Solidi
  (b)}}} \textbf{\bibinfo{volume}{6}}, \bibinfo{pages}{409}.

\bibitem[{\citenamefont{Zawadzki} \emph{et~al.}(1965)\citenamefont{Zawadzki,
  Kowalczyk, and Kolodziejczak}}]{Zawadzki1965}
\bibinfo{author}{\bibnamefont{Zawadzki}, \bibfnamefont{W.}},
  \bibinfo{author}{\bibfnamefont{R.}~\bibnamefont{Kowalczyk}}, and
  \bibinfo{author}{\bibfnamefont{J.}~\bibnamefont{Kolodziejczak}},
  \bibinfo{year}{1965}, \bibinfo{journal}{{{Phys. Status Solidi
  (b)}}} \textbf{\bibinfo{volume}{10}}, \bibinfo{pages}{513}.

\bibitem[{\citenamefont{Zawadzki and Lax}(1966)}]{Zawadzki1966}
\bibinfo{author}{\bibnamefont{Zawadzki}, \bibfnamefont{W.}}, and
  \bibinfo{author}{\bibfnamefont{B.}~\bibnamefont{Lax}}, \bibinfo{year}{1966},
  \bibinfo{journal}{{{Phys. Rev. Lett.}}}
  \textbf{\bibinfo{volume}{16}}, \bibinfo{pages}{1001}.

\bibitem[{\citenamefont{Zawadzki and Rusin}(2010)}]{Zawadzki2010}
\bibinfo{author}{\bibnamefont{Zawadzki}, \bibfnamefont{W.}}, and
  \bibinfo{author}{\bibfnamefont{T.~M.} \bibnamefont{Rusin}},
  \bibinfo{year}{2010}, \bibinfo{journal}{{{Phys. Lett. A}}}
  \textbf{\bibinfo{volume}{374}}, \bibinfo{pages}{3533}.

\bibitem[{\citenamefont{Zawadzki and Rusin}(2011)}]{Zawadzki2011}
\bibinfo{author}{\bibnamefont{Zawadzki}, \bibfnamefont{W.}}, and
  \bibinfo{author}{\bibfnamefont{T.~M.} \bibnamefont{Rusin}},
  \bibinfo{year}{2011}, \bibinfo{journal}{{{J. Phys.
  Condens. Matter}}} \textbf{\bibinfo{volume}{23}}, \bibinfo{pages}{143201}.

\bibitem[{\citenamefont{Zawadzki} \emph{et~al.}(1966)\citenamefont{Zawadzki,
  Vrehen, and Lax}}]{Zawadzki1966a}
\bibinfo{author}{\bibnamefont{Zawadzki}, \bibfnamefont{W.}},
  \bibinfo{author}{\bibfnamefont{Q.~H.~F.} \bibnamefont{Vrehen}}, and
  \bibinfo{author}{\bibfnamefont{B.}~\bibnamefont{Lax}}, \bibinfo{year}{1966},
  \bibinfo{journal}{{{Phys. Rev.}}}
  \textbf{\bibinfo{volume}{148}}, \bibinfo{pages}{849}.

\bibitem[{\citenamefont{Zhang} \emph{et~al.}(2005)\citenamefont{Zhang, Tan,
  Stormer, and Kim}}]{Zhang2005}
\bibinfo{author}{\bibnamefont{Zhang}, \bibfnamefont{Y.}},
  \bibinfo{author}{\bibfnamefont{Y.-W.} \bibnamefont{Tan}},
  \bibinfo{author}{\bibfnamefont{H.~L.} \bibnamefont{Stormer}}, and
  \bibinfo{author}{\bibfnamefont{P.}~\bibnamefont{Kim}}, \bibinfo{year}{2005},
  \bibinfo{journal}{{Nature}} \textbf{\bibinfo{volume}{438}},
  \bibinfo{pages}{201}.

\bibitem[{\citenamefont{{\'{Z}}ukoty{\'{n}}ski and
  Kolodziejczak}(1963)}]{Zukotynski1963}
\bibinfo{author}{\bibnamefont{{\'{Z}}ukoty{\'{n}}ski}, \bibfnamefont{S.}}, and
  \bibinfo{author}{\bibfnamefont{J.}~\bibnamefont{Kolodziejczak}},
  \bibinfo{year}{1963}, \bibinfo{journal}{{{Phys. Status Solidi
  (b)}}} \textbf{\bibinfo{volume}{3}}, \bibinfo{pages}{990}.

\end{thebibliography}
